\documentclass[]{IEEEtran}
\usepackage{mathrsfs}
\usepackage{amsfonts}
\usepackage{graphicx,cite,epsfig,amssymb,amsmath}
\usepackage{color,xcolor}
\definecolor{red}{rgb}{1.00, 0.00, 0.00}
\usepackage{pifont}
\usepackage{stmaryrd}
\usepackage{setspace}
\usepackage{cite}
\usepackage{array}
\usepackage{float}
\usepackage{verbatim}
\usepackage{multirow}
\usepackage[linesnumbered, ruled]{algorithm2e}
\usepackage{epstopdf}
\usepackage{mathtools}
\usepackage{diagbox}
\usepackage{cases}
\usepackage{makecell}

\usepackage{tabularx}

\usepackage{url}
\usepackage[colorlinks,linkcolor = black,anchorcolor = black,citecolor = black,urlcolor=black]{hyperref}

\providecommand{\algorithmname}{Algorithm}
\floatname{algorithm}{\protect\algorithmname}
\newcommand{\bm}[1]{\mbox{\boldmath{$#1$}}}

\usepackage{arydshln}
\allowdisplaybreaks[4]
\usepackage{booktabs}
\usepackage{svg} 
\usepackage{subcaption}
\usepackage{stfloats}

\usepackage[labelformat=simple]{subcaption}

\ifodd 1

\else

\fi 

\newcommand{\cmark}{\ding{51}}
\newcommand{\xmark}{\ding{55}}

\def\BibTeX{{\rm B\kern-.05em{\sc i\kern-.025em b}\kern-.08em T\kern-.1667em\lower.7ex\hbox{E}\kern-.125emX}}

\begin{document}
\newcommand{\name}{RespirFi\xspace}
\title{High-Fidelity and Location-Robust Respiratory Waveform Monitoring with Single-Antenna WiFi}

\author{Hefei Wang, Jianwei Liu, Yinghui He, \IEEEmembership{Member,~IEEE},\\Guanding Yu, \IEEEmembership{Senior Member,~IEEE}, and Jinsong Han, \IEEEmembership{Senior Member,~IEEE}

\thanks{
Manuscript received 22 January 2026; revised 31 March 2026; accepted 20 April 2026.
This work was supported in part by the National Natural Science Foundation of China under Grant No.~62372400, in part by the Postdoctoral Fellowship Program of CPSF under Grant No.~GZC20241488, in part by China Postdoctoral Science Foundation under Grant No.~2025M781520, in part by the ``Pioneer'' and ``Leading Goose'' R\&D Program of Zhejiang under Grant No.~2026LDC01029(JRB), and in part by the Postdoctoral Research Excellence Funding Project of Zhejiang Province under Grant No.~ZJ2025024. \textit{(Corresponding authors: Yinghui He and Jianwei Liu.)}

H. Wang, Y. He, and G. Yu are with the College of Information Science and Electronic Engineering, Zhejiang University, Hangzhou 310027, China, and also with the Zhejiang Key Laboratory of Multimodal Communication Networks and Intelligent Information Processing, Hangzhou 310027, China (email: \{22460271,  2014hyh, yuguanding\}@zju.edu.cn).

J. Liu is with the College of Information Science and Electronic Engineering, Zhejiang University, Hangzhou 310027, China, and also with the School of Information and Electrical Engineering, Hangzhou City University, Hangzhou 310015, China (email: jianweiliu@zju.edu.cn).

J. Han is with the College of Computer Science and Technology, Zhejiang University, Hangzhou 310027, China (e-mail: hanjinsong@zju.edu.cn).

Copyright (c) 2026 IEEE. Personal use of this material is permitted. However, permission to use this material for any other purposes must be obtained from the IEEE by sending a request to pubs-permissions@ieee.org.
}}

\maketitle

\begin{abstract}
In recent years, WiFi sensing has been recognized as a promising technology to bring respiratory monitoring into everyday homes, thanks to its contactless nature and ubiquitous availability. However, existing WiFi-based respiratory monitoring systems still fall short of deployment-oriented performance: they suffer from restrained hardware scalability, limited accuracy, and are highly sensitive to user location.
To overcome these limitations and push WiFi sensing towards clinically meaningful precision, we propose \name, a novel system that robustly delivers high-fidelity respiratory waveforms with WiFi Channel State Information (CSI), thereby enabling accurate estimation of key physiological biomarkers. 
At the core of \name is a theoretical human reflection model, through which we perform an in-depth characterization of how CSI variations are shaped by both subcarrier frequency and spatial user location.
Guided by these insights, we develop a location-robust waveform construction method that adaptively selects high-quality subcarriers and aligns their waveform trends, ensuring accurate waveform recovery. Furthermore, we propose a breathing phase identification method that leverages inter-subcarrier CSI differences to reliably distinguish inhalation from exhalation.
We implement \name over commodity WiFi devices, and extensive experiments demonstrate that it outperforms state-of-the-art approaches across a wide range of clinically relevant respiratory metrics.
\end{abstract}

\section{Introduction}

Respiration, a fundamental physiological process, encodes vital information about an individual's physical health and emotional well-being~\cite{ashhad2022breathing,cretikos2008respiratory}. Variations in respiratory rate, depth, and rhythm serve as important biomarkers for a broad spectrum of conditions, including cardiac disorders, pneumonia, fever, psychological stress, and cognitive overload~\cite{harrison2021interoception,tsai2011interaction}. Accordingly, accurate and reliable monitoring of respiration is essential across both clinical and everyday health contexts. Although wearable devices remain the prevailing clinical standard, their practical deployment, particularly among elderly or vulnerable populations, faces persistent challenges related to comfort, compliance, and long-term usability~\cite{DBLP:journals/health/ZhangCXZ25,zeng2019farsense}. Furthermore, in applications involving emotional or psychological assessment, the presence of wearable sensors can induce the Hawthorne effect~\cite{adair1984hawthorne}, inadvertently altering natural behavior and compromising data integrity. These limitations have spurred growing interest in non-contact sensing technologies that are less intrusive and more suitable for continuous, real-world monitoring. Among these, WiFi-based sensing emerges as a particularly promising solution as it leverages existing infrastructure, enables passive and unobtrusive operation while preserving privacy~\cite{he2023integrated}.

Over the past decade, extensive research has investigated the feasibility of respiration detection using WiFi signals. These approaches can be broadly categorized based on the signal indicator they leverage: Received Signal Strength (RSS)-based methods~\cite{abdelnasser2015ubibreathe} and Channel State Information (CSI)-based methods~\cite{liu2014wi,zeng2019farsense,wang2024wiresp,wang2016human,gui2025ralisense}. RSSI-based techniques, such as UbiBreathe~\cite{abdelnasser2015ubibreathe}, estimate respiratory rates by capturing fluctuations in signal strength caused by human respiration. However, due to the coarse measurement granularity and low-dimensional nature of RSSI, these methods struggle to support accurate long-range sensing. In contrast, CSI-based approaches offer both higher precision and extended detection range, owing to their fine-grained measurements and richer data dimensionality, encompassing both signal amplitude and phase. This enables CSI to capture subtle signal variations more effectively, thereby improving the performance of respiration detection systems. For example, Wi-Sleep~\cite{liu2014wi} and WiResP~\cite{wang2024wiresp} utilize CSI amplitude alone for monitoring respiration, while PhaseBeat~\cite{DBLP:conf/icdcs/WangYM17} and TR-BREATH~\cite{chen2017tr} exploit phase differences across receiving antennas. FarSense~\cite{zeng2019farsense} further enhances detection robustness over long distances by combining amplitude and phase information received from
multiple antennas to compute CSI ratios.


Despite the promising potential of WiFi CSI-based respiration monitoring, the practical deployment of existing methods remains severely constrained by several critical shortcomings: \textbf{1) Limited scalability.} Distinguishing between inhalation and exhalation forms the foundation for acquiring many critical respiratory biomarkers. Existing systems that achieve this, such as the CSI ratio method in FarSense~\cite{zeng2019farsense}, typically rely on multi-antenna setups to mitigate CSI phase error. However, many IoT devices, such as those based on ESP32~\cite{esp32}, lack multi-antenna capability, which significantly limits the scalability of current respiration monitoring solutions.
\textbf{2) Poor robustness.} Existing methods often require a specific location, such as within a Fresnel zone~\cite{wang2016human}, to ensure reliable performance. This constraint greatly limits deployment flexibility. Deep learning-based solutions are particularly fragile; even minor shifts in user location relative to the training setup may cause significant degradation in respiratory signal estimation~\cite{kontou2023contactless}. \textbf{3) Insufficient accuracy.} Accurate estimation of respiratory rates and waveforms is essential for meaningful health assessment. Traditional techniques typically average CSI data across all subcarriers or rely on basic dimensionality reduction methods like Principal Component Analysis (PCA)~\cite{soto2022survey}. However, subcarriers that are less sensitive to respiratory motion may contribute noise, undermining the signal quality derived from more informative subcarriers. This indiscriminate processing leads to reduced efficiency and precision in both RR estimation and respiratory waveform reconstruction.


\begin{figure}[t]
    \centering
    \setlength{\abovecaptionskip}{6pt}
    \includegraphics[width=0.95\linewidth, trim={1cm 0cm 0cm 5cm},clip]{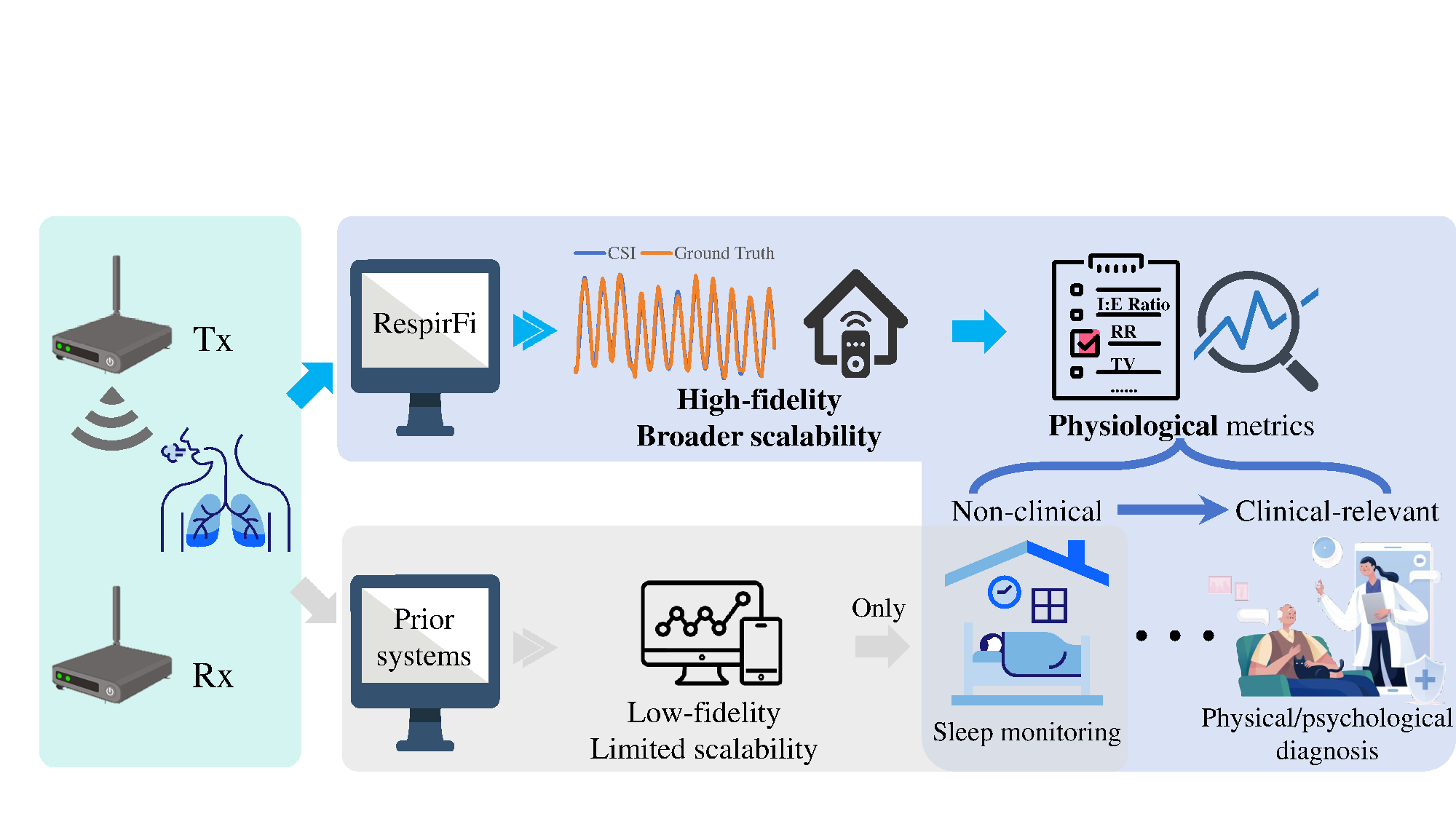}
    \vspace{-0.8em}
    \caption{\name can achieve robust and accurate respiratory biomarker measurement with broad scalability.}
    \label{fig:intro}
    \vspace{-1.5em}
\end{figure}

For effective and practical implementation of \name, we resolve the following critical challenges: \\
\noindent \textbf{1) How to improve the device scalability without introducing extra hardware costs?} The displacements caused by in/exhalation induce subtle yet intricate variations in CSI, making it challenging to differentiate between the two phases without using multi-antenna solutions to recover both CSI amplitude and phase. To address this challenge, we develop a human reflection model that explicitly characterizes how chest motion affects the CSI amplitude, 
grounded in the physical propagation principles of WiFi signals. Guided by this theoretical framework, we reveal the underlying pattern of CSI variation corresponding to the transition between inhalation and exhalation. Building on this insight, we propose a respiration phase identification algorithm that can distinguish breathing phases based on amplitude variation trends across subcarriers accurately without relying on multi-antenna,
thereby enabling the estimation of diverse physiological metrics.\\
\noindent \textbf{2) How to enable robust large-area respiration monitoring?} Traditional approaches often lack a rigorous characterization of the relationship between respiration and CSI, limiting their adaptability to variations in user location. Building on our theoretical insights, we shift the robustness challenge
from the spatial domain to the frequency domain and propose a location-robust subcarrier selection method based on spectral properties. 
By modeling how different subcarriers encode respiration, it
systematically filters out unreliable subcarriers and identifies high-quality ones across a wide frequency range, enabling robust, high-fidelity respiratory waveform recovery regardless of user position—paving the way for room-scale, location-agnostic monitoring.\\
\noindent \textbf{3) How to achieve accurate RR estimation and precise waveform reconstruction using massive, noisy, and inconsistent CSI subcarriers?} Due to hardware imperfections and environmental interference, raw CSI data often contains significant noise and irrelevant components, which hinder reliable respiration monitoring. To mitigate these issues, we first apply denoising techniques to obtain cleaner CSI signals and extract coarse respiratory waveforms. Subsequently, we align the waveform trends and integrate them through a multi-round subcarrier selection process, which systematically filters and combines high-quality subcarriers. This approach substantially improves both the accuracy and efficiency of respiratory waveform reconstruction.

We build a prototype of \name using Commercial Off-The-Shelf (COTS) WiFi devices and evaluate its performance with 30 participants. Extensive experiments demonstrate that \name not only outperforms State-Of-The-Art (SOTA) methods in RR estimation and respiration waveform reconstruction, but also enables the measurement of additional physiological biomarkers (I:E ratio, TV variability, and ApEn) beyond the capabilities of existing approaches. Robustness evaluations further show that \name can accurately monitor respiration even at distances up to seven meters, without being constrained by user location. In summary, the contributions of this paper are as follows:
\begin{itemize}
    \item We propose \name, a robust and accurate respiration monitoring system capable of comprehensively measuring respiration-related physiological biomarkers. \name maintains high performance across a wide sensing range, enabling flexible and practical deployment.
    \item We develop a signal propagation model grounded in the physics of WiFi signal for human reflection analysis. This theoretical model uncovers the relationship between user location, signal frequency, and CSI, offering potential for broader applications in human-centered wireless sensing.
    \item We prototype \name using COTS WiFi devices and evaluate it through real-world experiments involving 30 participants. Experiment results demonstrate that \name achieves accurate and comprehensive estimation of respiration metrics while enabling long-range and location-robust monitoring.
\end{itemize}

The paper is structured as follows. Related works are briefly captured in Section~\ref{sec:related_work}. Section~\ref{sec:bre_model} introduces the developed human-reflection model and highlights the insights derived from it. Section~\ref{sec:design} details the design of \name, including preliminary waveform extraction, location-robust waveform construction, and breathing phase identification. Section~\ref{sec:eva} reports the experiment setting and performance evaluation results of \name, and the whole paper is concluded in Section~\ref{sec:conclusion}.




\section{Related Work} \label{sec:related_work}
This work is primarily related to two areas of research: non-contact respiration monitoring and WiFi-based sensing.

\begin{table*}[t] 
\centering
\small
\caption{Comparison of \name with representative WiFi-based respiration monitoring systems.}
\label{tab:related}
\setlength{\tabcolsep}{6pt}
\begin{tabularx}{\textwidth}{c *{6}{>{\centering\arraybackslash}X}}
\toprule
& \multicolumn{1}{c}{\textbf{Hardware}} & \multicolumn{5}{c}{\textbf{Functions}} \\
\cmidrule(lr){2-2}\cmidrule(lr){3-7}
\textbf{Solution} 
& \textbf{Single-antenna Rx} 
& \textbf{Respiratory rate} 
& \textbf{Waveform reconstruction} 
& \textbf{Phase distinguishing} 
& \textbf{Further biomarkers} 
& \textbf{Location robustness} \\
\midrule
FarSense~\cite{zeng2019farsense}      & \xmark & \cmark & \xmark & \cmark & \xmark & \cmark\\
SMARS~\cite{zhang2019smars}           & \cmark & \cmark & \xmark & \xmark & \xmark & \cmark\\
WiResP~\cite{wang2024wiresp}          & \xmark & \cmark & \xmark & \xmark & \xmark & \cmark\\
M$^2$Fi~\cite{hu2024m}                  & \cmark & \cmark & \cmark & \xmark & \xmark & \xmark\\
\textbf{RespirFi (ours)}              & \cmark & \cmark & \cmark & \cmark & \cmark & \cmark\\
\bottomrule
\end{tabularx}
\end{table*}

\noindent \textbf{Non-contact respiration monitoring.} Conventional non-contact respiration monitoring techniques can be categorized based on the sensing medium they utilize, including camera-based~\cite{bartula2013camera,braun2018contactless}, ultra-wideband (UWB)-based~\cite{yang2019multi,DBLP:conf/sensys/ZhengCZCL21}, radar-based~\cite{wang2021driver}, and acoustic/vibration-based methods~\cite{wang2023multiresp}. However, each modality presents inherent limitations. Camera-based approaches raise privacy concerns due to visual recording. UWB and radar systems require additional dedicated hardware, increasing deployment complexity and cost. 
Acoustic approaches are intrinsically sensitive to ambient sound interference, which compromises their reliability in real-world environments.
To overcome these limitations, researchers have turned to ubiquitous WiFi signals as a promising medium for privacy-preserving, low-cost, and noise-resilient respiration monitoring. 
Existing WiFi-based methods (e.g., FarSense~\cite{zeng2019farsense}, SMARS~\cite{zhang2019smars}, WiResP~\cite{wang2024wiresp}, and M$^2$Fi~\cite{hu2024m}) exploit CSI/BFI perturbations induced by subtle chest motion to estimate respiratory rate or recover respiration waveforms.
Nevertheless, as summarized in Table~\ref{tab:related}, they exhibit gaps along hardware/configuration constraints and functional robustness.
Specifically, FarSense and WiResP rely on multi-antenna reception, which limits deployability on low-cost single-antenna receivers.
Moreover, most prior systems provide limited physiological interpretability, which often focus on respiratory rate or coarse patterns, and are less robust to user-location variations, which can compromise reliability in unconstrained real-world settings. To address these challenges, \name introduces refined signal propagation modeling to inform novel subcarrier selection and respiratory phase identification algorithms, enabling robust, high-accuracy respiration monitoring with extended support for diverse physiological biomarkers.

\noindent \textbf{WiFi-based sensing.}
With the advancement of integrated communication and sensing (ISAC)~\cite{he2023sencom,pei2025distributed,he2024forward,he2025task}, WiFi-based sensing has attracted increasing attention in recent years. The human body is often the primary target of WiFi sensing~\cite{ye2025practical}. Based on the magnitude of human motion, existing human-centric WiFi sensing tasks can be categorized into three classes: coarse-grained motion sensing~\cite{huang2018widet}, fine-grained gesture sensing~\cite{DBLP:conf/sensys/XiaoLH021}, and micro-motion sensing~\cite{wang2016human}. Coarse-grained motion sensing (e.g., fall detection~\cite{DBLP:conf/mobicom/JiangMMYWYXSMKX18} and activity recognition~\cite{DBLP:journals/tmc/ChenLJMX25}) is relatively easier to achieve, as large-scale body movements induce significant distortions in WiFi signals. In contrast, fine-grained sensing such as hand sign~\cite{DBLP:journals/imwut/MaZWZJ18} and gesture recognition~\cite{DBLP:conf/infocom/Zhao00H24} poses greater challenges due to the relatively subtle signal variations involved. Micro-motion sensing typically targets physiological monitoring applications such as heartbeat~\cite{10386433} and respiration detection~\cite{liu2014wi}. These applications are particularly difficult to implement with high performance because the displacements they rely on induce only minimal perturbations in WiFi signals. This paper focuses on respiration monitoring, which often suffers from limited metric coverage, high sensitivity to location, and poor accuracy. To address these challenges, we develop a human reflection model that captures the impact of chest movement on WiFi signals. Guided by this model, we reconstruct high-fidelity respiratory waveforms and enable accurate, location-robust estimation of multiple respiratory biomarkers.

\section{Modeling Respiratory Reflection} \label{sec:bre_model}
Understanding the impact of chest-induced reflections on CSI is essential for enabling accurate respiration monitoring. In this section, we develop a human-reflection model based on the physical principles of WiFi signal propagation. Our modeling is partly consistent with the Fresnel zone theory commonly used to interpret respiration-induced CSI variations in prior WiFi sensing studies~\cite{wang2016human}. However, from a different perspective, our model uncovers the underlying relationship between user location, signal frequency, and CSI amplitude, offering a theoretical explanation for the limitations of existing methods, specifically, their sensitivity to user location and their inability to differentiate between inhalation and exhalation phases. We further validate our theoretical insights through a series of experiments, demonstrating their validity.

\begin{figure}[t]
    \centering
    \vspace{-0.5em}
    \includegraphics[trim={0cm 11cm 17cm 0cm},clip,width=0.78\linewidth]{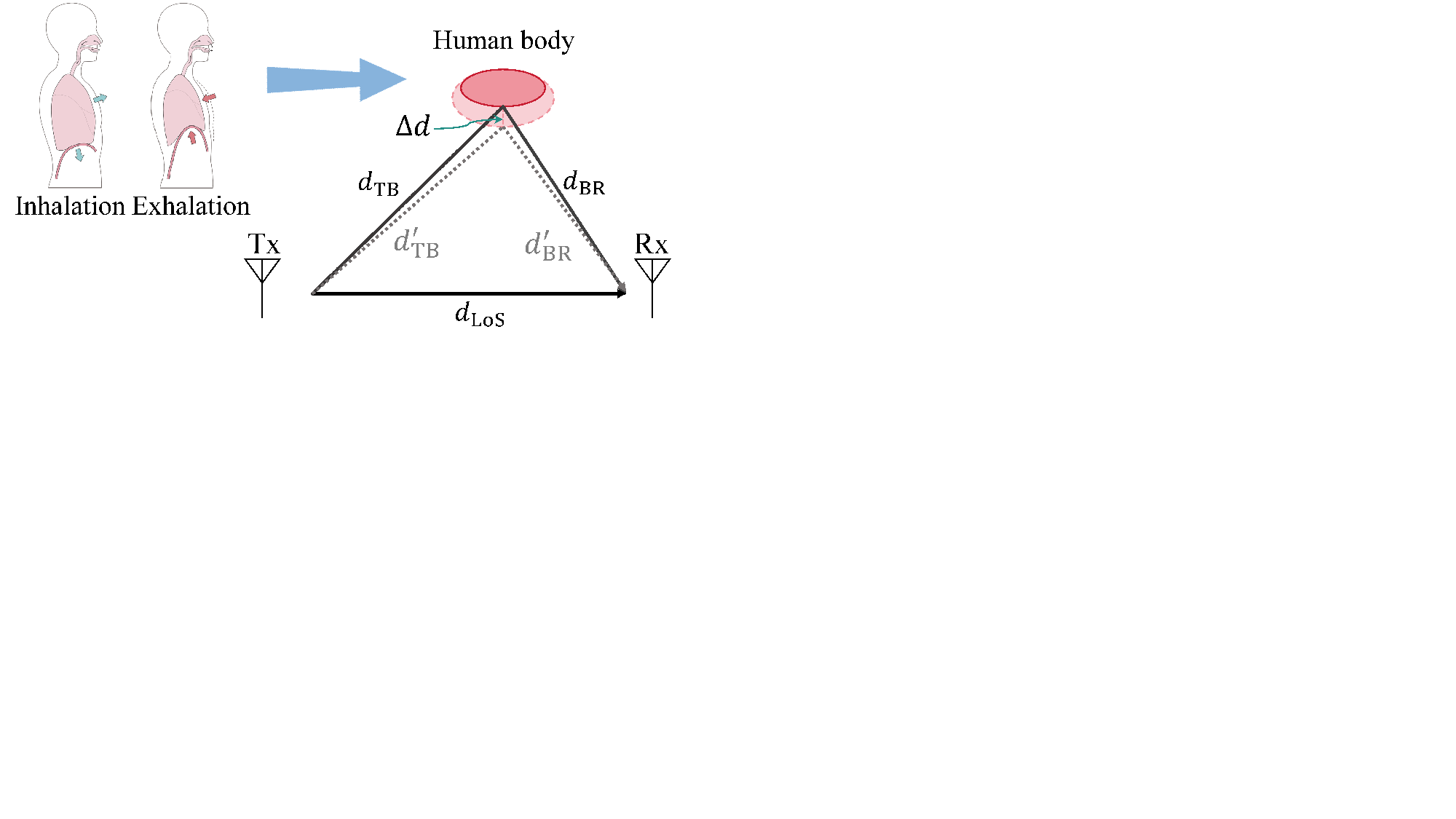}
    \vspace{-0.5em}
    \caption{Human reflection model for respiration monitoring.}
    \label{fig:model_breath}
    \vspace{-1.5em}
\end{figure}
\begin{figure*}[!t]
    \centering
    \setlength{\abovecaptionskip}{6pt}
    \begin{subfigure}{0.32\linewidth}
        \centering
        \includegraphics[width=0.95\linewidth, trim={0cm 10.5cm 21cm 0cm},clip]{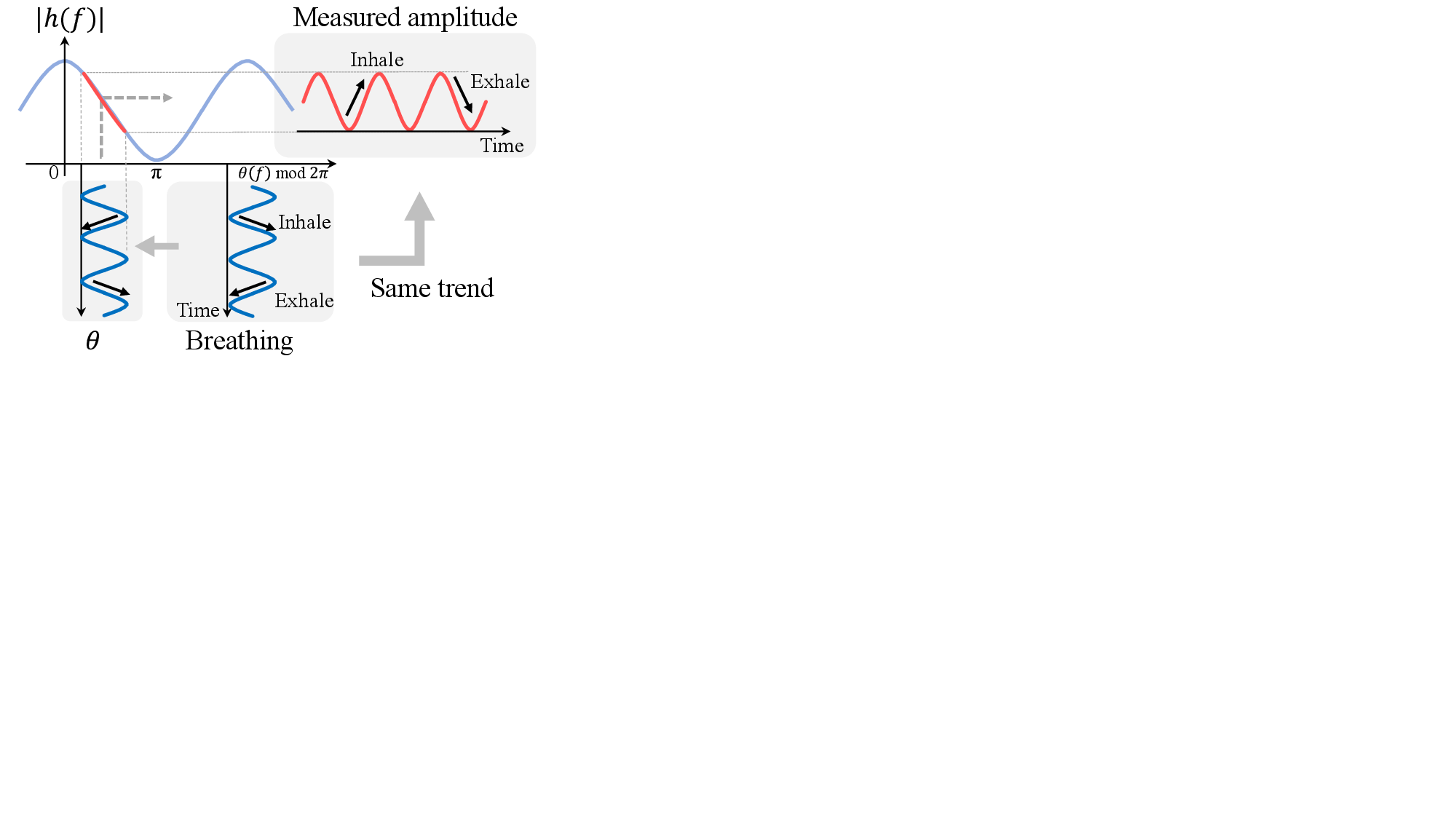}
        \vspace{-0.3em}
        \caption{Case 1.}
        \label{fig:model_case1}
    \end{subfigure}
    \begin{subfigure}{0.32\linewidth}
        \centering
        \includegraphics[width=0.95\linewidth, trim={0cm 10.5cm 21cm 0cm},clip]{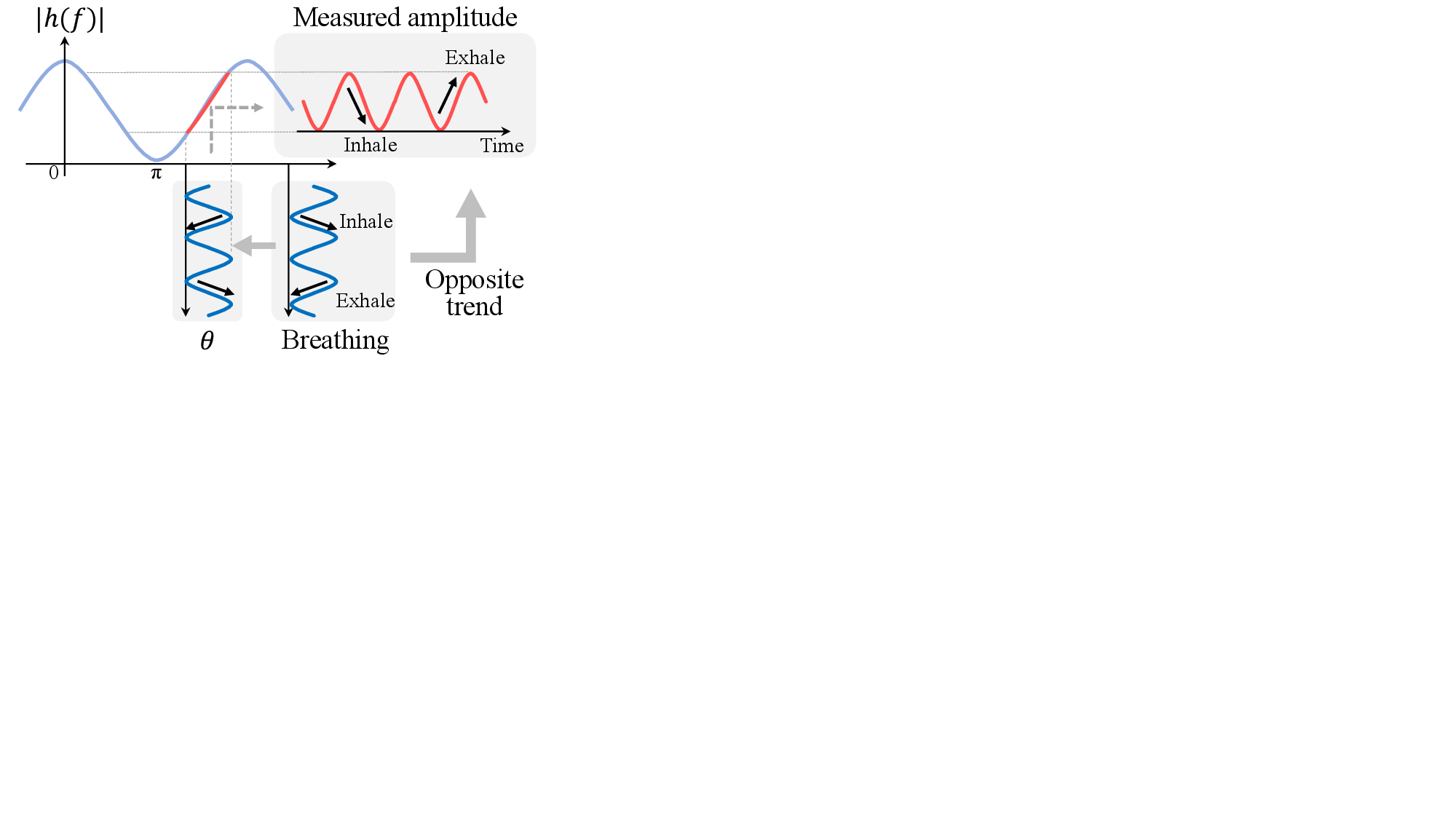}
        \vspace{-0.3em}
        \caption{Case 2.}
        \label{fig:model_case2}
    \end{subfigure}
    \begin{subfigure}{0.32\linewidth}
        \centering
        \includegraphics[width=0.95\linewidth, trim={0cm 10.5cm 21cm 0cm},clip]{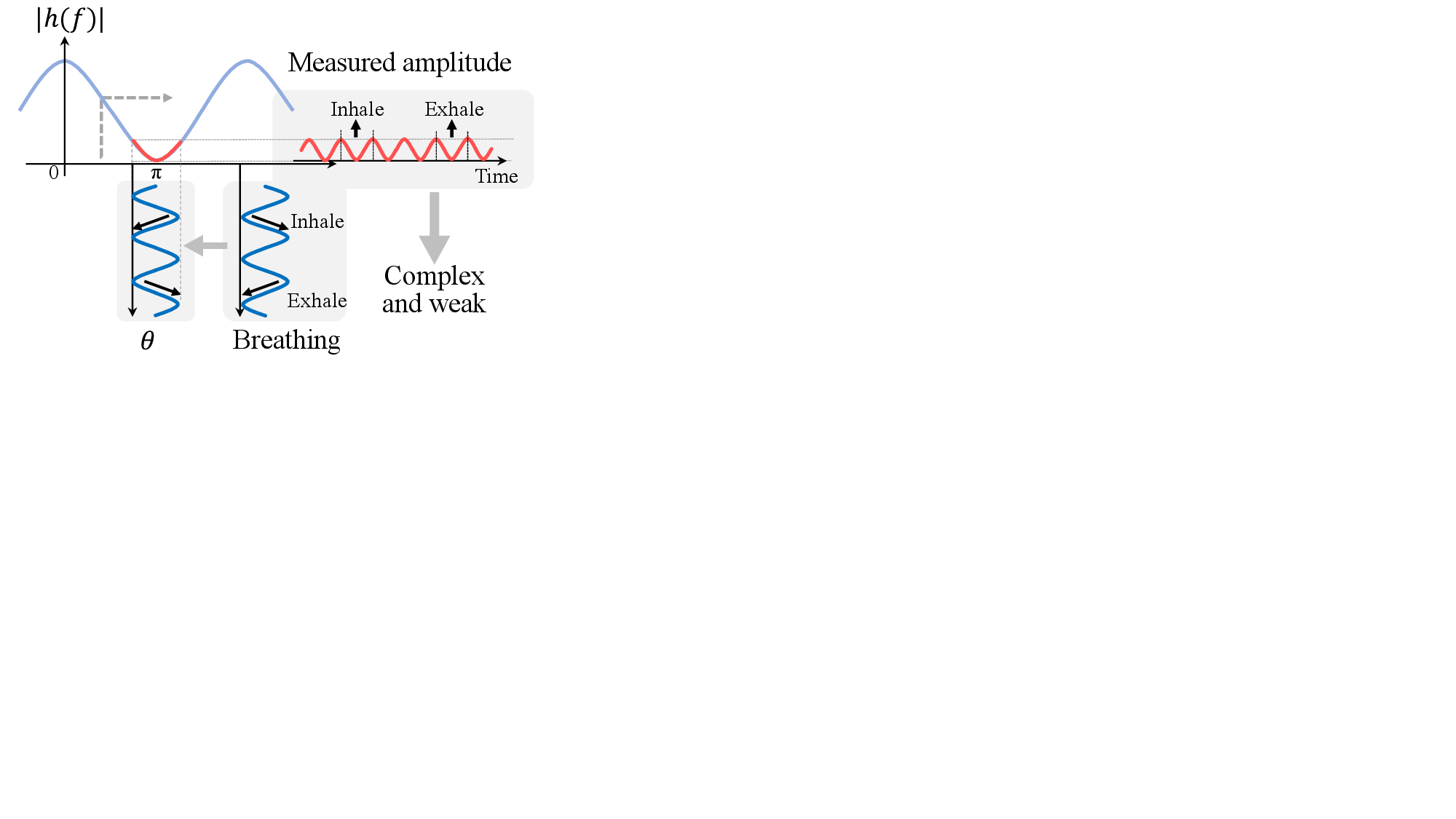}
        \vspace{-0.3em}
        \caption{Case 3.}
        \label{fig:model_case3}
    \end{subfigure}
    \vspace{-1em}
    \caption{Three cases of CSI amplitude variation caused by chest-reflected path length changes during in/exhalation.}
    \label{fig:model_relationship}
    \vspace{-3ex}
\end{figure*}

\subsection{Theoretical Model and Problem Statement} \label{sec:model}
A typical WiFi-based respiration monitoring scenario is illustrated in Figure~\ref{fig:model_breath}. The system employs a transmitter (Tx) to emit wireless signals, which propagate through multiple paths in the environment before being captured by a receiver (Rx). By comparing the known transmitted signal with the received signal, the system can estimate the CSI. Since CSI encapsulates both amplitude attenuation and phase shifts along all propagation paths (including those reflected by the chest), it can be leveraged to infer respiratory characteristics. Mathematically, with the frequency $f$, CSI can be expressed as:
\begin{equation} \label{eq:csi}
h(f) = \sum_{n=1}^{N} \alpha_n e^{-j2\pi d_nf/c},
\end{equation}
where $\alpha_n$ is the complex path attenuation of the $n$-th path, $d_n$ is the propagation distance of the $n$-th path, $c$ is the speed of light, and $N$ is the total number of paths.
In a respiration monitoring scenario, aside from the line-of-sight (LoS),  characterized by ($\alpha_{\mathrm{LoS}}, d_{\mathrm{LoS}}$), the most significant contribution to the received signal typically comes from the chest-reflected path, characterized by ($\alpha_{\mathrm{bre}}, d_{\mathrm{bre}}$), due to the relatively high permittivity of human tissue. Thus, $h(f)$ can be simplified to: $h(f)\approx \alpha_{\mathrm{LoS}} e^{-j2\pi d_{\mathrm{LoS}}f/c} + \alpha_{\mathrm{bre}} e^{-j2\pi d_{\mathrm{bre}}f/c} $, and its amplitude\footnote{It is worth noting that, for single-antenna systems, CSI phase is highly susceptible to hardware-induced impairments such as carrier frequency offset and sampling frequency offset~\cite{zhang2019smars,wang2024wiresp}, which makes it unreliable for practical use. Therefore, \name focuses exclusively on CSI amplitude for robust and scalable respiration sensing.} can be derived as 
\begin{align}
     &|h(f)|\nonumber \\ =& \sqrt{|\alpha_{\mathrm{LoS}}|^2 +|\alpha_{\mathrm{bre}}|^2 + 2 \mathcal{R}\{\alpha_{\mathrm{LoS}}\alpha^*_{\mathrm{bre}} e^{j2\pi (d_{\mathrm{bre}}-d_{\mathrm{LoS}})f/c} \} } \nonumber \\
     = & \sqrt{|\alpha_{\mathrm{LoS}}|^2 +|\alpha_{\mathrm{bre}}|^2 + 2 |\alpha_{\mathrm{LoS}}| |\alpha_{\mathrm{bre}}| \cos\theta(f)  }, \label{eq:h_abs}
\end{align}
where $|\cdot|$ denotes the absolute value operation, $\mathcal{R}\{\cdot\}$ represents the real part extraction operation, and $\theta$ represents the phase difference between the LoS path and chest-reflected path. Moreover, $\theta(f)$ can be expressed as 
\begin{equation} \label{eq:theta}
    \theta(f) = \frac{2\pi f (d_{\mathrm{bre}}-d_{\mathrm{LoS}}) }{c}+ \vartheta, 
\end{equation}
where $\vartheta = \angle{\alpha_{\mathrm{LoS}}\alpha^*_{\mathrm{bre}}}$ represents the phase difference caused by factors other than the path propagation distance, and is treated as a constant during our modeling.

As shown in Figure~\ref{fig:model_breath}, the breathing process causes movement in the chest cavity, which in turn leads to changes in the propagation path. This variation results in a phase shift, ultimately causing fluctuations in the amplitude of the received CSI signal, according to equation~\eqref{eq:h_abs}. However, since the cosine function is not monotonic, a deeper analysis of the mathematical relationship between CSI amplitude and the breathing phases (inhalation/exhalation) requires us to first determine the variation range of $\theta(f)$. According to equation~\eqref{eq:theta}, this range is determined by $2\pi f d_{\mathrm{bre}} /{c}$. 
As illustrated in Figure~\ref{fig:model_breath}, during exhalation, the movement of the chest cavity causes the distance from the Tx to the chest to change from $d_{\mathrm{TB}}$ to $d_{\mathrm{TB}}'$. These distances, along with the chest displacement $\Delta d$, form a triangular relationship. This implies that the change in distance satisfies $0<d_{\mathrm{TB}} - d_{\mathrm{TB}}' < \Delta d$. A similar conclusion can be drawn for the distance variation between the chest and the Rx. Therefore, the total path distance change caused by chest reflection is bounded by $2\Delta d$. This implies that the maximum variation of $\theta$, denoted as $\Delta \theta$, is bounded by the following range:
\begin{equation}
    \Delta \theta < 4\pi \Delta d  f/c.
\end{equation}
As reported in~\cite{lowanichkiattikul2016impact}, the chest displacement during respiration is typically 4.2$\sim$5.4~\!mm. Given that the maximum WiFi frequency reaches up to 7.1~\!GHz in the latest WiFi~7~\cite{wikilistwlanchannels}, the corresponding maximum value of $\Delta \theta $ is 0.51~\!$\pi$. 

Revisiting equation~\eqref{eq:h_abs}, when $\Delta \theta $ is smaller than the half-period $\pi$ of the cosine function, the following three cases can be observed during respiration:\\
\ding{182} $\theta(f) \bmod 2\pi$ always falls within the range $[0, \pi)$. As shown in Figure~\ref{fig:model_case1}, during exhalation, the increased chest displacement causes $d_{\text{bre}}$ to increase, leading to a larger $\theta$, and consequently a decrease in CSI amplitude $|h(f)|$, due to the monotonically decreasing nature of the cosine function in this interval. Conversely, inhalation results in an increase in $|h(f)|$.\\
\ding{183} $\theta(f) \bmod 2\pi$ always falls within the range $(\pi, 2\pi]$. As shown in Figure~\ref{fig:model_case2}, exhalation increases $d_{\text{bre}}$, which increases $\theta$, resulting in an increase in $|h(f)|$, since the cosine function is monotonically increasing in this interval. Inhalation, on the other hand, causes $|h(f)|$ to decrease. \\
\ding{184} $\theta(f) \bmod 2\pi$ spans across $\pi$ or $2\pi$. As shown in Figure~\ref{fig:model_case3}, $\cos\theta(f)$ is no longer monotonic with respect to $\theta(f)$. Additionally, the rate of change around $\pi$ is minimal. These factors introduce complex amplitude variations during respiration, making the resulting waveform harder to interpret.

Based on the above theoretical analysis, we derive several key insights that explain why prior studies struggle to distinguish between inhalation and exhalation phases solely based on single-antenna CSI amplitude and why their performance is sensitive to user location:\\
\noindent \textbf{Insight 1 — Can amplitude reveal respiratory patterns?} Chest movements during respiration introduce corresponding variations in signal amplitude. Thus, amplitude inherently encodes respiratory information and can serve as a reliable indicator for respiratory biomarker estimation. \\
\noindent \textbf{Insight 2 — Why is the performance impacted by user location?}
Changes in user location alter the phase difference between the LoS path and the chest-reflected one. For a given subcarrier, this may lead to strong, monotonic amplitude variations that clearly reflect respiratory patterns, or to weak, non-monotonic fluctuations that are easily masked by noise. Prior works often mitigate this issue by confining the user to specific spatial regions (e.g., the Fresnel zone boundary~\cite{wang2016human}). However, once the user's location changes, the sensing model may lose its effectiveness. Moreover, blindly using all subcarriers for waveform reconstruction or applying simple dimensionality reduction techniques can allow low-quality subcarriers to interfere with high-quality ones, thereby increasing the estimation error.\\
\noindent \textbf{Insight 3 — Why is it technically challenging to distinguish between inhalation and exhalation with single antenna?}
Due to the lack of synchronization between the respiration cycle and WiFi signals, the amplitude fluctuation may align with, oppose, or even mix both trends of the breathing waveform, even when the user remains stationary. This non-deterministic behavior significantly hampers the ability to infer inhalation or exhalation based solely on amplitude trends.

In this paper, we propose \name, a respiration monitoring approach that is robust to user location and capable of distinguishing between inhalation and exhalation phases. Based on the aforementioned theoretical model, \name incorporates a selection-based waveform fusion algorithm to achieve large-area, location-robust respiration monitoring (Section~\ref{sec:waveform_fusion}), and further introduces a phase identification algorithm to enable precise segmentation of inhalation and exhalation phases (Section~\ref{sec:phase_iden}), ultimately supporting comprehensive estimation of respiration-related metrics.

\begin{figure}[t]
    \centering
    \setlength{\abovecaptionskip}{6pt}
    \includegraphics[width=0.75\linewidth, trim={4cm 5cm 4cm 4cm},clip]{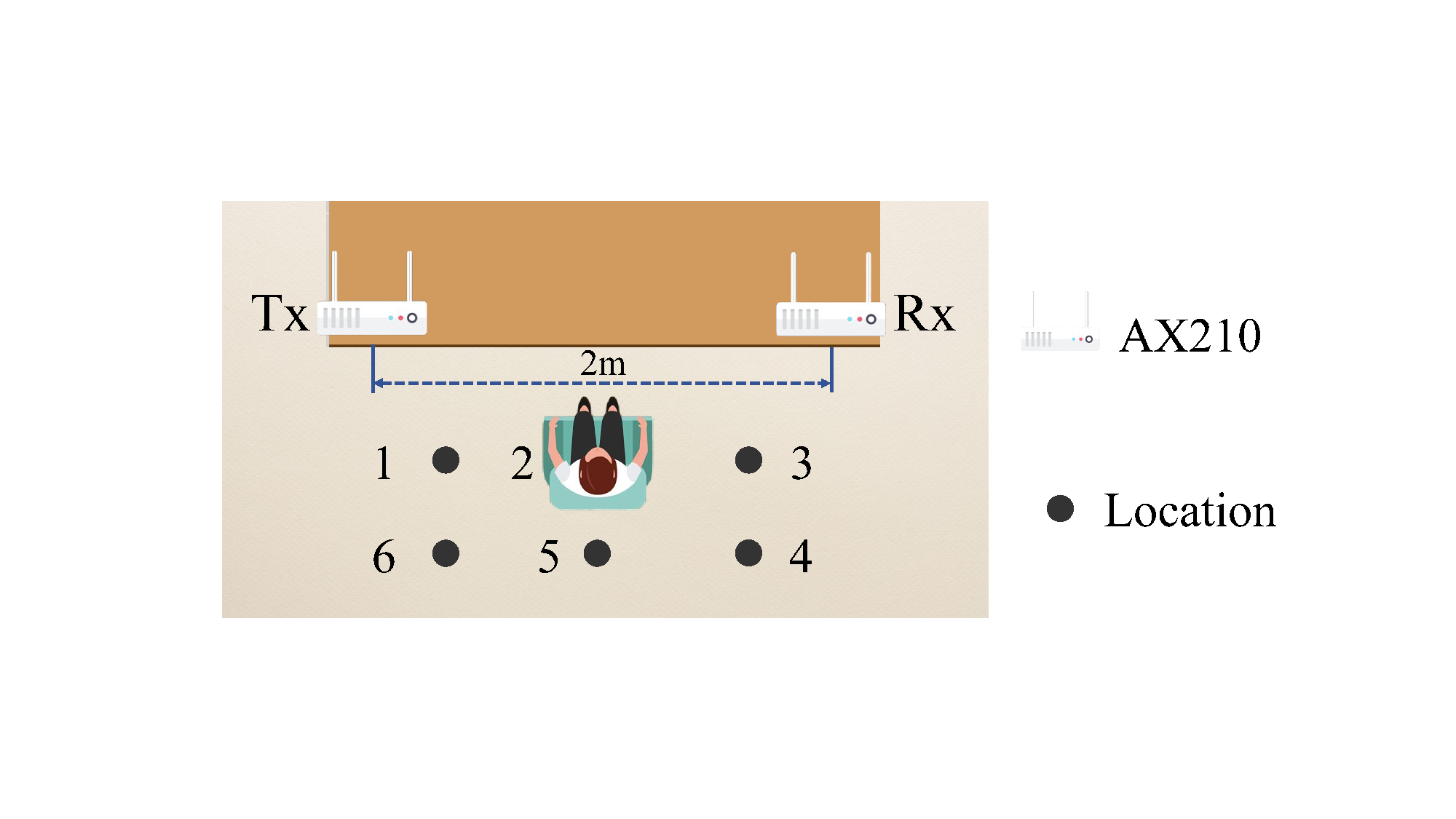}
    \vspace{-0.5em}
    \caption{Preliminary experiment with six different locations.}
    \label{fig:Preliminary}
    \centering
    \begin{subfigure}{0.49\linewidth}
        \centering
        \includegraphics[width=1\linewidth]{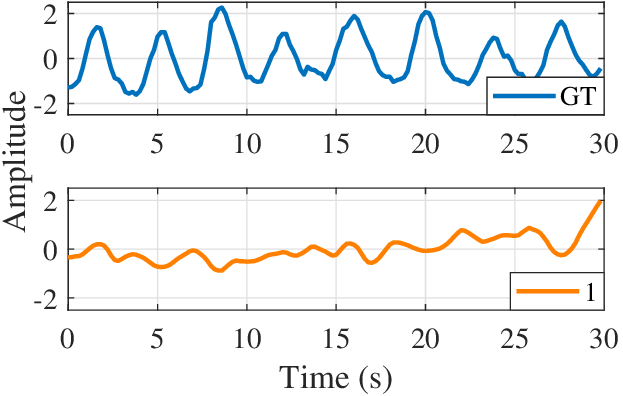}
        \vspace{-1.5em}
        \caption{Waveforms of Location 1.}
        \label{fig:pos1}
    \end{subfigure}
    \begin{subfigure}{0.49\linewidth}
        \centering
        \includegraphics[width=1\linewidth]{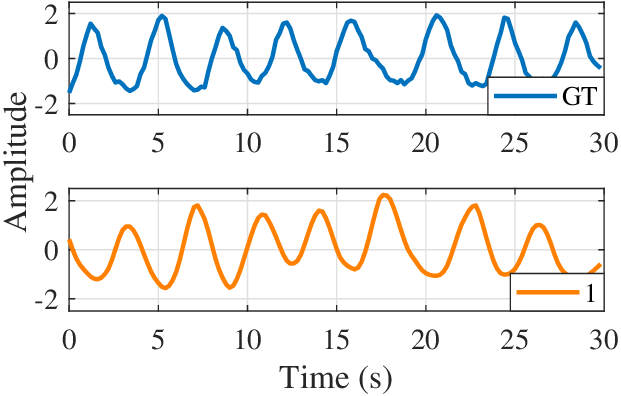}
        \vspace{-1.5em}
        \caption{Waveforms of Location 2.}
        \label{fig:pos2}
    \end{subfigure}
    \begin{subfigure}{0.49\linewidth}
        \centering
        \includegraphics[width=1\linewidth]{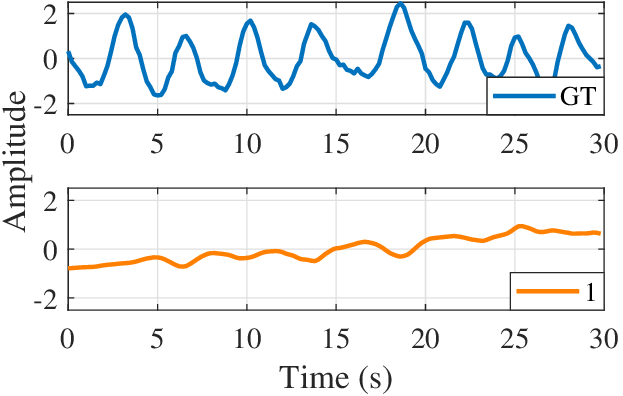}
        \vspace{-1.5em}
        \caption{Waveforms of Location 3.}
        \label{fig:pos3}
    \end{subfigure}
    \begin{subfigure}{0.49\linewidth}
        \centering
        \includegraphics[width=1\linewidth]{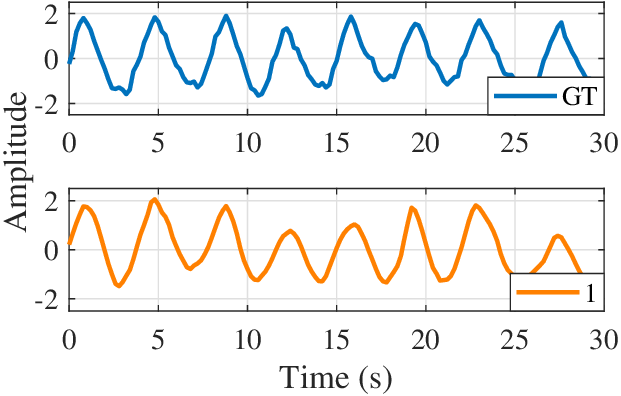}
        \vspace{-1.5em}
        \caption{Waveforms of Location 4.}
        \label{fig:pos4}
    \end{subfigure}
    \begin{subfigure}{0.49\linewidth}
        \centering
        \includegraphics[width=1\linewidth]{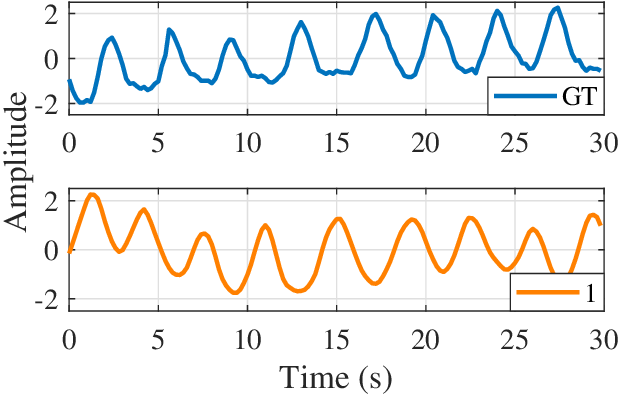}
        \vspace{-1.5em}
        \caption{Waveforms of Location 5.}
        \label{fig:pos5}
    \end{subfigure}
    \begin{subfigure}{0.49\linewidth}
        \centering
        \includegraphics[width=1\linewidth]{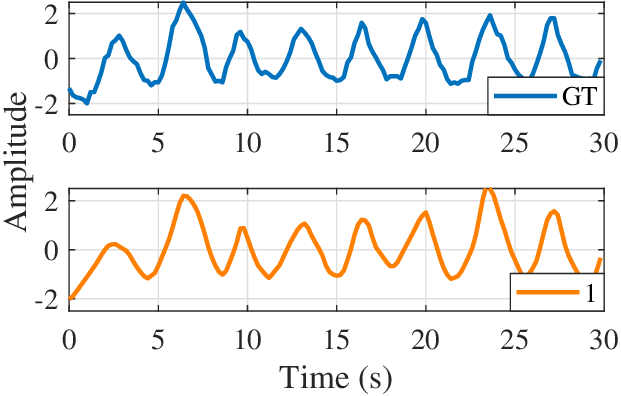}
        \vspace{-1.5em}
        \caption{Waveforms of Location 6.}
        \label{fig:pos6}
    \end{subfigure}
    \begin{subfigure}{1\linewidth}
        \centering
        \includegraphics[width=1\linewidth]{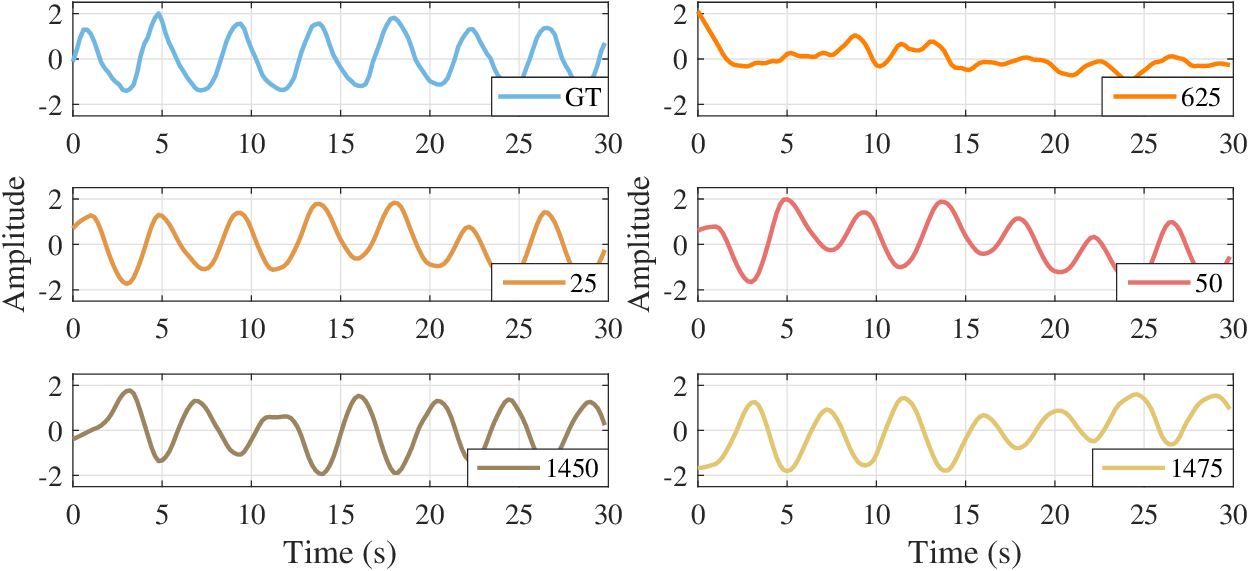}
        \vspace{-1.5em}
        \caption{Several typical subcarriers.}
        \label{fig:pre_exp_freq}
    \end{subfigure}
    \vspace{-0.7em}
    \caption{(a)-(f) Ground truth (GT) vs. waveforms measured from Subcarrier 1 in six different locations. (g) GT vs. waveforms from different subcarriers.}
    \label{fig:pre_experiment_1}
    \vspace{-1.8em}
\end{figure}

\subsection{Preliminary Experiment}
To validate the correctness of our theoretical analysis, we conduct two preliminary experiments.

As shown in Figure~\ref{fig:Preliminary}, we set up a typical respiration monitoring scenario using two Lenovo ThinkPad X201 laptops equipped with Intel AX210 network interface cards, serving as the Tx and Rx, respectively. The central frequency of the WiFi signal is set to 6025~\!MHz under WiFi~6 standard. CSI data is collected using the PicoScenes platform~\cite{jiang2021eliminating}. The Tx and Rx are placed two meters apart on a desk, and a subject is seated at one of six predefined locations, breathing normally. A respiratory belt is used to record the ground-truth respiratory waveform. At each location, we collect WiFi signals for a fixed duration and extract a coarse respiratory waveform 
(detailed in Section~\ref{sec:pre_processing}). To ensure a fair comparison, we select the subcarrier with the lowest frequency (5955~\!MHz) for waveform visualization. The extracted waveforms for all six locations are shown in Figure~\ref{fig:pre_experiment_1}. The results demonstrate that at locations 1, 3, 4, and 6, the estimated waveforms exhibit variation trends that are synchronized with the ground truth, confirming \textbf{Insight 1}—that CSI captures respiratory patterns effectively. Moreover, the polarity of waveform variations differs across locations: at locations 2 and 5 (Figures~\ref{fig:pos2} and~\ref{fig:pos5}), the estimated waveforms are inversely correlated with the ground truth; at locations 4 and 6 (Figures~\ref{fig:pos4} and~\ref{fig:pos6}), the trends align with the ground truth; while at locations 1 and 3 (Figures~\ref{fig:pos1} and~\ref{fig:pos3}), the waveforms exhibit weak or non-monotonic fluctuations. These observations correspond to the scenarios described in Case~1, Case~2, and Case~3, respectively, and preliminarily validate \textbf{Insight 2}—that user location can impact the performance of respiratory monitoring.

In the second experiment, we fix the subject's location at location~2 and utilize a 160~\!MHz bandwidth to capture CSI across a broad range of subcarrier frequencies. We select five subcarriers distributed across the channel bandwidth to illustrate the variability. As shown in Figure~\ref{fig:pre_exp_freq}, all the three cases of waveform behavior
are observed at different subcarriers, even under the same spatial configuration. Additionally, there are gradual transitions between different cases across adjacent subcarriers. These findings confirm the uncertainty in the alignment between inhalation–exhalation phases and amplitude variation trends, supporting \textbf{Insight 3}.

Taken together, these two experiments confirm the validity of our theoretical model and the correctness of the three key insights. The results also highlight the urgent need for a location-robust respiration monitoring system capable of distinguishing between inhalation and exhalation, an essential requirement for achieving high-fidelity respiratory assessment.

\section{\name Design} \label{sec:design}
In this section, we first present the overall workflow of \name and then provide an in-depth introduction to its technical implementation.

\begin{figure}[t]
    \centering
    \includegraphics[width=0.85\linewidth]{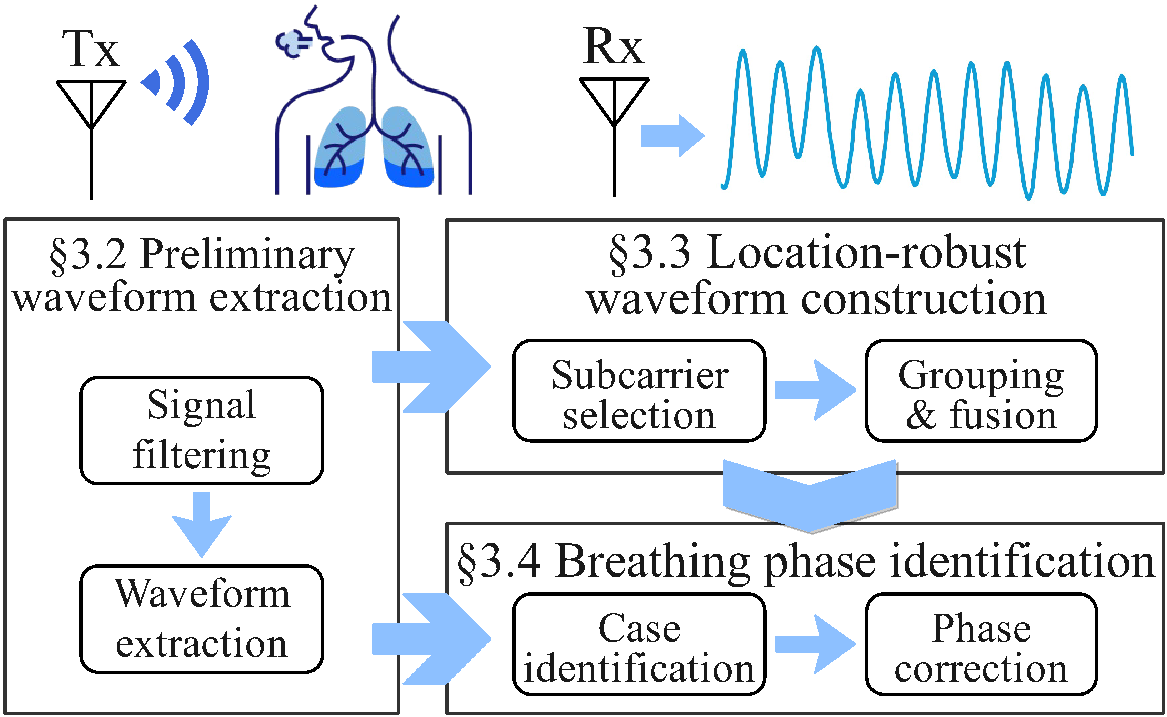}
    \vspace{-1ex}
    \caption{Overview of \name.}
    \label{fig:overview}
    \vspace{-1em}
\end{figure}

\subsection{System Overview}

As illustrated in Figure~\ref{fig:overview}, \name takes CSI amplitude time series as input, processes them through a series of stages, and ultimately outputs an accurate respiratory waveform. Specifically, the framework consists of three key components: preliminary respiratory waveform extraction, waveform fusion over subcarriers, and breathing phase identification.


\name first filters out noise components outside the respiratory frequency band from the raw CSI amplitude and extracts an initial approximation of the breathing waveform. Then, to address the unpredictable distributions of signal quality (Cases~1 to~3) caused by user location variations, 
\name identifies and removes low-quality subcarriers—most likely corresponding to Case~3—based on their spectral energy profiles, thereby preventing these components from distorting the reconstructed waveform. The remaining subcarriers are then clustered into two groups, which are assumed to correspond to Case~1 and Case~2, although the specific mapping is initially unknown. During this clustering process, additional low-quality subcarriers may also be excluded to improve signal integrity. Next, \name aligns the phase trends between the two subcarrier groups, enabling the fusion of high-quality subcarriers into a precise and location-robust respiratory waveform. Finally, to determine the breathing phase—i.e., inhalation vs. exhalation—\name analyzes the amplitude trends across contiguous subcarriers.
By leveraging these trends, it maps each group to its corresponding case (Case~1 or~2) and assigns the appropriate phase label to each segment of the waveform, such that rising segments represent inhalation and falling segments represent exhalation.



\begin{figure*}[t]
    \centering
    \setlength{\abovecaptionskip}{6pt}
\begin{minipage}[b]{0.67\textwidth}
    \begin{subfigure}{0.325\linewidth}
        \centering
        \includegraphics[width=0.98\linewidth]{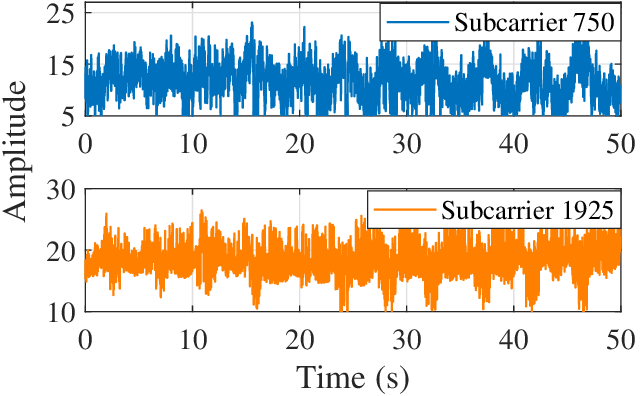}
        \vspace{-1em}
        \caption{Raw signal.}
        \label{fig:raw}
    \end{subfigure}
    \begin{subfigure}{0.325\linewidth}
        \centering
        \includegraphics[width=0.98\linewidth]{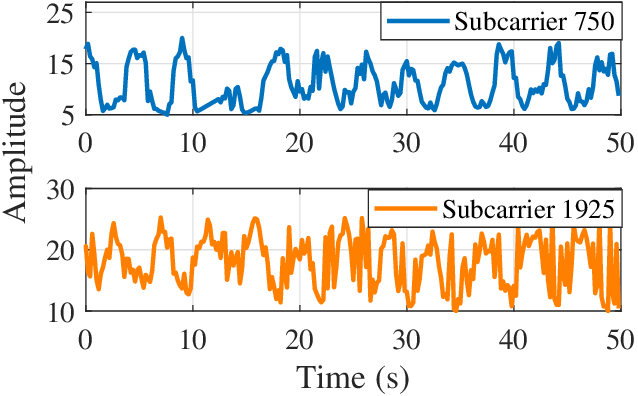}
        \vspace{-1em}
        \caption{Signal filtering.}
        \label{fig:filtering}
    \end{subfigure}
    \begin{subfigure}{0.325\linewidth}
        \centering
        \includegraphics[width=0.98\linewidth]{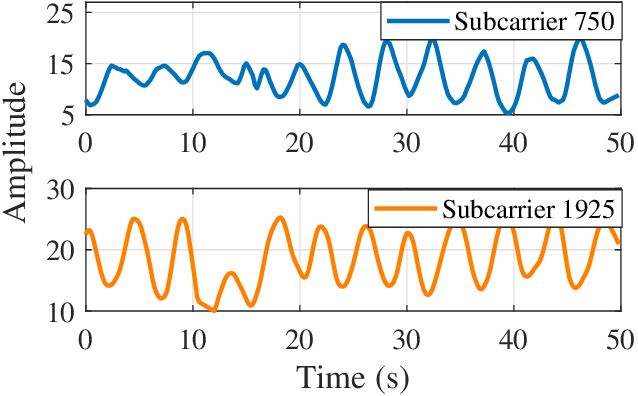}
        \vspace{-1em}
        \caption{Waveform extraction.}
        \label{fig:extraction}
    \end{subfigure}
    \vspace{-0.3em}
    \caption{Preliminary waveform extraction: (a) raw CSI signal, (b) signal after filtering, and (c) extracted preliminary breathing waveform.}
    \label{fig:preliminary_process}
    \vspace{-0.5em}
\end{minipage}
\begin{minipage}[b]{0.32\textwidth}
    \centering
    \vspace{-1.0em}
    \includegraphics[width=0.82\linewidth,clip]{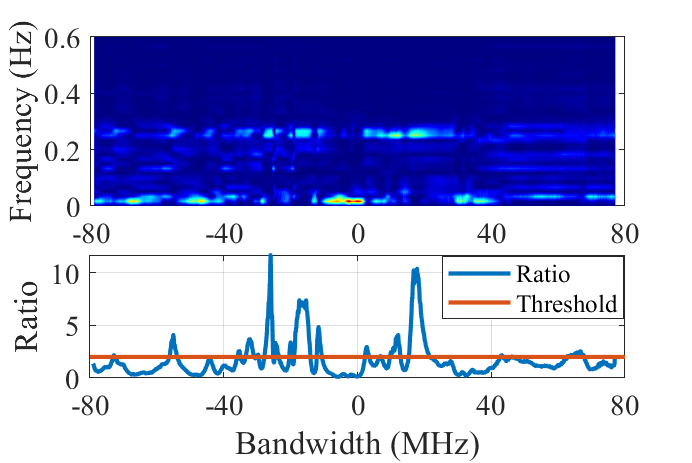}
    \vspace{0em}
    \caption{Spectral power $|H(F)|^2$ (top) and BNR (bottom) on different subcarriers.}
    \label{fig:spectral_subcarrier}
    \vspace{-0.5em}
\end{minipage}
\vspace{-1.5em}
\end{figure*}

\subsection{Preliminary Respiratory Waveform Extraction} \label{sec:pre_processing}
To obtain preliminary respiratory waveforms, we first apply filtering to remove noise unrelated to respiration. Subsequently, we smooth the amplitude profile of each subcarrier to extract coarse representations of the breathing signals.

\noindent \textbf{Signal filtering.} As shown in Figure~\ref{fig:raw}, raw amplitude streams are often contaminated by environmental and hardware-induced noise, necessitating effective filtering to preserve the low-frequency components associated with respiration (typically 0.16$\sim$0.5~\!Hz) while eliminating irrelevant fluctuations. To address this, we adopt Savitzky-Golay (SG) filtering~\cite{schafer2011savitzky}, which applies polynomial least-squares fitting to optimize the signal, with a window length of 11 samples and a polynomial order of 3. This approach effectively suppresses high-frequency noise while retaining low-frequency variations such as those induced by breathing. As shown in Figure~\ref{fig:filtering}, the application of SG filtering strikes a balance between noise suppression and signal preservation, producing cleaner amplitude sequences across subcarriers for subsequent analysis.

\noindent \textbf{Waveform extraction.} Even after filtering, residual stochastic variations and fine-grained fluctuations may obscure the macro-level temporal dynamics critical for sensing chest movements. To further refine the signal and extract a preliminary respiratory waveform for each subcarrier, we apply locally weighted scatterplot smoothing (LOWESS)~\cite{cleveland1979robust} with a span of 0.05. As a flexible non-parametric regression technique, LOWESS captures local trends without assuming a global functional form. By fitting low-order polynomials within dynamically adjusted sliding windows over each subcarrier's time series, LOWESS produces smooth and continuous waveforms that highlight the temporal evolution of the amplitude. This process effectively captures the salient time-varying signatures of chest motion embedded in the amplitude stream, resulting in a preliminary respiratory waveform as illustrated in Figure~\ref{fig:extraction}.

\subsection{Location-Robust Waveform Construction} \label{sec:waveform_fusion}
In this section, we introduce a subcarrier selection-based method to achieve location-robust waveform fusion, enabling the reconstruction of high-precision respiratory waveforms.

\subsubsection{Rationale behind Our Approach}


As revealed by Insight 2 in Section~\ref{sec:model}, the CSI fluctuations induced by respiration can vary significantly across user locations due to changes in the phase range $\theta(f)$. For a given subcarrier at a specific frequency $f_\text{spe}$, if the corresponding $\theta(f)$ range does not lie entirely within either $[0, \pi]$ or $[\pi, 2\pi]$, the resulting amplitude variation may become non-monotonic or negligible, making it difficult to capture meaningful respiratory patterns. Does this imply that respiration monitoring becomes infeasible under such conditions?

Fortunately, WiFi is inherently a wideband system with bandwidth $B$, offering multiple subcarriers across the spectrum. This enables us to shift the effective phase range $\theta(f)$ without changing the user's physical location (i.e., $d_{\mathrm{bre}}$), as indicated by the expression of $\theta(f)$ in equation~\eqref{eq:theta}. By selecting a subcarrier with a frequency offset of $\Delta f$ from $f_\text{spe}$, the entire phase range $\theta(f)$ is shifted by:
\begin{equation}
    \Delta \theta^{\mathrm{f}} = \frac{2\pi \Delta f (d_{\mathrm{bre}} - d_{\mathrm{LoS}})}{c}.
\end{equation}
This phase shift enables \name to select subcarriers whose $\theta(f)$ falls within a favorable interval (i.e., Case~1 or 2 in Figure~\ref{fig:model_case1}/~\ref{fig:model_case2}), effectively avoiding the undesirable Case~3 scenario shown in Figure~\ref{fig:model_case3}. In doing so, we can achieve location-robust monitoring by leveraging subcarrier diversity.

However, due to the limited bandwidth of WiFi, it is important to assess whether the achievable frequency shift is sufficient to enable such a phase transition in most real-world scenarios. To this end, we analyze a worst-case setting. Specifically, consider the central subcarrier, where the maximum possible frequency shift is $B/2$. As derived in Section~\ref{sec:model}, the phase variation range caused by respiration in this case is $[0.745\pi, 1.255\pi]$, with a span of $0.51\pi$. To ensure that the amplitude-based CSI can produce a high-quality respiratory waveform, we need the frequency shift to move the phase range by at least $0.255\pi$, i.e., $\Delta \theta^{\mathrm{f}} \ge 0.255\pi$. This shift would relocate the effective phase interval to either $[0.49\pi, \pi]$ or $[\pi, 1.51\pi]$, where the cosine function exhibits monotonic behavior. This requirement imposes the following constraint on the bandwidth $B$ and the geometric configuration of the user:
\begin{equation}
    B(d_{\mathrm{bre}} - d_{\mathrm{LoS}}) \ge 0.255c.
\end{equation}
Given that modern WiFi standards, such as WiFi~7, support bandwidths up to 160~\!MHz and even 320~\!MHz, this inequality indicates that as long as the path length difference between the chest-reflected and LoS components exceeds 0.47~\!m, there will exist at least one group of subcarriers capable of yielding high-quality CSI for accurate respiratory waveform recovery.

In the remainder of this subsection, we first perform an initial selection of high-quality subcarriers. Considering that the selected subcarriers may fall under different cases—some exhibiting the amplitude-increasing pattern (Case~1) and others the amplitude-decreasing pattern (Case~2)—we then unify these variations by partitioning the remaining subcarriers into two groups. The final set of subcarriers is subsequently fused to reconstruct a high-precision respiratory waveform.

\subsubsection{Subcarrier Selection}
\label{sec:subcarrier_selection}
Here, we first perform an initial selection of subcarriers to eliminate those that fall into Case~3. As analyzed in Figure~\ref{fig:model_relationship}, compared to CSI signals associated with Case~1 or Case~2, those in Case~3 exhibit significantly weaker amplitude fluctuations. 
Moreover, due to the symmetry of the cosine function, both inhalation and exhalation produce similar peaks in Case~3, resulting in a respiration-induced variation frequency that is twice the actual respiratory rate.
Consequently, CSI signals corresponding to Case~3 tend to exhibit lower energy concentration within the respiration frequency band.
We leverage this observation to filter out low-quality subcarriers. Specifically, we first normalize each subcarrier's CSI time series to have zero mean and unit variance. Then, we apply a Fast Fourier Transform (FFT) to obtain its spectral distribution $H(F)$. We adopt the breathing-to-noise ratio (BNR)~\cite{yue2018extracting} as our selection criterion, defined as the ratio between the total power within the respiration frequency band ($0.16\sim 0.5$~\!Hz) and the total power outside this band. Formally, BNR is computed as:
\begin{equation}
    P_{\mathrm{bre}} = \frac{\sum_{F \in [0.16, 0.5]} |H(F)|^2}{\sum_{F \notin [0.16, 0.5]} |H(F)|^2}.
\end{equation}
Subcarriers with $P_{\mathrm{bre}}$ exceeding a predefined threshold $\eta$ are retained as valid candidates for waveform reconstruction. Specifically, we adopt a hybrid adaptive threshold to ensure robustness:
\begin{equation}
    \eta = \max(P_{80}, 2).
\end{equation}
The $80^{th}$ percentile ($P_{80}$) allows the system to prioritize the most responsive subcarriers relative to the current environment, while the constant quality floor of $2$ prevents the inclusion of unreliable Case~3 subcarriers even in low-SNR scenarios.
As illustrated in Figure~\ref{fig:spectral_subcarrier}, different subcarriers exhibit distinct spectral profiles, and some show minimal energy within the respiration band. By removing these subcarriers, likely corresponding to Case~3, we mitigate their negative impact on the final breathing waveform reconstruction.


\subsubsection{Subcarrier Grouping and Fusion} \label{sssec:trend}
After the initial screening described in Section~\ref{sec:subcarrier_selection}, most subcarriers corresponding to Case~3 have been removed. The remaining subcarriers predominantly belong to Case~1 or Case~2, with a small number located near the boundary between Case~3 and the other two cases. In this part, we aim to achieve two goals: 1) further eliminating boundary-case subcarriers; 2) partitioning the remaining subcarriers into two groups corresponding to Case~1 and Case~2, without presupposing which group corresponds to which case. 
The first goal helps reduce residual noise and interference, while the second is critical because, as illustrated in Figure~\ref{fig:model_relationship}, subcarriers in Case~1 and Case~2 typically exhibit opposing trends with respect to actual respiration. 
Directly aggregating these subcarriers—as done in some prior works—may attenuate, rather than strengthen, the extracted breathing waveform.
Since no external ground truth is available for subcarrier labeling, we leverage \textit{inter-subcarrier similarity} to achieve the two objectives.
As analyzed in Figure~\ref{fig:model_relationship}, due to the nature of the cosine function, when $\theta(f)$ is centered around $\pi/2$ or $3\pi/2$ (i.e., within the central region of Case~1 or Case~2), the respiration-induced amplitude fluctuation reaches its peak. Assuming that noise is uniformly distributed across subcarriers, CSI waveforms in these regions are less affected by noise. 
As a result, high-quality subcarriers, which belong to Case~1 or Case~2, tend to exhibit strong correlation with one another: subcarriers within the same case are positively correlated, whereas those belonging to different cases are negatively correlated. In contrast, low-quality subcarriers, corresponding to Case~3, exhibit weak correlation with the vast majority of other subcarriers, with correlation values close to zero.

\begin{figure}[t]
    \centering
    \setlength{\abovecaptionskip}{6pt}
    \begin{subfigure}{0.49\linewidth}
        \centering
        \includegraphics[width=1\linewidth, trim={0.3cm 0.1cm 0.3cm 0.2cm},clip]{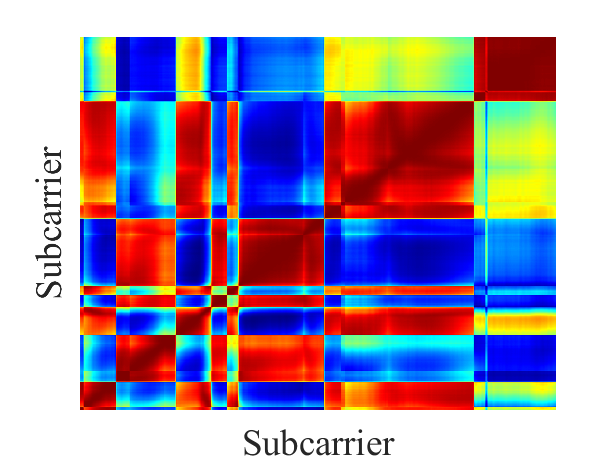}
        \vspace{-1.5em}
        \caption{Before grouping.}
        \label{fig:sim_before}
    \end{subfigure}
    \begin{subfigure}{0.49\linewidth}
        \centering
        \includegraphics[width=1\linewidth, trim={0.3cm 0.1cm 0.3cm 0.2cm},clip]{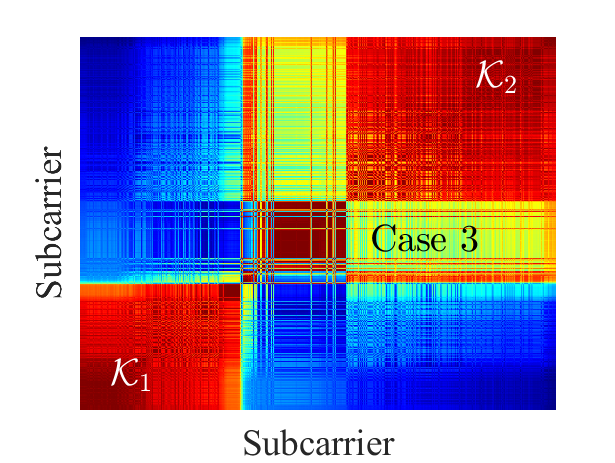}
        \vspace{-1.5em}
        \caption{After grouping.}
        \label{fig:sim_after}
    \end{subfigure}
    \vspace{-1em}
    \caption{Similarity matrix across all subcarriers (a) before and (b) after grouping.}
    \label{fig:similarity}
    \vspace{-1em}
\end{figure}

We leverage the above characteristics to cluster subcarriers into their respective cases. Since our primary goal is to mitigate random noise rather than address nonlinear interference, we adopt cosine similarity as the metric for quantifying the correlation between subcarriers. Specifically, for subcarrier $k$, let $\bm{g}(f_k)$ denote its time-domain CSI amplitude vector after preprocessing (as described in Section~\ref{sec:pre_processing}). The similarity between subcarriers $k_1$ and $k_2$ is expressed as:
\begin{equation}
w_{k_1, k_2} = \frac{\bm{g}(f_{k_1})^T \bm{g}(f_{k_2})}{||\bm{g}(f_{k_1})|| \cdot ||\bm{g}(f_{k_2})||}.
\end{equation}
By computing pairwise similarities across all subcarriers, we construct the similarity matrix $\bm{W} = [w_{k_1, k_2}]$, as illustrated in Figure~\ref{fig:sim_before}. In the figure, red regions indicate high similarity (likely same case), blue regions indicate strong negative correlation (likely different cases), and yellow regions suggest weak or no correlation (typically involving Case~3).
Given that subcarriers belonging to Case~3 are few, we aim to partition the subcarriers into two sets such that intra-group similarity is maximized and inter-group similarity is minimized. Let $x_k \in \{1, -1\}$ represent the group assignment of subcarrier $k$. The objective function is formulated as:
\begin{equation} \label{pb:max}
\max_{x_k} \sum_{k_1, k_2} \frac{x_{k_1}x_{k_2} + 1}{2} \cdot w_{k_1, k_2}
    - \sum_{k_1, k_2} \frac{1 - x_{k_1}x_{k_2}}{2} \cdot w_{k_1, k_2}.
\end{equation}
In the above, the first term rewards high similarity within groups ($x_{k_1} = x_{k_2}$), while the second penalizes high similarity across groups ($x_{k_1} \neq x_{k_2}$).

Optimization problem~\eqref{pb:max} is NP-hard, but can be efficiently approximated using heuristic methods. Given its formulation as a graph partitioning problem, we employ the widely-used spectral clustering algorithm~\cite{von2007tutorial,mercado2019spectral}, which has proven effective in domains such as social networks and knowledge graphs. Since spectral clustering requires ternary affinity values (positive, negative, or near-zero), we discretize the similarity matrix into three levels using a threshold $\eta$:
\begin{equation}
w_{k_1, k_2}' =
\begin{cases}
1, & \text{if } w_{k_1, k_2} \geq \eta, \\
-1, & \text{if } w_{k_1, k_2} \leq -\eta, \\
0, & \text{otherwise}.
\end{cases}
\end{equation}
Applying the method in~\cite{mercado2019spectral}, we partition the subcarriers into two groups, denoted as $\mathcal{K}_{1}$ and $\mathcal{K}_{2}$. However, since a few subcarriers of Case~3 may still remain, we further refine the selection. To this end, for each subcarrier $k$, we define a score $v_k$ that quantifies both its intra-group similarity and inter-group dissimilarity, a simplified form derived from the objective function~\eqref{pb:max}, as 
\begin{equation}
v_k =  \sum_{j \in \mathcal{K}_i } w_{k, j} -  \sum_{j \in \mathcal{K}_{3-i} } w_{k, j} , ~\mathrm{when}~k\in\mathcal{K}_i.
\end{equation}
Subcarriers in each group are then sorted by $v_k$, 
and only those with high similarity are retained for final waveform reconstruction. Figure~\ref{fig:sim_after} shows the similarity matrix after grouping and sorting. As can be observed, subcarriers within the same group exhibit high mutual similarity, while those between groups show negative similarity. Notably, yellow regions appear near the boundary between the two groups, corresponding to subcarriers from Case~3. These subcarriers are thus discarded to prevent contamination of the final waveform.

After the subcarriers are partitioned into two groups, their fluctuation trends are inherently opposite. Therefore, it is necessary to align their directions before aggregation.
Given that, in Section~\ref{sec:pre_processing}, we have normalized each subcarrier's CSI amplitude to have zero mean and unit variance, it suffices to simply flip the sign of all CSI amplitude sequences within the second group $\mathcal{K}_2$.
This operation ensures that all selected subcarriers reflect a unified respiratory trend prior to waveform fusion.

After the two-stage subcarrier screening and trend normalization, we obtain a set of high-quality subcarriers that exhibit consistent fluctuation trends. We then perform averaging across the subcarrier dimension to suppress noise, ultimately yielding a location-robust and high-accuracy waveform.

\subsection{Breathing Phase Identification} \label{sec:phase_iden}

Following the waveform fusion step in the previous subsection, we obtain a high-quality and location-robust respiratory waveform. However, the phase of the waveform—specifically, whether a peak corresponds to inhalation or exhalation—remains ambiguous. While prior works rely on multi-antenna to mitigate the adverse effects of noise-sensitive CSI phase, this subsection addresses the challenge of accurately identifying the breathing phase with the minimum facility cost, using only CSI amplitude from a single antenna.

As illustrated in Figure~\ref{fig:model_relationship}, due to the properties of the cosine function, an increase in $|h(f)|$ does not reveal whether $\theta(f)$ has increased or decreased. Consequently, we cannot infer whether the chest-reflected path has lengthened (exhalation) or shortened (inhalation). To resolve this ambiguity, we introduce an additional known dimension: the frequency over different subcarriers. From equation~\eqref{eq:theta}, $\theta(f)$ increases monotonically with frequency $f$, enabling us to infer the underlying case (Case~1 or~2) by examining how $|h(f)|$ varies with $f$. In WiFi systems, typical subcarrier spacing is 312.5~\!kHz (WiFi~5 and earlier) or 78.125~\!kHz (WiFi~6 and WiFi~7). A one-subcarrier shift at 312.5~\!kHz induces a phase change of approximately $0.00654 \cdot (d_{\mathrm{bre}} - d_{\mathrm{LoS}})$ radians. Even for path length differences of several meters, shifting a few subcarriers induces a small change in phase, ensuring the waveform remains within the same case. This confirms the feasibility of our proposal.

\begin{figure}[t]
    \centering
    \setlength{\abovecaptionskip}{6pt}
    \begin{subfigure}{0.49\linewidth}
        \centering
        \includegraphics[width=1\linewidth]{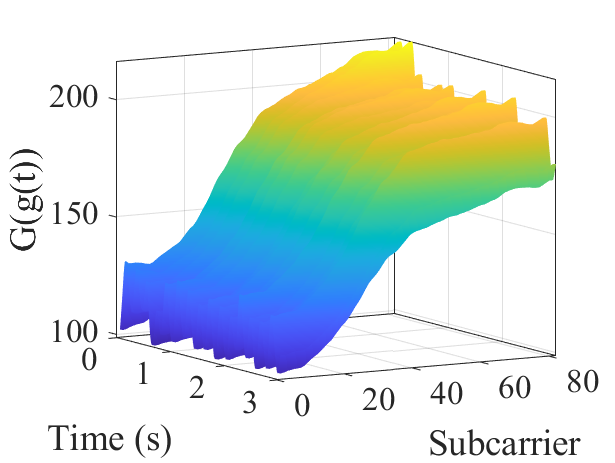}
        \vspace{-1.5em}
        \caption{$G(\bm{g}(t))$.}
        \label{fig:G_gt}
    \end{subfigure}
    \begin{subfigure}{0.49\linewidth}
        \centering
        \includegraphics[width=0.9\linewidth]{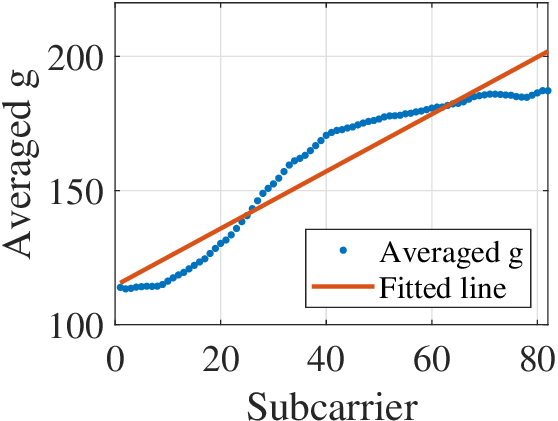}
        \caption{$\bar{\bm{g}}$.}
        \label{fig:bar_g}
    \end{subfigure}
    \caption{Breathing phase identification: (a) $\bm{g}(t)$ after Gaussian smoothing for each time point $t$ and (b) averaged $\bar{\bm{g}}$.}
    \vspace{-1em}
\end{figure}

As noted in Section~\ref{sssec:trend}, we have identified two disjoint subcarrier groups corresponding to Case~1 and Case~2, although their exact mapping remains undetermined.
Since each group may contain multiple sub-regions from the same case (under modular ambiguity), directly using the entire group for identification may be unreliable. Instead, we identify the longest contiguous subcarrier subset $\mathcal{K}^{\mathrm{s}}$ from one of the groups.
At each time point $t$, we construct a vector across subcarriers in $\mathcal{K}^{\mathrm{s}}$: $\bm{g}(t) = [g(f_k)]_{k \in K}$. 
To suppress noise, we first apply Gaussian smoothing to each $\bm{g}(t)$, as shown in Figure~\ref{fig:G_gt}, and subsequently perform temporal averaging across all time points:
\begin{equation}
\bar{\bm{g}} = \frac{1}{T} \sum\nolimits_{t=1}^{T} G(\bm{g}(t)),
\end{equation}
where $G(\cdot)$ denotes Gaussian smoothing. 
For subcarriers within Case~1 or Case~2, $|h(f)|$ (i.e., $g(f)$) at each time point $t$ follows a monotonic trend with frequency $f$. Therefore, the averaged vector $\bar{\bm{g}}$ should also exhibit a monotonic curve.
By analyzing its increasing or decreasing trend, we can determine whether the subcarrier set $\mathcal{K}^{\mathrm{s}}$ belongs to Case~1 or Case~2. Specifically, according to Figure~\ref{fig:model_relationship}, if $\bar{\bm{g}}$ is increasing with frequency (and corresponding $\theta(f)$), it corresponds to Case~2; otherwise, it corresponds to Case~1.
To robustly estimate the trend and mitigate the influence of outliers, we employ a least-squares linear fitting approach. That is, we fit a line $g= a_1 k +a_2$ to $\bar{\bm{g}}$, where $k$ indexes the subcarriers in $\mathcal{K}^{\mathrm{s}}$.
If $a_1 > 0$, the trend is increasing, and $\mathcal{K}^{\mathrm{s}}$ falls into Case~2; if $a_1 < 0$, the trend is decreasing, and $\mathcal{K}^{\mathrm{s}}$ falls into Case~1. Figure~\ref{fig:bar_g} illustrates this process, and the overall algorithm is summarized in Algorithm~\ref{alg:phase}. Once the case is identified, the final respiratory waveform obtained in Section~\ref{sec:waveform_fusion} is adjusted by flipping its sign (if needed), thus yielding a waveform in which upward trends correspond to inhalation and downward trends correspond to exhalation.

\begin{algorithm}[t]
\caption{Breathing Phase Identification Algorithm}\label{alg:phase}
\DontPrintSemicolon
\KwIn{Two groups ($\mathcal{K}_1$ and $\mathcal{K}_2$), corresponding CSI $g(f)$ obtained in Section~\ref{sec:pre_processing}, and breathing waveform $w^{\mathrm{i}}(t)$ obtained in Section~\ref{sssec:trend};}
\KwOut{Final breathing waveform $w^{\mathrm{o}}(t)$.}
Identify the longest contiguous subcarrier subset $\mathcal{K}^{\mathrm{s}}$;\;
\For{each time point $t$}{
    Construct $\bm{g}(t) = [g(f_k)]_{k \in \mathcal{K}^{\mathrm{s}}}$;\;
}
Apply Gaussian smoothing and temporal averaging:
 $\bar{\bm{g}} = \frac{1}{T} \sum_{t=1}^{T} G(\bm{g}(t))$;\;
Perform least-squares linear regression on $\bar{\bm{g}}$:
~~~~~~~~ ~~~~~~~~ Fit $\bar{\bm{g}} \approx a_1 k + a_2$;\;
\eIf{($a_1>0$ and $\mathcal{K}^{\mathrm{s}} \subset \mathcal{K}_1 $) or ($a_1<0$ and $\mathcal{K}^{\mathrm{s}} \subset \mathcal{K}_2 $)}{$w^{\mathrm{o}}(t)=-w^{\mathrm{i}}(t)$.}{$w^{\mathrm{o}}(t)=w^{\mathrm{i}}(t)$.}
\end{algorithm}

\subsection{Computational Complexity}
As \name is intended for deployment on low-cost IoT devices, we now examine its computational complexity. Let $K$ denote the number of subcarriers, $L$ denote the processing window length for each subcarrier, and $K'$ denote the number of subcarriers retained after BNR-based subcarrier selection. The computational complexity of \name consists of pre-processing with $\mathcal{O}(KL)$, BNR-based subcarrier selection with $\mathcal{O}(KL\log L)$ due to the FFT operations, and similarity matrix construction with $\mathcal{O}(K'^2L)$. The subsequent spectral clustering and waveform fusion stages incur complexities of $\mathcal{O}(K'^3)$ and $\mathcal{O}(K'^2 + K'\log K' + K'L)$, respectively. Since BNR-based subcarrier selection significantly reduces the number of subcarriers ($K' \ll K$), the overall complexity is effectively $\mathcal{O}(KL\log L + K'^2L + K'^3)$. In practice, this computational complexity is low enough for real-time deployment on resource-constrained platforms. For instance, on a Raspberry Pi~4B, the entire processing pipeline for a 15-second window can be completed in less than 500~\!ms. Given that the human respiratory rate is relatively low (typically 0.16$\sim$0.5~Hz), this latency satisfies the requirement of real-time respiratory monitoring and physiological feedback.

\section{Evaluations} \label{sec:eva}
This section first presents the implementation of \name, followed by a detailed evaluation of its performance.

\subsection{\name Implementation}
\begin{figure}[t]
    \centering
    \setlength{\abovecaptionskip}{6pt}
    \includegraphics[width=0.98\linewidth]{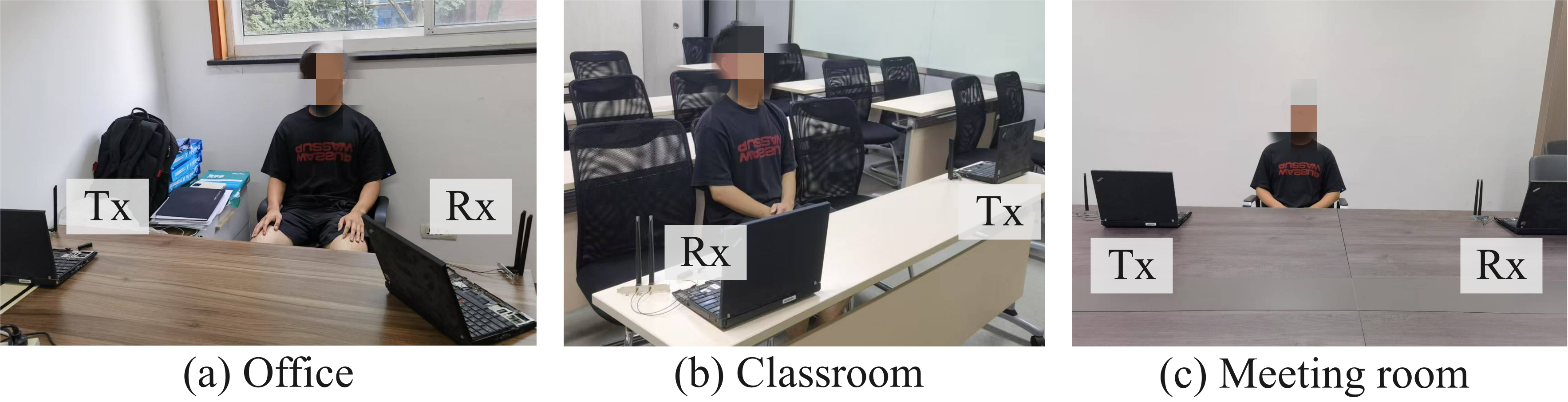}
    \vspace{-0.5em}
    \caption{Experiment setup in three environments.}
    \label{fig:layout}
    \vspace{-1em}
\end{figure}

\noindent \textbf{Experiment setup.} As shown in Figure~\ref{fig:layout}, we employ two Lenovo ThinkPad X201 laptops, each equipped with an Intel AX210 network interface card (NIC), to serve as the Tx and Rx, respectively. The system operates under the 802.11ax standard at a center frequency of 6025~\!MHz with a 160~\!MHz bandwidth, transmitting packets at a rate of 100 packets per second using one transmit antenna and two receive antennas. 
To validate the scalability of \name for single-antenna IoT devices, we implement software-based data isolation. Unless otherwise specified, \name randomly selects the CSI amplitude from a single receive antenna, making its processing pipeline functionally equivalent to that of a single-receive-antenna system. The other baselines operate on all available streams in accordance with their original design requirements.
PicoScenes~\cite{jiang2021eliminating} is used to configure communication parameters and measure CSI. By default, the Tx-Rx distance is 2~\!m, while the user is 1.5~\!m away from the LoS path. Each WiFi packet contains around 2,000 subcarriers. 
Note that for ease of experimentation, we used a WiFi NIC as the Tx in our prototype. \name can be readily used with a commodity home router as the Tx without affecting its functionality, since it fundamentally relies on CSI estimated from physical-layer pilot signals.
All the experiments are conducted by adhering to the approval of our university’s Institutional Review Board (IRB).

\noindent \textbf{Data collection.} We recruit 30 participants (22 males and 8 females) to participate in experiments conducted across three different environments: an office, a classroom, and a meeting room. Each participant sits on a chair and wears a smart respiratory belt to record ground-truth respiration parameters. For each subject, we collect at least 180~seconds of data. Unless otherwise specified, the data is collected in the office. In total, we obtain over 12,000~seconds of 
CSI and ground-truth respiration data for evaluation and achieve a high-precision synchronization by connecting both the PCIe-based WiFi NIC and the respiratory belt to the same host system, utilizing a unified system clock to eliminate clock drift.

\subsection{Metrics and Biomarkers}

We define three evaluation metrics—Mean Absolute Error (MAE), Pearson Correlation Coefficient (PCC), and Mean Squared Error (MSE)—to assess the performance of \name\ in estimating five key respiratory biomarkers: RR, respiratory waveform, I:E ratio, TV variability, and ApEn. These biomarkers provide comprehensive insights into respiratory health, ranging from basic ventilation rate to deeper autonomic regulation. Below, we describe how each metric is computed and 
employed to evaluate the 
biomarkers.

\subsubsection{Mean Absolute Error}
MAE is used to assess \name's capability in estimating the RR, defined as the number of breaths per minute (BPM). RR is a fundamental vital sign and an essential diagnostic indicator for a variety of conditions, including respiratory infections, cardiac failure, and metabolic imbalances~\cite{cretikos2008respiratory,huang2020clinical}. Once the respiration waveform is obtained, we identify the peaks corresponding to breathing cycles and compute the average peak-to-peak interval $\overline{t}_{pp}$ (in seconds) with a defined window length of 60~\!s. Peak detection parameters include a minimum distance of 2~\!s and an adaptive prominence threshold set at 10~\!\% of the windowed signal's amplitude. The estimated RR (in BPM) is then calculated as $
\hat{\text{RR}} = \frac{60}{\overline{t}_{pp}}$. The MAE (in BPM) is computed using the ground-truth RR and the estimated $\hat{\text{RR}}$ as follows:
\begin{equation}
    \text{MAE} = \frac{1}{n} \sum\nolimits_{i=1}^{n} |{\text{RR}_i} - \hat{\text{RR}_i}|,
\end{equation}

\subsubsection{Pearson Correlation Coefficient}
PCC is used to evaluate the accuracy and reliability of the reconstructed respiratory waveform. An accurate waveform is crucial for clinical applications such as detecting abnormal breathing patterns (e.g., Cheyne-Stokes respiration~\cite{sahlin2005cheyne}), estimating tidal volume dynamics, and supporting ventilator calibration~\cite{chen2024continuous}. PCC quantifies the linear correlation between the estimated and ground-truth waveforms and is calculated as:
\begin{equation}
\rho_{XY} = \frac{\sum_{i=1}^{n} (x_i - \bar{x})(y_i - \bar{y})}{\sqrt{\sum_{i=1}^{n} (x_i - \bar{x})^2} \sqrt{\sum_{i=1}^{n} (y_i - \bar{y})^2}},
\end{equation}
where $x$ and $y$ are the reconstructed and ground-truth respiratory waveforms, respectively.

\subsubsection{Mean Squared Error}
MSE is employed to evaluate \name's accuracy in estimating three advanced biomarkers: the I:E ratio, TV variability, and ApEn. \textbf{I:E ratio} is defined as the ratio of inhalation duration to exhalation duration. It serves as a critical index for assessing respiratory function and autonomic nervous system balance, particularly vagal tone. Abnormal I:E ratios are associated with respiratory conditions such as chronic obstructive pulmonary disease (COPD) and may guide the tuning of ventilation therapies~\cite{murphy2005prolongation}. \textbf{TV variability} reflects the variation in air volume exchanged during each breath. It is essential for diagnosing respiratory muscle disorders, evaluating lung compliance, and monitoring patients with respiratory failure or restrictive lung diseases. Tracking its fluctuations helps detect pathophysiological deterioration early~\cite{worsham2021dyspnea,barbour2004increased}. \textbf{ApEn} measures the irregularity and complexity of the breathing signal. It captures the degree of autonomic control and adaptability of the respiratory system. High ApEn values indicate erratic or dysregulated breathing patterns, often seen in anxiety disorders or sleep apnea, whereas low ApEn values are characteristic of stable, healthy breathing~\cite{pincus1991approximate,caldirola2004approximate}. For details on the calculation methods of these biomarkers using inhalation-exhalation-phase-aware waveforms, please refer to~\cite{addison2024non,fekr2015design}. MSE is calculated as:
\begin{equation}
    \text{MSE} = \frac{1}{n} \sum\nolimits_{i=1}^{n} (y_i - \hat{y}_i)^2,
\end{equation}
where $\hat{y}_i$ and $y_i$ represent the estimated and ground-truth biomarker values, respectively.

\subsubsection{Bland-Altman Analysis}
The Bland-Altman analysis~\cite{altman1983measurement} provides a robust statistical framework for evaluating the agreement between two quantitative measurement techniques—namely, respiration belt-based and WiFi-based methods in our case. Demonstrating satisfactory agreement indicates that the two methods can be considered clinically interchangeable within predefined tolerance limits. In this work, we employ the Bland-Altman analysis to assess whether the measurements of the I:E ratio, TV variability, and ApEn obtained from our WiFi-based system reach robust agreement with respiratory belts.



\begin{figure*}[t]
    \centering
    \setlength{\abovecaptionskip}{6pt}
    \begin{subfigure}{0.31\linewidth}
        \centering
        \includegraphics[width=0.98\linewidth]{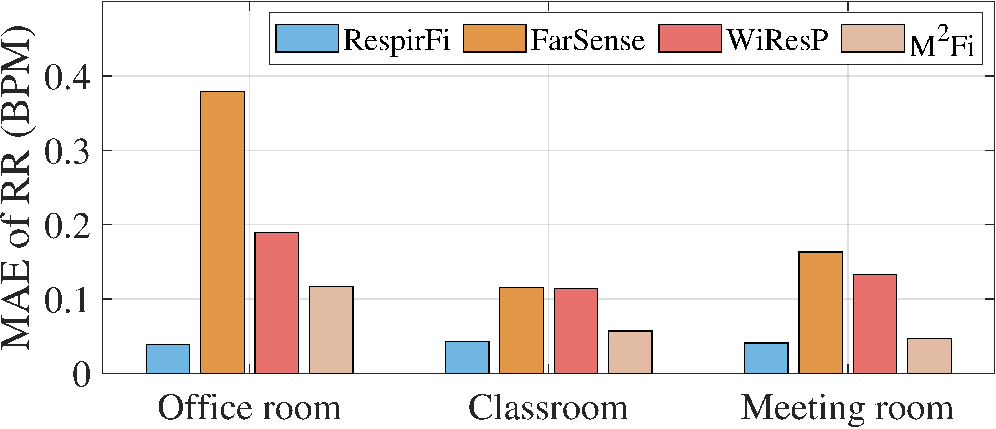}
        \vspace{-0.5em}
        \caption{Overall MAEs of RR.}
        \label{fig:02_RR}
    \end{subfigure}
    \begin{subfigure}{0.34\linewidth}
        \centering
        \includegraphics[width=0.98\linewidth]{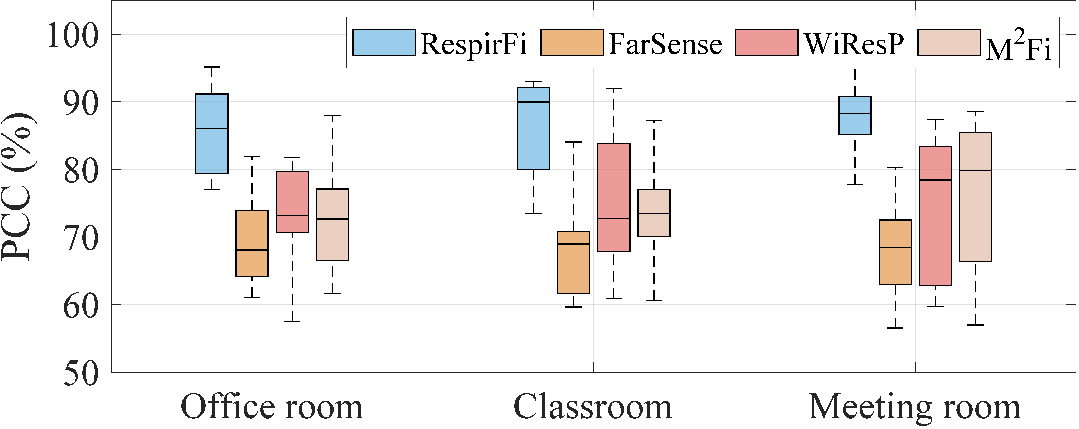}
        \vspace{-0.55em}
        \caption{Overall PCCs of respiratory waveform.}
        \label{fig:01_PCC}
    \end{subfigure}
    \begin{subfigure}{0.32\linewidth}
        \centering
        \includegraphics[width=0.98\linewidth]{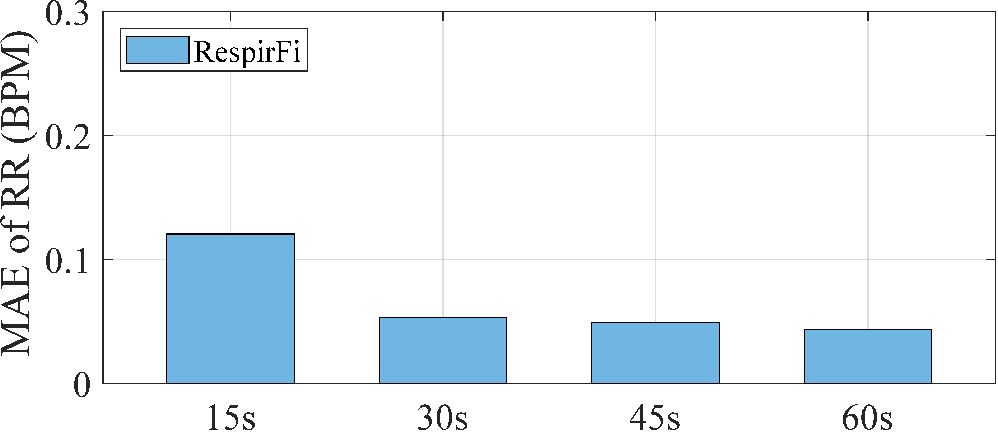}
        \vspace{-0.5em}
        \caption{Effect of sliding-window length on MAE.}
        \label{fig:sli}
    \end{subfigure}
    \begin{subfigure}{0.32\linewidth}
        \centering
        \includegraphics[width=0.98\linewidth]{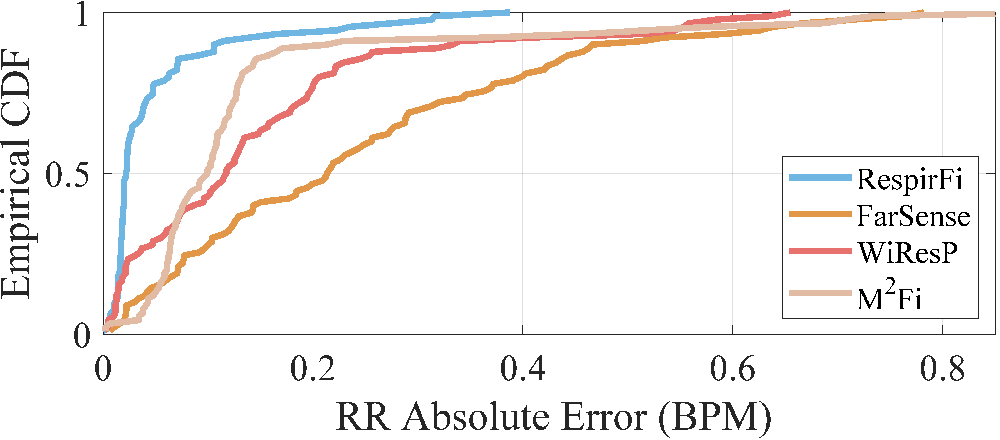}
        \vspace{-0.5em}
        \caption{CDF comparison of RR Absolute Error.}
        \label{fig:cdf}
    \end{subfigure}
    \begin{subfigure}{0.31\linewidth}
        \centering
        \includegraphics[width=0.98\linewidth]{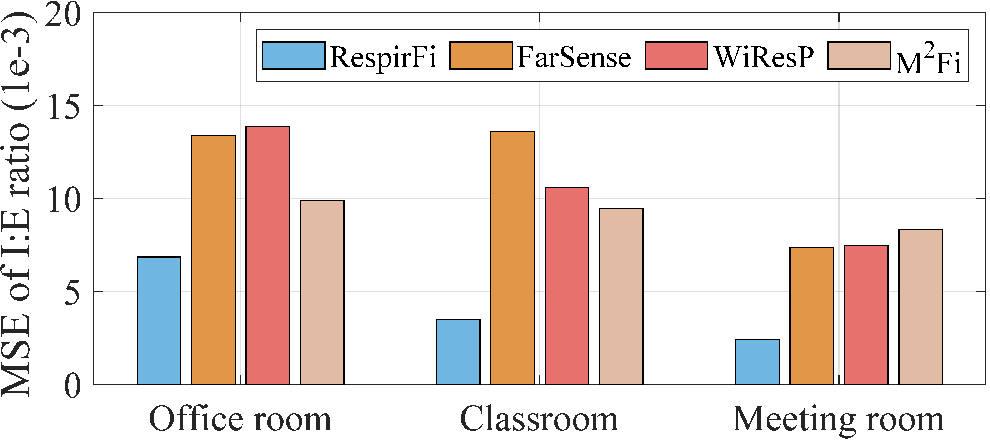}
        \vspace{-0.5em}
        \caption{Overall I:E ratio.}
        \label{fig:03_IE}
    \end{subfigure}
    \begin{subfigure}{0.32\linewidth}
        \centering
        \includegraphics[width=0.98\linewidth]{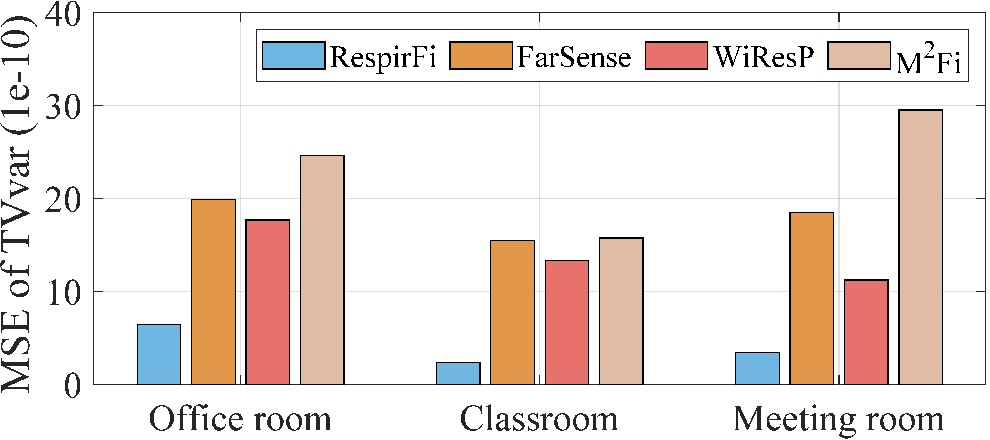}
        \vspace{-0.5em}
        \caption{Overall TV variability.}
        \label{fig:04_TV}
    \end{subfigure}
   \begin{subfigure}{0.32\linewidth}
        \centering
        \includegraphics[width=0.98\linewidth]{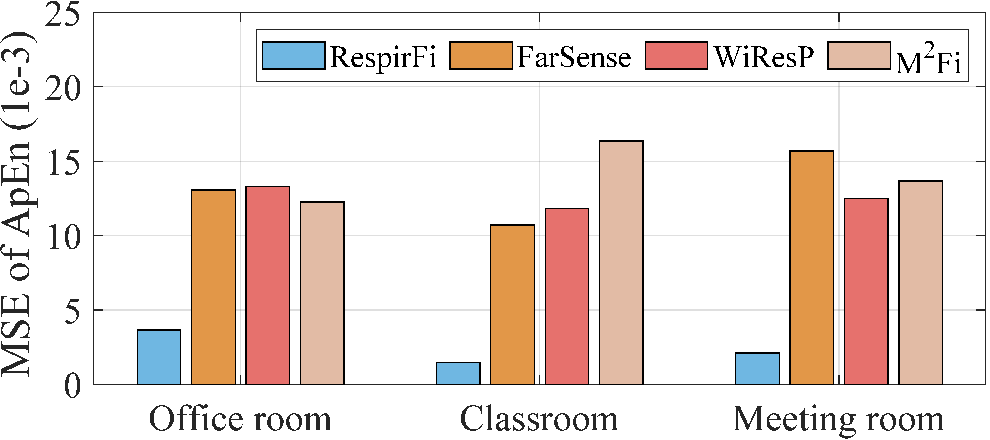}
        \vspace{-0.5em}
        \caption{Overall ApEn.}
        \label{fig:05_APEN}
    \end{subfigure}
    \begin{subfigure}{0.32\linewidth}
        \centering
        \includegraphics[width=0.98\linewidth]{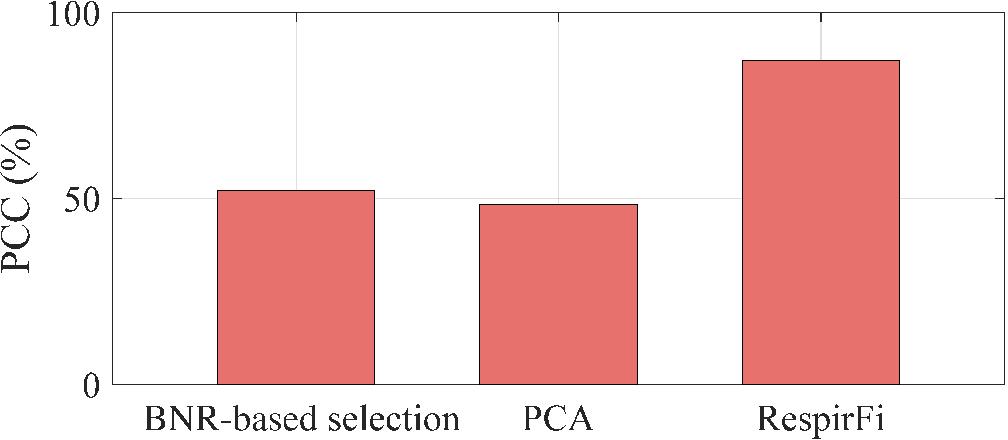}
        \vspace{-0.5em}
        \caption{PCCs of \name and two alternatives.}
        \label{fig:17_METHOD}
    \end{subfigure}
    \begin{subfigure}{0.32\linewidth}
        \centering
        \includegraphics[width=0.98\linewidth]{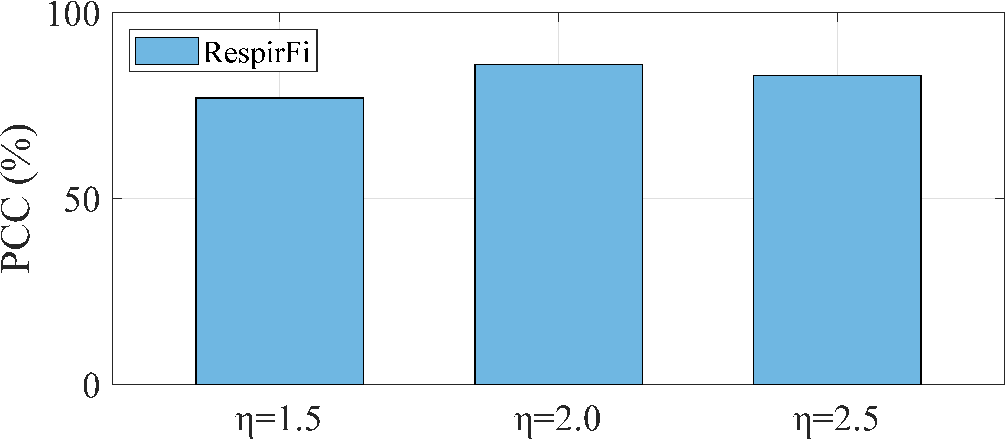}
        \vspace{-0.5em}
        \caption{Sensitivity analysis of BNR threshold $\eta$.}
        \label{fig:sen_ana}
    \end{subfigure}
    \\
    \begin{subfigure}{0.32\linewidth}
        \centering
        \includegraphics[width=0.98\linewidth]{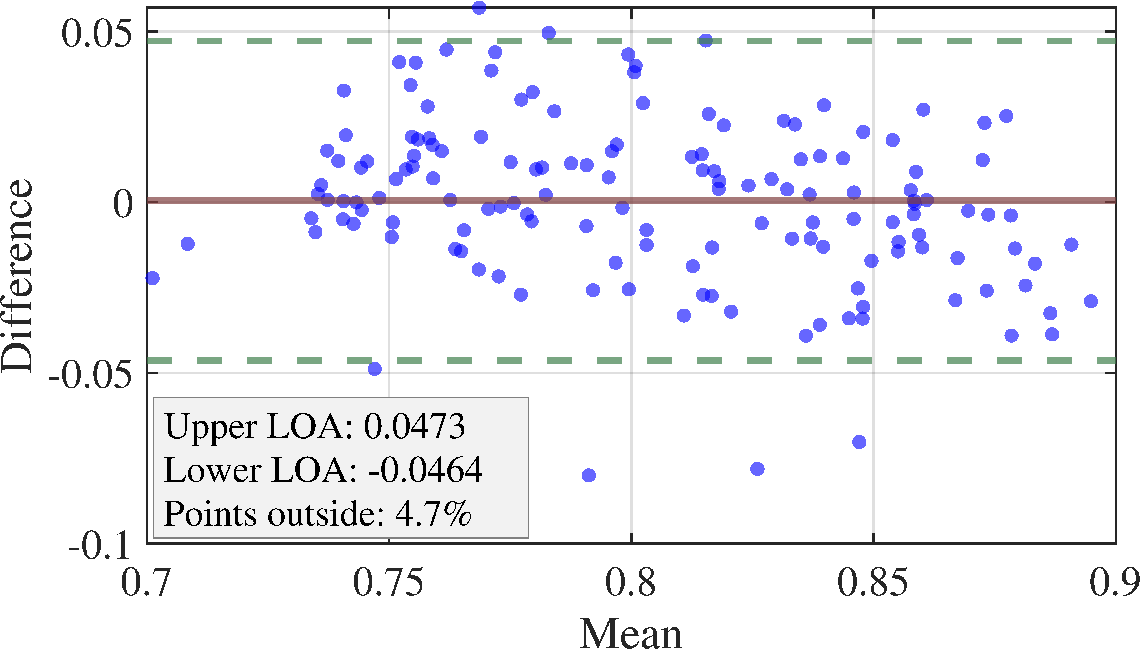}
        \vspace{-0.5em}
        \caption{Bland-Altman plot: I:E ratio.}
        \label{fig:07_BA_IE}
    \end{subfigure}
    \begin{subfigure}{0.32\linewidth}
        \centering
        \includegraphics[width=0.94\linewidth]{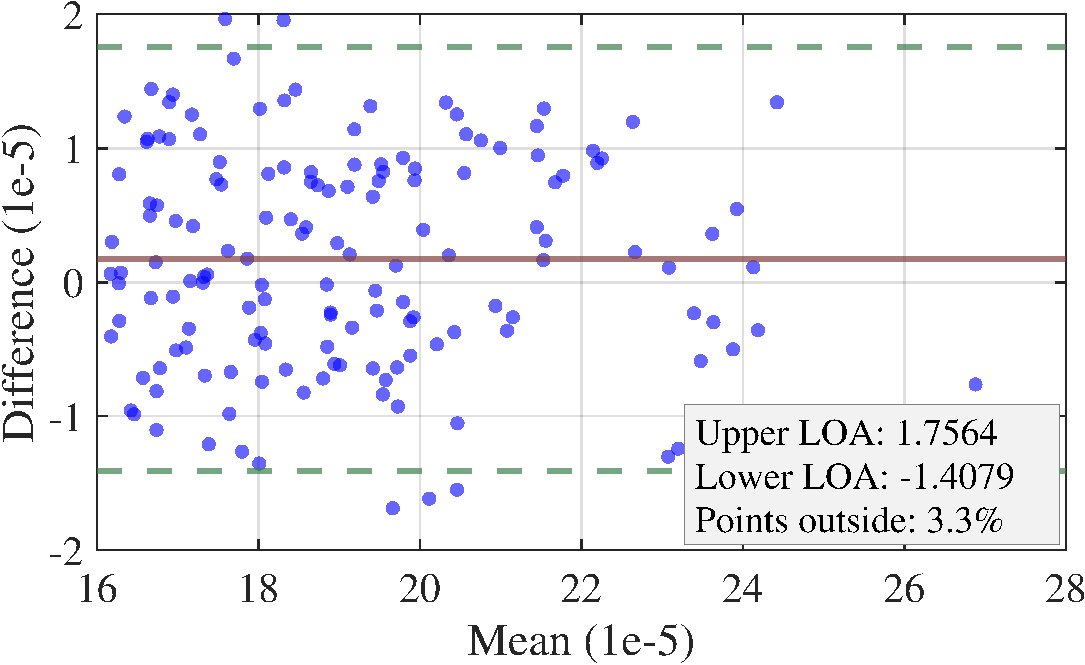}
        \vspace{-0.7em}
        \caption{Bland-Altman plot: TV variability.}
        \label{fig:06_BA_TV}
    \end{subfigure}
    \begin{subfigure}{0.32\linewidth}
        \centering
        \includegraphics[width=0.98\linewidth]{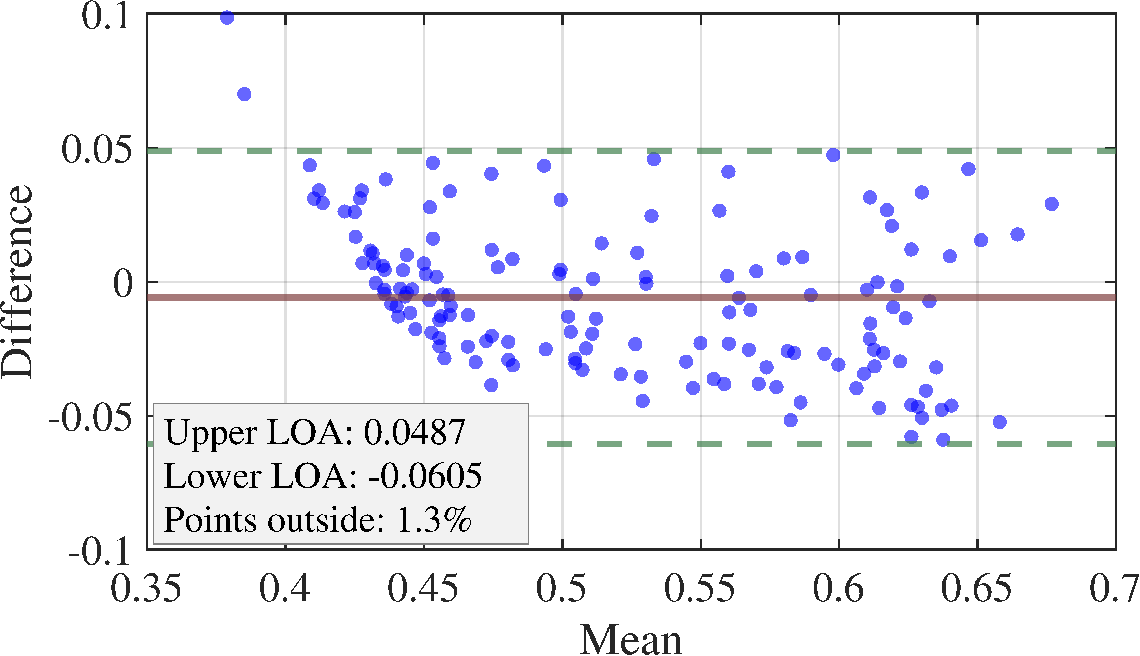}
        \vspace{-0.5em}
        \caption{Bland-Altman plot: ApEn.}
        \label{fig:08_BA_APEN}
    \end{subfigure}
    \vspace{-0.5em}
    \caption{Overall performance of \name
    .}
    \label{fig:overall_performance}
    \vspace{-1.3em}
\end{figure*}

\subsection{Overall Performance}
\begin{figure}[t]
    \centering
    \setlength{\abovecaptionskip}{6pt}
    \begin{subfigure}{0.56\linewidth}
        \centering
        \includegraphics[width=1\linewidth]{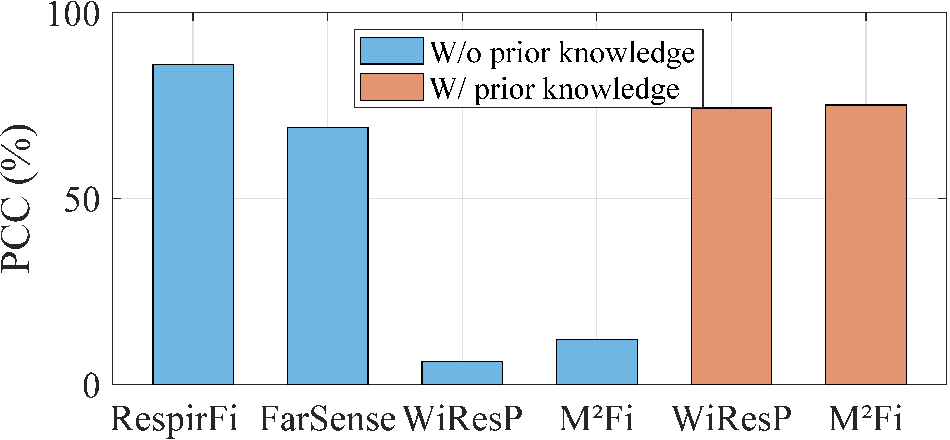}
        \vspace{-1.5em}
        \caption{Overall comparison of waveform PCCs.}
        \vspace{-0.5em}
        \label{fig:pcc_bi}
    \end{subfigure}
    \begin{subfigure}{0.42\linewidth}
        \centering
        \includegraphics[width=0.9\linewidth]{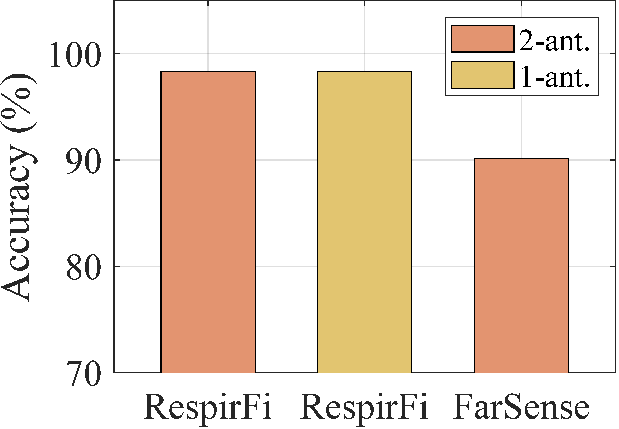}
        \vspace{-0.5em}
        \caption{Accuracy of breathing phase identification.}
        \vspace{-0.5em}
        \label{fig:22_PH_ID}
    \end{subfigure}
    \caption{Effectiveness of breathing phase identification.}
    \vspace{-1.3em}
\end{figure}

To comprehensively evaluate the performance of \name, we conduct experiments in three distinct indoor environments—an office, a classroom, and a meeting room—as illustrated in Figure~\ref{fig:layout}. To demonstrate its superiority, we compare \name with three SOTA baselines: FarSense~\cite{zeng2019farsense}, WiResP~\cite{wang2024wiresp}, and M$^2$Fi~\cite{hu2024m}. FarSense detects respiration by jointly analyzing the amplitude and phase components of CSI, specifically leveraging the temporal dynamics of CSI ratios. WiResP enhances the CSI amplitude spectrum to improve the detectability of respiration signals under low Signal-to-Noise Ratio (SNR) conditions, and applies autocorrelation analysis for signal extraction. M$^2$Fi relies on Beamforming Feedback Information (BFI) to capture respiratory patterns, implementing a deployable framework on handheld devices and employing Generative Adversarial Networks (GANs) to reconstruct high-fidelity respiratory waveforms. Since BFI can be computationally derived from CSI measurements~\cite{DBLP:conf/sensys/WangW00ZYYY024}, we implement the M$^2$Fi baseline using BFI data computed from our collected CSI to ensure consistency in evaluation following the WiFi standards~\cite{ieee2016ieee}. 
It should be noted that, among the three baselines, only FarSense is able to identify inhalation-exhalation phases, while WiResP and M$^2$Fi lack this capability. Nevertheless, to make the comparison as comprehensive as possible, unless otherwise specified, we assume perfect knowledge of the inhalation-exhalation phases for all baselines, obtained from the ground truth. Such prior knowledge is not available in practical deployment, and thus, this assumption is highly idealized for the baseline methods. Furthermore, the performance of all methods without ground-truth phase information is separately evaluated and discussed at the end of this section (see Figure~\ref{fig:pcc_bi}).

\noindent \textbf{Performance of RR and waveform estimation.} The results for RR MAE and waveform PCC are shown in Figures~\ref{fig:02_RR} and~\ref{fig:01_PCC}, respectively. As illustrated, \name achieves RR estimation errors consistently within 0.05~\!BPM across all tested environments. In contrast, the average RR MAEs for FarSense, WiResP, and M$^2$Fi are 0.22, 0.14, and 0.07~\!BPM, respectively. 
To further examine the robustness of RR estimation, we evaluate \name using sliding windows ranging from 15~\!s to 60~\!s, thereby avoiding overly optimistic long-term averaging. As shown in Figure~\ref{fig:sli}, \name maintains stable performance even under shorter window lengths. 
In addition, we apply non-parametric bootstrapping~\cite{efron1993bootstrap} to estimate the 95~\!\% confidence interval of \name's RR error, yielding a range of $[0.0175, 0.0834]$~BPM. Furthermore, Wilcoxon signed-rank tests show that the performance improvement of \name over all baselines is statistically significant ($p < 0.001$).
To provide a more comprehensive view of the error distribution, we also plot the cumulative distribution function (CDF) in Figure~\ref{fig:cdf}, from which it can be observed that more than 90~\!\% of the evaluation segments achieve an absolute error below 0.1~\!BPM.
For waveform reconstruction, \name maintains a PCC exceeding 85~\!\% in all environments, while the average PCCs for the three baselines are 69.01~\!\%, 74.29~\!\%, and 75.10~\!\%, respectively. These results clearly demonstrate that \name outperforms existing methods in both RR estimation accuracy and respiratory waveform fidelity. The superior performance of \name can be attributed to its selective use of high-quality subcarriers, which effectively avoids signal degradation caused by noisy or low-quality subcarriers.

\noindent \textbf{Performance of biomarker estimation.} In this experiment, we evaluate \name's capability to estimate three key respiratory biomarkers: I:E ratio, TV variability, and ApEn, comparing its performance against the three baseline methods. Notably, accurate computation of certain biomarkers (e.g., I:E ratio) requires the ability to distinguish between inhalation and exhalation phases. The MSEs of the three biomarkers are shown in Figures~\ref{fig:03_IE},~\ref{fig:04_TV}, and~\ref{fig:05_APEN}, respectively. For the I:E ratio (scaled by $10^{-3}$), the average MSEs for \name, FarSense, WiResP, and M$^2$Fi are 4.70, 11.41, 10.64, and 9.39, respectively. Regarding TV variability (scaled by $10^{-10}$), the average MSEs are 4.05, 17.94, 14.08, and 23.30, respectively. For ApEn (scaled by $10^{-3}$), the MSEs are 2.41, 13.15, 12.54, and 14.09, respectively. Clearly, \name consistently achieves significantly lower MSEs than all baselines across all three biomarkers. In addition, its performance remains more stable across different environments. These results demonstrate that \name produces biomarker estimates more closely aligned with ground truth, highlighting its superior accuracy and robustness. \textit{The comprehensive capability of \name in biomarker estimation represents a substantial step toward achieving an exceptional level respiration monitoring with WiFi.}

\noindent \textbf{Bland-Altman analysis of biomarkers.} To further validate whether \name achieves comparable accuracy to ground truth in measuring these three biomarkers, we aggregated data from all three environments and performed Bland-Altman analysis. Figures~\ref{fig:07_BA_IE}, \ref{fig:06_BA_TV}, and \ref{fig:08_BA_APEN} represent Bland-Altman plots for the three biomarkers. It can be observed that \name consistently exhibits lower systematic bias and a higher proportion of data points within the limits of agreement (LOA). For I:E ratio, TV variability, and ApEn respectively,  95.3~\!\%, 96.7~\!\%, and 98.7~\!\% of measurement differences fall within $\pm$1.96~\!$\times$standard deviation of the mean difference for \name, each satisfying the acceptability criterion where $<$5~\!\% of points may lie beyond the LOA. \textit{This demonstrates robust agreement between \name's measurements and specialized reference sensors.} 

\noindent \textbf{Effectiveness of subcarrier selection and fusion.} 
To demonstrate the stability of the subcarrier selection strategy, we evaluate the sensitivity of the hybrid BNR threshold $\eta$ within the range [1.5, 2.5]. The results observed in Figure~\ref{fig:sen_ana} show that \name is inherently robust to the exact threshold choice.
Moreover, we assess the impact of our subcarrier selection and fusion method (Section~\ref{sec:waveform_fusion}) by comparing \name with two alternative strategies: one relying solely on BNR-based selection, and another employing PCA. The PCCs for respiratory waveform reconstruction under each method are presented in Figure~\ref{fig:17_METHOD}. As shown, \name consistently achieves higher PCCs than both alternatives, demonstrating that our subcarrier selection and fusion strategy significantly enhances the accuracy of waveform reconstruction.

\noindent \textbf{Effectiveness of breathing phase identification.} 
We next evaluate the effectiveness of the proposed breathing phase identification algorithm. To this end, we consider a practical setting in which no prior knowledge of the inhalation-exhalation phases is available. Figure~\ref{fig:pcc_bi} presents the waveform PCC results of the four methods under this setting. As can be observed, WiResP and M$^2$Fi suffer severe performance degradation, and their reconstructed waveforms become largely inaccurate, since they are unable to determine the breathing phases. In contrast, \name maintains a stable waveform PCC above 85~\!\%, demonstrating the effectiveness of the proposed phase identification algorithm under a fully blind setting.
We further conduct a targeted comparison between \name and FarSense~\cite{zeng2019farsense}, since FarSense is the only baseline that also supports inhalation/exhalation phase identification through CSI ratio analysis. Because FarSense requires WiFi devices to be equipped with multiple Rx antennas, we evaluate both methods using dual-antenna CSI. In addition, \name is independently evaluated with single-antenna CSI. In this comparison, we focus exclusively on phase identification accuracy, without considering waveform fidelity or other respiratory parameters. As illustrated in Figure~\ref{fig:22_PH_ID}, \name achieves higher phase identification accuracy than FarSense. Moreover, \name avoids the use of CSI phase and maintains stable performance with only a single Rx antenna, demonstrating greater practicality and deployment flexibility.


In conclusion, the above evaluations demonstrate that \name outperforms SOTA approaches in both fundamental respiratory monitoring tasks, namely waveform reconstruction and RR estimation, and in the quantification of advanced respiratory biomarkers, including the I:E ratio, TV variability, and ApEn. This superiority is reflected in higher accuracy and greater precision. Furthermore, Bland–Altman analysis offers rigorous statistical validation, showing that \name achieves satisfactory agreement with common research-standard reference sensors for all three biomarkers.

\begin{figure*}[t]
    \centering

    \begin{subfigure}{0.31\linewidth}
        \centering
        \vspace{-0.5em}
        \includegraphics[width=0.98\linewidth,trim={0cm 0cm 0cm 0cm},clip]{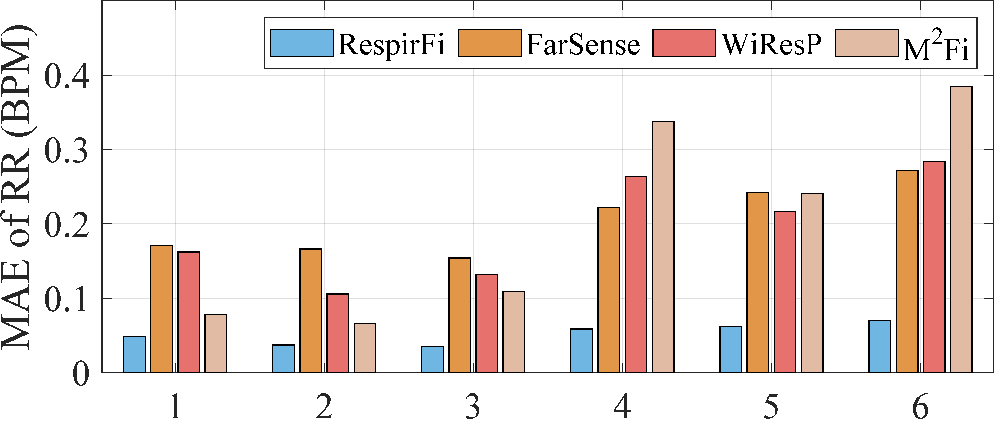}
        \vspace{-0.5em}
        \caption{Effect of user locations on MAE.}
        \label{fig:21_RR_POS}
    \end{subfigure}
    \begin{subfigure}{0.31\linewidth}
        \centering
        \includegraphics[width=0.98\linewidth]{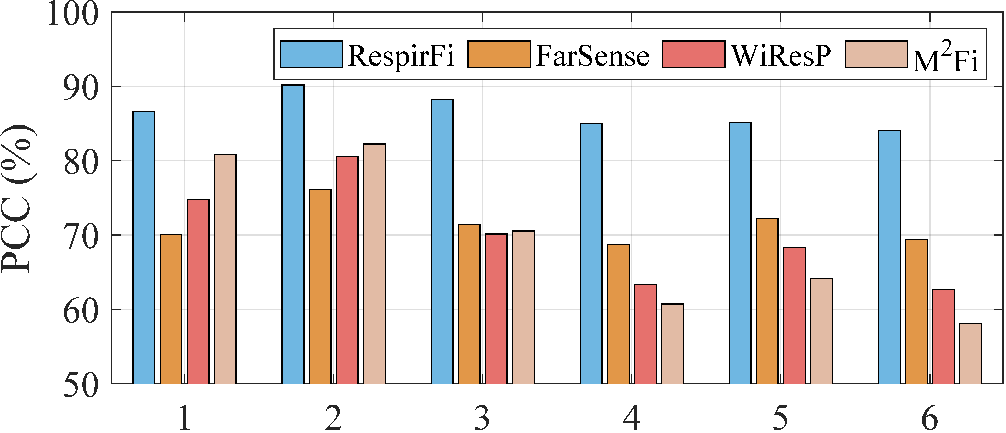}
        \vspace{-0.5em}
        \caption{Effect of user locations on PCC.}
        \label{fig:20_PCC_POS}
    \end{subfigure}
    \begin{subfigure}{0.34\linewidth}
        \centering
        \includegraphics[width=0.98\linewidth]{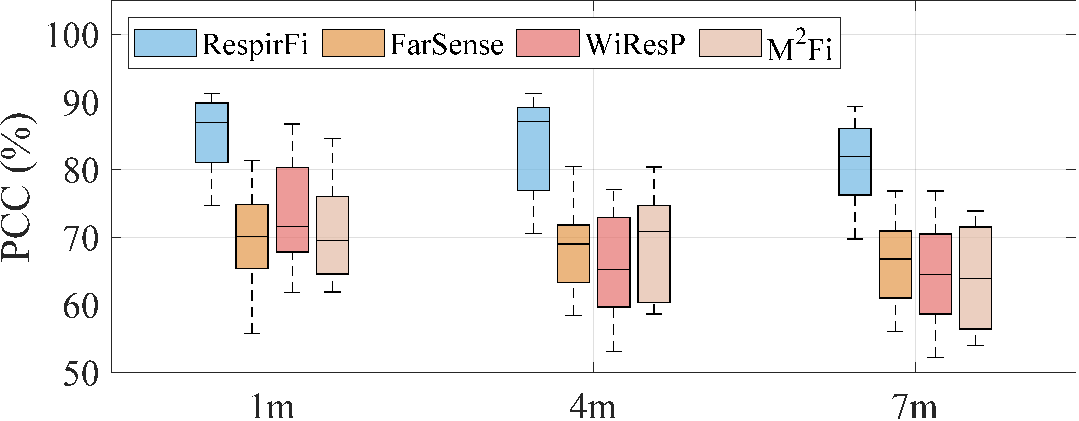}
        \vspace{-0.5em}
        \caption{Effect of Tx-Rx distance on PCC.}
        \label{fig:09_PCC_DIS1}
    \end{subfigure}
     \begin{subfigure}{0.31\linewidth}
        \centering
        \includegraphics[width=0.98\linewidth]{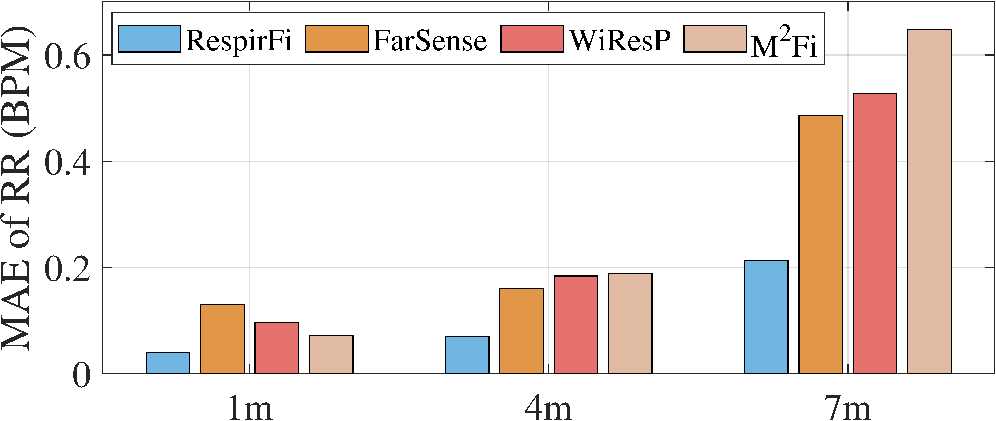}
        \vspace{-0.5em}
        \caption{Effect of Tx-Rx distance on MAE.}
        \label{fig:10_RR_DIS1}
    \end{subfigure}
    \begin{subfigure}{0.31\linewidth}
        \centering
        \includegraphics[width=0.98\linewidth]{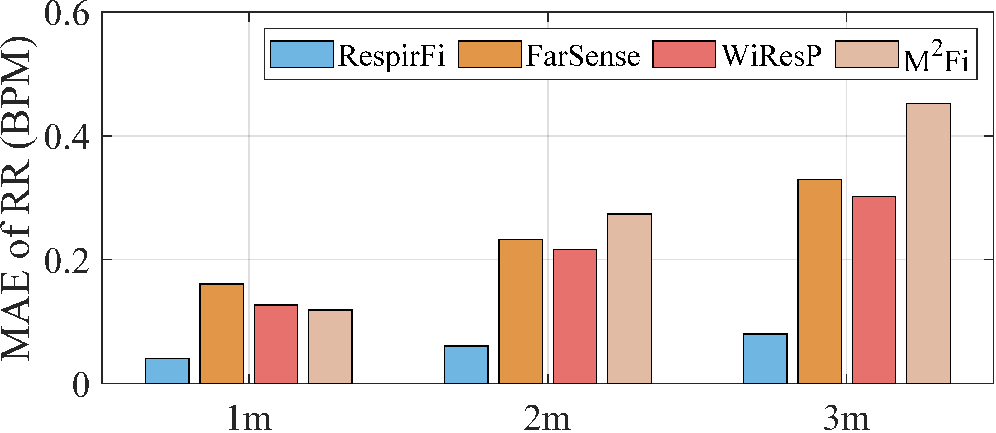}
        \vspace{-0.5em}
        \caption{\mbox{Effect of user-LoS path distance on MAE.}}
        \label{fig:12_RR_DIS2}
    \end{subfigure} 
    \begin{subfigure}{0.34\linewidth}
        \centering
        \includegraphics[width=0.98\linewidth]{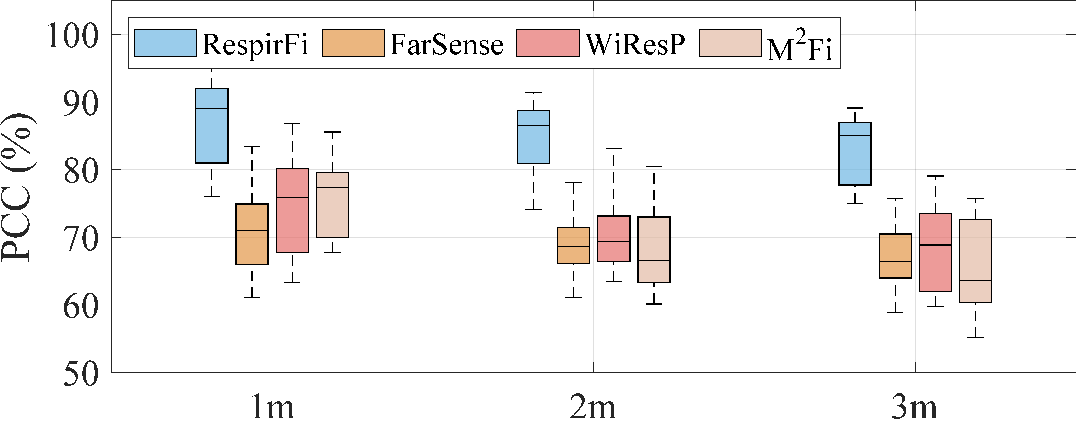}
        \vspace{-0.5em}
        \caption{Effect of user-LoS path distance on PCC.}
        \label{fig:11_PCC_DIS2}
    \end{subfigure}
     \vspace{-0.5em}
    \caption{Effect of user location, Tx-Rx distance and user-LoS path distance on MAE of RR and PCC of waveform.}
    \label{fig:Effect of distance}
    \vspace{-1.3em}
\end{figure*}

\subsection{Effect of User Location}
Imposing no constraints on user location is critically important for enhancing the flexibility and usability of a monitoring system. To evaluate whether the performance of \name is affected by variations in user location, we conduct experiments at six distinct locations, as illustrated in Figure~\ref{fig:Preliminary}, and recalculate the performance metrics accordingly. The results, presented in Figures~\ref{fig:21_RR_POS} and~\ref{fig:20_PCC_POS}, demonstrate that \name consistently maintains high stability and accuracy across all tested locations. In comparison, FarSense exhibits moderate consistency but with generally lower performance, while M$^2$Fi shows significant fluctuations in performance depending on the user's location. Their performance degradation and instability are likely due to their specific design for different deployment scenarios. In contrast, \name benefits from its subcarrier selection and fusion strategy, which enhances robustness against spatial variability, thereby ensuring reliable performance for all user locations.

\subsection{Effect of Distance}
In practical deployments, system constraints may necessitate changes in the spacing between antennas or the distance between the user and the LoS path. In this subsection, we vary both parameters to examine whether such changes impact the performance of \name.

\noindent \textbf{Effect of Tx-Rx distance.} We vary the distance between Tx and Rx from 1~\!m to 7~\!m in increments of 3~\!m. The resulting RR MAEs and waveform PCCs are shown in Figures~\ref{fig:10_RR_DIS1} and~\ref{fig:09_PCC_DIS1}, respectively. As illustrated, for RR estimation, \name consistently maintains MAEs around 0.2~\!BPM, representing over a 60~\!\% error reduction compared to the three baselines, whose MAEs exceed 0.5~\!BPM. Furthermore, \name demonstrates robust waveform reconstruction performance, maintaining high PCCs despite fluctuations caused by significant signal attenuation at longer distances. Notably, at 7~\!m, \name achieves a median PCC exceeding 80~\!\%, while all baselines fall below 70~\!\%. These results confirm \name’s resilience to variations in Tx-Rx distance.

\noindent \textbf{Effect of user-LoS path distance.} We vary the lateral distance between the user and the LoS path from 1~\!m to 3~\!m in 1~\!m increments and recalculate the performance metrics. The resulting RR MAEs and waveform PCCs are presented in Figures~\ref{fig:12_RR_DIS2} and~\ref{fig:11_PCC_DIS2}, respectively. The results demonstrate that \name exhibits strong robustness to lateral displacement, maintaining high performance even as the offset increases. Specifically, at the maximum 3~\!m offset, \name achieves a median PCC exceeding 80~\!\% and sustains an RR MAE of just 0.1~\!BPM. In contrast, M$^2$Fi experiences significant performance degradation beyond a 2~\!m offset. These findings conclusively establish \name's superior resilience to user-LoS path variations, underscoring its suitability for practical non-LoS deployment scenarios.

\begin{figure*}[t]
    \centering
    \setlength{\abovecaptionskip}{6pt}
    \begin{subfigure}{0.16\linewidth}
        \centering
        \includegraphics[width=0.98\linewidth]{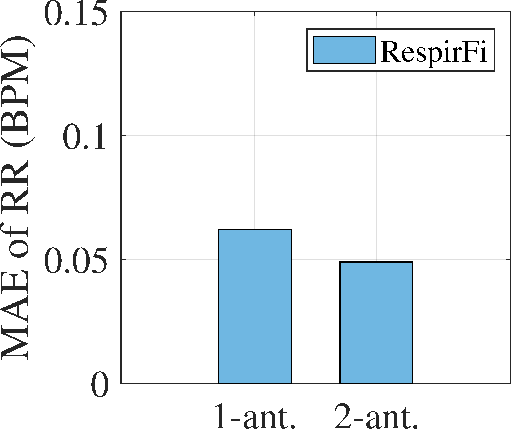}
        \vspace{-1.5em}
        \caption{Effect of Rx configuration on MAE.}
        \label{fig:14_RR_ANT}
    \end{subfigure}
    \begin{subfigure}{0.17\linewidth}
        \centering
        \includegraphics[width=0.98\linewidth]{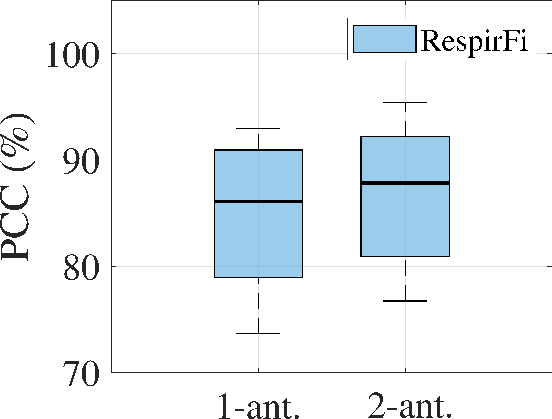}
        \vspace{-1.5em}
        \caption{Effect of Rx configuration on PCC.}
        \label{fig:13_PCC_ANT}
    \end{subfigure}
    \begin{subfigure}{0.315\linewidth}
        \centering
        \includegraphics[width=0.98\linewidth]{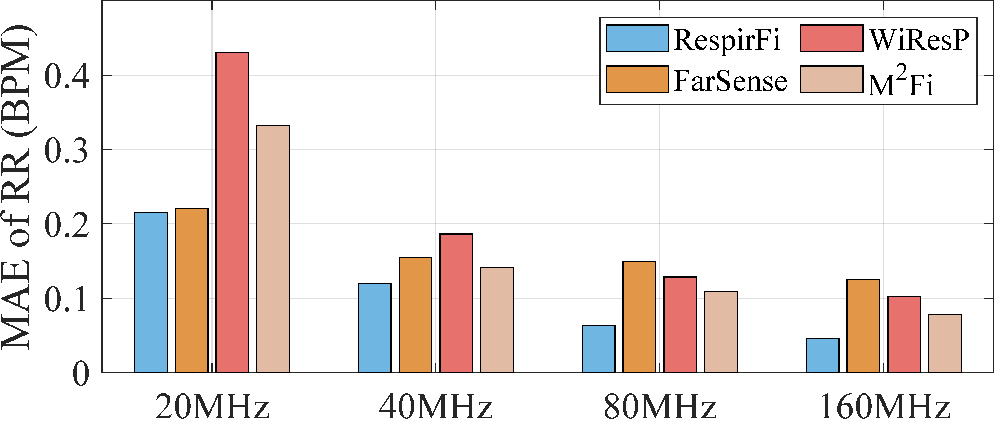}
        \vspace{-0.5em}
        \caption{Effect of bandwidth on MAE.}
        \vspace{1em}
        \label{fig:16_RR_BW}
    \end{subfigure}
    \begin{subfigure}{0.325\linewidth}
        \centering
        \includegraphics[width=0.98\linewidth]{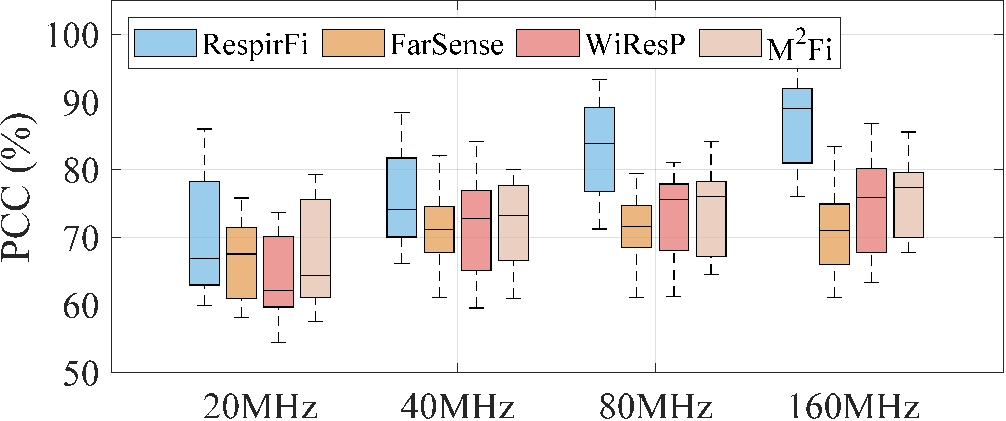}
        \vspace{-0.5em}
        \caption{Effect of bandwidth on PCC.}
        \vspace{1em}
        \label{fig:15_PCC_BW}
    \end{subfigure}
    \begin{subfigure}{0.16\linewidth}
        \centering
        \includegraphics[width=0.98\linewidth]{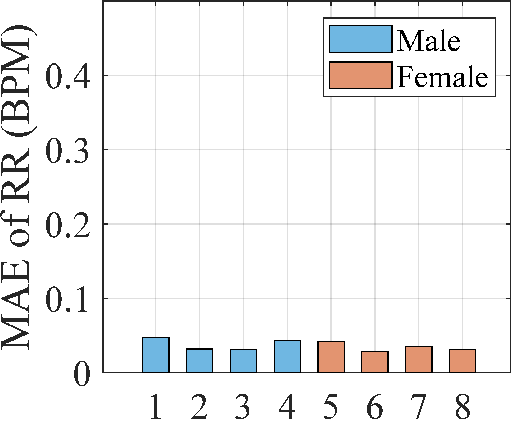}
        \vspace{-1.5em}
        \caption{MAEs of different individuals.}
        \vspace{-1em}
        \label{fig:19_GENDER_RR}
    \end{subfigure}
    \begin{subfigure}{0.16\linewidth}
        \centering
        \includegraphics[width=0.98\linewidth]{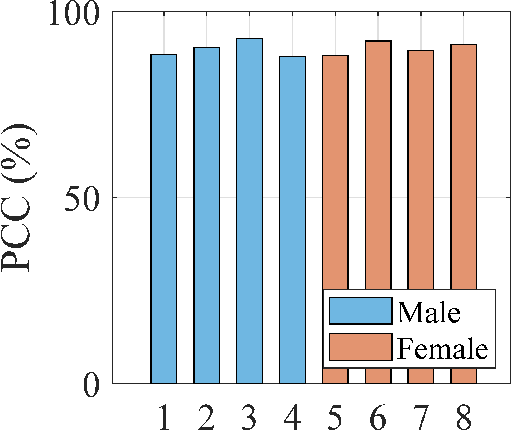}
        \vspace{-1.5em}
        \caption{PCCs of different individuals.}
        \vspace{-1em}
        \label{fig:18_GENDER_PCC}
    \end{subfigure}
    \begin{subfigure}{0.317\linewidth}
        \centering
        \includegraphics[width=0.98\linewidth]{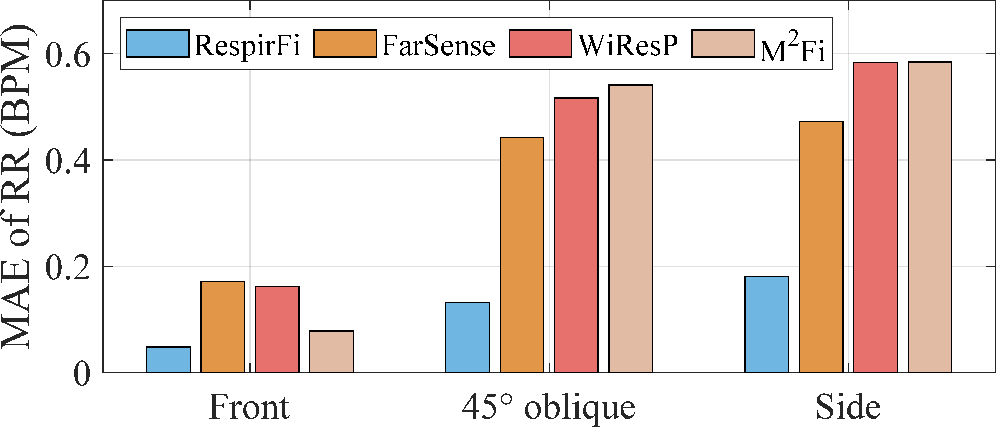}
        \vspace{-0.5em}
        \caption{\mbox{Effect of user orientation on MAE.}}
        \label{fig:ori_rr}
    \end{subfigure} 
    \begin{subfigure}{0.34\linewidth}
        \centering
        \includegraphics[width=0.98\linewidth]{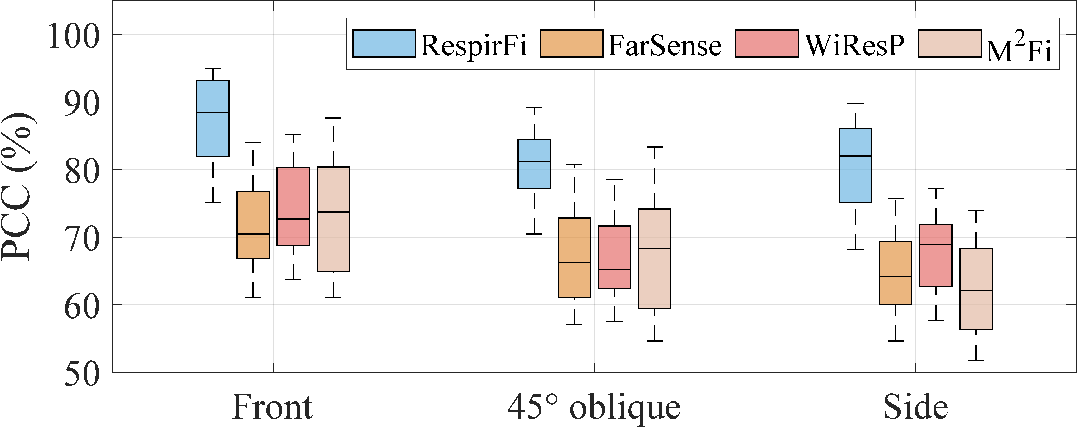}
        \vspace{-0.5em}
        \caption{Effect of user orientation on PCC.}
        \label{fig:ori_pcc}
    \end{subfigure}
    \vspace{-0.3em}
    \caption{Effect of Rx antenna configuration, bandwidth, individual differences and user orientation on MAE of RR and PCC of waveform. }
    \label{fig:Effect of NBD}
    \vspace{-1.3em}
\end{figure*}

\subsection{Effect of Number of Rx Antennas}
In the default configuration, the receiver is equipped with two antennas for capturing WiFi packets. However, practical IoT deployments often face cost and hardware constraints{\cite{he2025rf}}, rendering Multi-Input Multi-Output (MIMO) configurations infeasible. To investigate the impact of antenna configuration on system performance, we evaluate \name under both single- and dual-antenna settings. It is important to note that the three baseline methods are excluded from this comparison due to their inherent reliance on multi-antenna setups: FarSense requires two antennas to compute CSI ratios; WiResP leverages spatial-frequency diversity from multiple antennas; and M$^2$Fi fundamentally depends on beamforming feedback which cannot be applied in a $1\times 1$ antenna configuration. \textit{Consequently, only \name remains operational under single-antenna configurations—highlighting a key advantage of our approach.} We compare the RR estimation MAEs and waveform PCCs of \name under both antenna configurations, using identical experimental conditions. The results in Figures~\ref{fig:14_RR_ANT} and~\ref{fig:13_PCC_ANT} indicate that while dual-antenna configurations yield marginal performance gains, the single-antenna setup does not suffer any statistically significant degradation. This robustness arises from \name's ability to fully exploit both amplitude and frequency-domain features within individual CSI streams, thereby eliminating reliance on antenna multiplicity. \textit{In summary, \name uniquely preserves high-fidelity respiratory monitoring performance under single-antenna constraints—an essential advantage for cost-effective and scalable deployments that competing methods fail to offer.}

\vspace{-0.2em}
\subsection{Effect of Bandwidth}
By default, our experiments employ CSI compliant with the 802.11ax (WiFi~6) standard, utilizing a 160~\!MHz bandwidth—an increasingly common configuration due to its high-resolution channel information. To evaluate \name's robustness under bandwidth-limited conditions, we compare its performance across four typical bandwidth settings: 20~\!MHz, 40~\!MHz, 80~\!MHz, and 160~\!MHz. The corresponding RR MAEs and waveform PCCs are shown in Figures~\ref{fig:16_RR_BW} and~\ref{fig:15_PCC_BW}. At 20~\!MHz, \name demonstrates slightly reduced but still highly robust performance, with a median PCC exceeding 60~\!\% and an MAE around 0.2~\!BPM—outperforming the baselines by a small margin. Noticeable performance gains appear at 40~\!MHz, while significant superiority is observed at 80~\!MHz and above, with optimal results sustained at 160~\!MHz. These findings confirm that broader bandwidths enhance sensing quality, but even under narrow-band (20~\!MHz) conditions, \name maintains practical effectiveness. Given that many current WiFi~6 deployments operate at 80~\!MHz or 160~\!MHz and that future standards are increasingly adopting wider bandwidths, \name is well-positioned for compatibility with both existing and next-generation COTS WiFi.

\vspace{-0.3em}
\subsection{Performance on Different Individuals}
Individual physiological variability poses a potential confounding factor in respiration monitoring. To evaluate the fairness and generalizability of \name across diverse users, we randomly select eight participants (four male and four female) and report their individual performance metrics. As shown in Figures~\ref{fig:19_GENDER_RR} and~\ref{fig:18_GENDER_PCC}, across users~1 to 8, respiratory monitoring demonstrates exceptional stability with RR MAE variations remaining within 0.02~\!BPM and waveform PCC fluctuations confined to $\le$4~\!\%. Notably, no significant performance disparity is observed between genders. These results indicate that \name maintains consistent and equitable sensing accuracy across different individuals, underscoring its robustness to inter-subject variability.

\subsection{Performance on Different Orientations}
Theoretically, our model suggests that CSI amplitude remains responsive to respiratory displacement even when the torso is not directly facing the devices. To validate the orientation robustness of \name, we conducted experiments under three representative body headings: Front (facing the devices), 45$^\circ$ (oblique), and Side (90$^\circ$ away). The results, shown in Figures~\ref{fig:ori_rr} and~\ref{fig:ori_pcc}, demonstrate that \name maintains high-fidelity respiratory monitoring across all tested orientations. This indicates that the proposed frequency-domain grouping and fusion algorithm can still effectively extract breathing signatures under oblique and side-facing conditions, where the reflected respiratory signal becomes weaker.


\section{Conclusion} \label{sec:conclusion}
To address the limitations of existing WiFi-based respiration monitoring systems in terms of scalability, robustness, and accuracy, this paper proposes a novel system, \name. By constructing a human reflection model, \name provides theoretical insights into the impact of chest movements on WiFi CSI. Guided by this model, \name introduces a location-robust, high-quality subcarrier selection algorithm and develops a breathing phase identification strategy based on frequency-domain characteristics, enabling accurate respiratory monitoring on common single-antenna devices. Together, these innovations enable accurate, location-resilient measurement of respiratory biomarkers. Extensive real experiments involving 30 participants demonstrate that \name outperforms SOTA approaches in scalability, robustness, and accuracy.



\bibliographystyle{IEEEtran}

\bibliography{reference_R1}

@article{lowanichkiattikul2016impact,
  title={Impact of chest wall motion caused by respiration in adjuvant radiotherapy for postoperative breast cancer patients},
  author={Lowanichkiattikul, C and Dhanachai, M and Sitathanee, C and Khachonkham, S and Khaothong, P},
  journal={SpringerPlus},
  volume={5},
  number={1},
  pages={144},
  year={2016},
  month={Feb.}
}

@article{wikilistwlanchannels,
  title={{IEEE 802.11be Wi-Fi 7: New challenges and opportunities}},
  author={Deng, Cailian and Fang, Xuming and Han, Xiao and Wang, Xianbin and Yan, Li and He, Rong and Long, Yan and Guo, Yuchen},
  journal={IEEE Commun. Surveys Tuts.},
  volume={22},
  number={4},
  pages={2136--2166},
  year={2020}
}

@misc{esp32,
  author = {{Espressif Systems}},
  title = {{ESP32} overview},
  year = {2025},
  url = {https://www.espressif.com.cn/en/products/socs/esp32}
}

@article{schafer2011savitzky,
  title={What is a {Savitzky-Golay} filter?[lecture notes]},
  author={Schafer, Ronald W},
  journal = {IEEE Signal Process. Mag.},
  volume={28},
  number={4},
  month={Jul.},
  pages={111--117},
  year={2011},
}

@article{cleveland1979robust,
  title={Robust locally weighted regression and smoothing scatterplots},
  author={Cleveland, William S},
  journal={J. Amer. Stat. Assoc.},
  volume={74},
  number={368},
  pages={829--836},
  year={1979},
  month={Apr.}
}

@inproceedings{wang2016human,
  title={Human respiration detection with commodity {WiFi} devices: Do user location and body orientation matter?},
  author={Wang, Hao and Zhang, Daqing and Ma, Junyi and Wang, Yasha and Wang, Yuxiang and Wu, Dan and Gu, Tao and Xie, Bing},
  booktitle={Proc. ACM Int. Joint Conf. Pervasive Ubiquitous Comput. (UbiComp)},
  year={2016},
  month={Sep.},
  pages={25--36}
}

@inproceedings{mercado2019spectral,
  author       = {Pedro Mercado and
                  Francesco Tudisco and
                  Matthias Hein},
  title        = {Spectral Clustering of Signed Graphs via Matrix Power Means},
  booktitle    = {Proc. Int. Conf. Mach. Learn. (ICML)},
  year         = {2019},
  month        ={Jun.},
  pages        ={4526--4536}
}

@article{cretikos2008respiratory,
  title={Respiratory rate: the neglected vital sign},
  author={Cretikos, Michelle A and Bellomo, Rinaldo and Hillman, Ken and Chen, Jack and Finfer, Simon and Flabouris, Arthas},
  journal={Med. J. Aust.},
  volume={188},
  number={11},
  pages={657--659},
  year={2008},
  month={Jun.}
}

@article{huang2020clinical,
  title={Clinical features of patients infected with 2019 novel coronavirus in {Wuhan, China}},
  author={Huang, Chaolin and others},
  journal={Lancet},
  volume={395},
  number={10223},
  pages={497--506},
  year={2020},
  month={Jan.}
}

@article{chen2024continuous,
  title={Continuous Monitoring of Heart Rate Variability and Respiration for the Remote Diagnosis of Chronic Obstructive Pulmonary Disease: Prospective Observational Study},
  author={Chen, Xiaolan and others},
  journal={JMIR mHealth uHealth},
  volume={12},
  number={1},
  pages={e56226},
  year={2024},
  month={Jul.}
}

@article{murphy2005prolongation,
  title={Prolongation of the expiratory phase in chronic obstructive lung disease},
  author={Murphy, Raymond L and Vyshedskiy, Andrey and Paciej, Rozanne and Wong-Tse, Anna and Bana, Dhirendra},
  journal={Chest},
  volume={128},
  number={4},
  pages={251S},
  year={2005},
  month={Oct.}
}

@article{worsham2021dyspnea,
  title={Dyspnea, acute respiratory failure, psychological trauma, and post-{ICU} mental health: A caution and a call for research},
  author={Worsham, Christopher M and Banzett, Robert B and Schwartzstein, Richard M},
  journal={Chest},
  volume={159},
  number={2},
  pages={749--756},
  year={2021},
  month={Feb.}
}

@article{von2007tutorial,
  title={A tutorial on spectral clustering},
  author={Von Luxburg, Ulrike},
  journal={Stat. Comput.},
  volume={17},
  number={4},
  pages={395--416},
  year={2007},
  month={Aug.}
}

@article{pincus1991approximate,
  title={Approximate entropy as a measure of system complexity.},
  author={Pincus, Steven M},
  journal={Proc. Natl. Acad. Sci. U.S.A.},
  volume={88},
  number={6},
  pages={2297--2301},
  year={1991},
  month={Mar.}
}

@article{caldirola2004approximate,
  title={Approximate entropy of respiratory patterns in panic disorder},
  author={Caldirola, Daniela and Bellodi, Laura and Caumo, Andrea and Migliarese, Giovanni and Perna, Giampaolo},
  journal={Am. J. Psychiatry},
  volume={161},
  number={1},
  pages={79--87},
  year={2004},
  month={Jan.}
}

@article{addison2024non,
  title={Non-Contact Monitoring of Inhalation-Exhalation ({I:E}) Ratio in Non-Ventilated Subjects},
  author={Addison, Paul S and Antunes, Andre and Montgomery, Dean and Borg, Ulf R},
  journal={IEEE J. Transl. Eng. Health Med.},
  year={2024},
  pages = {721--726},
  volume = {12},
  month={Nov.}
}

@article{fekr2015design,
  title={Design and evaluation of an intelligent remote tidal volume variability monitoring system in e-health applications},
  author={Fekr, Atena Roshan and Radecka, Katarzyna and Zilic, Zeljko},
  journal={IEEE J. Biomed. Health Inform.},
  volume={19},
  number={5},
  pages={1532--1548},
  year={2015},
  month={Sep.},
}

@article{yue2018extracting,
  title={Extracting multi-person respiration from entangled {RF} signals},
  author={Yue, Shichao and He, Hao and Wang, Hao and Rahul, Hariharan and Katabi, Dina},
  journal={Proc. ACM Interact. Mob. Wearable Ubiquitous Technol.},
  volume={2},
  number={2},
  pages={1--22},
  year={2018},
  month={Jul.}
}

@article{altman1983measurement,
  title={Measurement in medicine: the analysis of method comparison studies},
  author={Altman, Douglas G and Bland, J Martin},
  journal={J. R. Stat. Soc., Ser. D (The Statistician)},
  volume={32},
  number={3},
  pages={307--317},
  year={1983},
  month={Sep.}
}

@article{zeng2019farsense,
  title={{FarSense}: Pushing the range limit of {WiFi}-based respiration sensing with {CSI} ratio of two antennas},
  author={Zeng, Youwei and Wu, Dan and Xiong, Jie and Yi, Enze and Gao, Ruiyang and Zhang, Daqing},
  journal={Proc. ACM Interact. Mob. Wearable Ubiquitous Technol.},
  volume={3},
  number={3},
  pages={1--26},
  year={2019},
  month={Sep.}
}

@inproceedings{hu2024m,
  title={{M$^2$-Fi}: Multi-person Respiration Monitoring via Handheld {WiFi} Devices},
  author={Hu, Jingyang and Jiang, Hongbo and Zheng, Tianyue and Hu, Jingzhi and Wang, Hongbo and Cao, Hangcheng and Chen, Zhe and Luo, Jun},
  booktitle={Proc. IEEE INFOCOM},
  year={2024},
  month={May},
  pages={1221--1230}
}

@article{wang2024wiresp,
  title={{WiResP}: A Robust {WiFi}-Based Respiration Monitoring via Spectrum Enhancement},
  author={Wang, Wei-Hsiang and Wang, Beibei and Zeng, Xiaolu and Liu, KJ Ray},
  journal={IEEE Sens. J.},
  year={2024},
  volume={24},
  number={13},
  pages={20999--21011},
  month={Jul.}
}

@article{barbour2004increased,
  title={Increased tidal volume variability in children is a better marker of opioid-induced respiratory depression than decreased respiratory rate},
  author={Barbour, Sean J and Vandebeek, Christine A and Ansermino, J Mark},
  journal={J. Clin. Monit. Comput.},
  volume={18},
  number={3},
  pages={171--178},
  year={2004},
  month={Jun.}
}

@article{jiang2021eliminating,
  title={Eliminating the barriers: Demystifying {Wi-Fi} baseband design and introducing the {PicoScenes Wi-Fi} sensing platform},
  author={Jiang, Zhiping and Luan, Tom H and Ren, Xincheng and Lv, Dongtao and Hao, Han and Wang, Jing and Zhao, Kun and Xi, Wei and Xu, Yueshen and Li, Rui},
  journal={IEEE Internet Things J.},
  volume={9},
  number={6},
  pages={4476--4496},
  year={2022},
  month={Mar.}
}

@misc{ieee2016ieee,
  title={{IEEE} Std 802.11-2016, {IEEE} Standard for Local and Metropolitan Area Networks—Part 11: Wireless {LAN} Medium Access Control ({MAC}) and Physical Layer ({PHY}) Specifications},
  author={{IEEE Standards Association and others}},
  year={2016}
}

@article{soto2022survey,
  title={A survey on vital signs monitoring based on {Wi-Fi CSI} data},
  author={Soto, Julio CH and Galdino, Iandra and Caballero, Egberto and Ferreira, Vinicius and Muchaluat-Saade, D{\'e}bora and Albuquerque, C{\'e}lio},
  journal={Comput. Commun.},
  volume={195},
  pages={99--110},
  year={2022},
  month={Aug.}
}

@inproceedings{DBLP:conf/sensys/XiaoLH021,
  author       = {Rui Xiao and
                  Jianwei Liu and
                  Jinsong Han and
                  Kui Ren},
  
  title        = {{OneFi}: One-Shot Recognition for Unseen Gesture via {COTS} {WiFi}},
  booktitle    = {Proc. ACM Conf. Embed. Netw. Sensor Syst. (SenSys)},
  year         = {2021},
  month={Nov.},
  pages={206--219}
}

@inproceedings{DBLP:conf/mobicom/JiangMMYWYXSMKX18,
  author       = {Wenjun Jiang and
                  Chenglin Miao and
                  Fenglong Ma and
                  Shuochao Yao and
                  Yaqing Wang and
                  Ye Yuan and
                  Hongfei Xue and
                  Chen Song and
                  Xin Ma and
                  Dimitrios Koutsonikolas and
                  Wenyao Xu and
                  Lu Su},
  title        = {Towards Environment Independent Device Free Human Activity Recognition},
  booktitle    = {Proc. ACM Annu. Int. Conf. Mobile Comput. Netw. (MobiCom)},
  year         = {2018},
  month={Oct.},
  pages={289--304}
}

@inproceedings{he2023sencom,
  title={{SenCom}: Integrated Sensing and Communication with Practical {WiFi}},
  author={He, Yinghui and Liu, Jianwei and Li, Mo and Yu, Guanding and Han, Jinsong and Ren, Kui},
  booktitle={Proc. ACM Annu. Int. Conf. Mobile Comput. Netw. (MobiCom)},
  pages={1--16},
  year={2023},
  month={Oct.}
}

@article{DBLP:journals/tmc/ChenLJMX25,
  author       = {Xingcan Chen and
                  Chenglin Li and
                  Chengpeng Jiang and
                  Wei Meng and
                  Wendong Xiao},
  title        = {{WiPhase}: {A} Human Activity Recognition Approach by Fusing of Reconstructed
                  {WiFi} {CSI} Phase Features},
  journal      = {IEEE Trans. Mobile Comput.},
  volume       = {24},
  number       = {1},
  pages        = {394--406},
  year         = {2025},
  month={Jan.}
}

@article{DBLP:journals/imwut/MaZWZJ18,
  author       = {Yongsen Ma and
                  Gang Zhou and
                  Shuangquan Wang and
                  Hongyang Zhao and
                  Woosub Jung},
  title        = {{SignFi}: Sign Language Recognition Using {WiFi}},
  journal      = {Proc. ACM Interact. Mob. Wearable Ubiquitous Technol.},
  volume       = {2},
  number       = {1},
  pages        = {1--21},
  year         = {2018},
  month={Mar.}
}

@inproceedings{DBLP:conf/infocom/Zhao00H24,
  author       = {Leqi Zhao and
                  Rui Xiao and
                  Jianwei Liu and
                  Jinsong Han},
  title        = {One is Enough: Enabling One-shot Device-free Gesture Recognition with
                  {COTS} {WiFi}},
  booktitle    = {Proc. IEEE INFOCOM},
  year         = {2024},
  month={May},
  pages={1231--1240}
}

@INPROCEEDINGS{10386433,

  author={Solgun, Hasan Ali and Ozcan, Alp and Pinarer, Ozgun},

  booktitle={Proc. IEEE Int. Conf. Big Data (BigData)}, 

  title={Real-Time Heart Rate Monitoring via {Wi-Fi} Signal}, 

  year={2023},
  month={Dec.},
  pages={4973--4978}
}

@inproceedings{bartula2013camera,
  title={Camera-based system for contactless monitoring of respiration},
  author={Bartula, Marek and Tigges, Timo and Muehlsteff, Jens},
  booktitle={Proc. Annu. Int. Conf. IEEE Eng. Med. Biol. Soc. (EMBC)},
  year={2013},
  month={Jul.},
  pages={2672--2675}
}

@inproceedings{braun2018contactless,
  title={Contactless respiration monitoring in real-time via a video camera},
  author={Braun, Fabian and Lemkaddem, Alia and Moser, Virginie and Dasen, Stephan and Grossenbacher, Olivier and Bertschi, Mattia},
  booktitle={Proc. Joint Conf. Eur. Med. Biol. Eng. (EMBEC) \& Nordic-Baltic Conf. Biomed. Eng. Med. Phys. (NBC)},
  year={2017},
  month={Jun.},
  pages={567--570}
}

@inproceedings{yang2019multi,
  title={{Multi-Breath}: Separate respiration monitoring for multiple persons with {UWB} radar},
  author={Yang, Yanni and Cao, Jiannong and Liu, Xiulong and Liu, Xuefeng},
  booktitle={Proc. IEEE Annu. Int. Comput. Softw. Appl. Conf. (COMPSAC)},
  year={2019},
  month={Jul.},
  pages={840--849}
}

@article{wang2021driver,
  title={Driver vital signs monitoring using millimeter wave radio},
  author={Wang, Fengyu and Zeng, Xiaolu and Wu, Chenshu and Wang, Beibei and Liu, KJ Ray},
  journal={IEEE Internet Things J.},
  volume={9},
  number={13},
  pages={11283--11298},
  year={2022},
  month={Jul.}
}

@article{wang2023multiresp,
  title={{MultiResp}: Robust respiration monitoring for multiple users using acoustic signal},
  author={Wang, Tianben and Li, Zhangben and Liu, Xiantao and Gu, Tao and Yan, HongHao and Lv, Jing and Hu, Jin and Zhang, Daqing},
  journal={IEEE Trans. Mobile Comput.},
  volume={23},
  number={5},
  pages={3785--3801},
  year={2024},
  month={May}
}

@article{zhang2019smars,
  title={{SMARS}: Sleep monitoring via ambient radio signals},
  author={Zhang, Feng and Wu, Chenshu and Wang, Beibei and Wu, Min and Bugos, Daniel and Zhang, Hangfang and Liu, KJ Ray},
  journal={IEEE Trans. Mobile Comput.},
  volume={20},
  number={1},
  pages={217--231},
  year={2021},
  month={Jan.}
}

@inproceedings{liu2014wi,
  title={{Wi-Sleep}: Contactless sleep monitoring via {WiFi} signals},
  author={Liu, Xuefeng and Cao, Jiannong and Tang, Shaojie and Wen, Jiaqi},
  booktitle={Proc. IEEE Real-Time Syst. Symp. (RTSS)},
  year={2014},
  month={Dec.},
  pages={346--355}
}

@article{harrison2021interoception,
  title={Interoception of breathing and its relationship with anxiety},
  author={Harrison, Olivia K and others},
  journal={Neuron},
  volume={109},
  number={24},
  pages={4080--4093},
  year={2021},
  month={Dec.}
}

@article{ashhad2022breathing,
  title={Breathing rhythm and pattern and their influence on emotion},
  author={Ashhad, Sufyan and Kam, Kaiwen and Del Negro, Christopher A and Feldman, Jack L},
  journal={Annu. Rev. Neurosci.},
  volume={45},
  number={1},
  pages={223--247},
  year={2022},
  month={Mar.}
}

@article{tsai2011interaction,
  title={Interaction between cardiovascular system and respiration},
  author={Tsai, Nan-Chyuan and Lee, Rong-Mao},
  journal={Appl. Math. Model.},
  volume={35},
  number={11},
  pages={5460--5469},
  year={2011},
  month={Nov.}
}

@article{adair1984hawthorne,
  title={The {Hawthorne} effect: a reconsideration of the methodological artifact.},
  author={Adair, John G},
  journal={J. Appl. Psychol.},
  volume={69},
  number={2},
  pages={334--345},
  year={1984}
}

@inproceedings{abdelnasser2015ubibreathe,
  title={{UbiBreathe}: A ubiquitous non-invasive {WiFi}-based breathing estimator},
  author={Abdelnasser, Heba and Harras, Khaled A and Youssef, Moustafa},
  booktitle={Proc. ACM MobiHoc},
  year={2015},
  month={Jun.},
  pages={277--286}
}

@article{kontou2023contactless,
  title={Contactless respiration monitoring using {Wi-Fi} and artificial neural network detection method},
  author={Kontou, Panagiota and Smida, Souheil Ben and Anagnostou, Dimitris E},
  journal={IEEE J. Biomed. Health Inform.},
  volume={28},
  number={3},
  pages={1297--1308},
  year={2024},
  month = {Mar.}
}

@inproceedings{DBLP:conf/icdcs/WangYM17,
  author       = {Xuyu Wang and
                  Chao Yang and
                  Shiwen Mao},
  booktitle    = {Proc. IEEE Int. Conf. Distrib. Comput. Syst. (ICDCS)},
  title        = {{PhaseBeat}: Exploiting {CSI} Phase Data for Vital Sign Monitoring with
                  Commodity {WiFi} Devices},
  year         = {2017},
  month        = {Jun.},
  pages        = {1230--1239}
}

@inproceedings{DBLP:conf/sensys/ZhengCZCL21,
  author       = {Tianyue Zheng and
                  Zhe Chen and
                  Shujie Zhang and
                  Chao Cai and
                  Jun Luo},
  title        = {{MoRe-Fi}: Motion-robust and Fine-grained Respiration Monitoring via
                  Deep-Learning {UWB} Radar},
  booktitle    = {Proc. ACM Conf. Embed. Netw. Sensor Syst. (SenSys)},
  year         = {2021},
  month={Nov.},
  pages={111--124}
}

@inproceedings{DBLP:conf/sensys/WangW00ZYYY024,
  author       = {Xuanzhi Wang and
                  Junzhe Wang and
                  Kai Niu and
                  Jie Xiong and
                  Fusang Zhang and
                  Enze Yi and
                  Anlan Yu and
                  Zhiyun Yao and
                  Daqing Zhang},
  title        = {{Wi2DMeasure}: {WiFi}-based {2D} Object Size Measurement},
  booktitle    = {Proc. ACM Conf. Embed. Netw. Sensor Syst. (SenSys)},
  year         = {2024},
  month        = {Nov.},
  pages        = {253--266}
}

@article{DBLP:journals/health/ZhangCXZ25,
  author       = {Daqing Zhang and
                  Yingying Chen and
                  Lei Xie and
                  Mingmin Zhao},
  title        = {Introduction to the Special Issue on Wireless Sensing for Health Monitoring
                  and Elderly Care},
  journal      = {ACM Trans. Comput. Healthcare},
  volume       = {6},
  number       = {1},
  pages        = {1--2},
  year         = {2025},
  month        = {Jan.}
}

@inproceedings{huang2018widet,
  title={{WiDet}: {Wi-Fi} based device-free passive person detection with deep convolutional neural networks},
  author={Huang, Hua and Lin, Shan},
  booktitle={Proc. ACM Int. Conf. Model., Anal., Simul. Wireless Mobile Syst. (MSWiM)},
  year={2018},
  pages={53--60},
  month={Oct.}
}

@article{chen2017tr,
  title={{TR-BREATH}: Time-reversal breathing rate estimation and detection},
  author={Chen, Chen and Han, Yi and Chen, Yan and Lai, Hung-Quoc and Zhang, Feng and Wang, Beibei and Liu, KJ Ray},
  journal={IEEE Trans. Biomed. Eng.},
  volume={65},
  number={3},
  pages={489--501},
  year={2018},
  month={Mar.}
}

@article{sahlin2005cheyne,
  title={{Cheyne-Stokes} respiration and supine dependency},
  author={Sahlin, Carin and Svanborg, Eva and Stenlund, Hans and Franklin, KA},
  journal={Eur. Respir. J.},
  volume={25},
  number={5},
  pages={829--833},
  year={2005},
  month={May},
  publisher={European Respiratory Society}
}

@article{gui2025ralisense,
  title={{RaliSense}: Extending {WiFi} Respiratory Detection Range by Rapid Alignment of Dynamic Components},
  author={Gui, Linqing and Zheng, Siyi and Guo, Zhengxin and Li, Zhetao and Gao, Ming and Dustdar, Schahram and Xiao, Fu},
  journal={IEEE Trans. Mobile Comput.},
  volume={24},
  number={9},
  pages={8119--8135},
  year={2025},
  month={Sep.},
  publisher={IEEE}
}

@article{he2025rf,
  title={{RF} computing: A new realm of {IoT} research},
  author={He, Yuan and Sun, Yi-Miao and Guo, Xiu-Zhen},
  journal={J. Comput. Sci. Technol.},
  volume={40},
  number={4},
  pages={941--956},
  year={2025},
  month={Jul.},
  publisher={Springer}
}

@article{pei2025distributed,
  title={Distributed large models training optimization with real-time wireless channel feedback},
  author={Pei, Jiaming and Frascolla, Valerio and Al-Dulaimi, Anwer and Liu, Wei and Aldhyani, Theyazn HH and Bashir, Ali Kashif and Mumtaz, Shahid},
  journal={IEEE J. Sel. Areas Commun.},
  volume={44},
  pages={2231--2243},
  year={2026},
  publisher={IEEE}
}

@book{efron1993bootstrap,
  title     = {An Introduction to the Bootstrap},
  author    = {Efron, Bradley and Tibshirani, Robert J.},
  series    = {Monographs on Statistics and Applied Probability},
  volume    = {57},
  publisher = {Chapman \& Hall},
  address   = {New York, NY, USA},
  year      = {1993}
}

@article{ye2025practical,
  title={Practical {WiFi} Indoor Localization: Unleashing the Potential of {GNNs} for Accuracy and Robustness},
  author={Ye, Ziqi and Xiao, Qiqi and Liu, Jianwei and He, Yinghui and Yu, Guanding and Han, Jinsong},
  journal={IEEE Trans. Mobile Comput.},
  volume={25},
  number={5},
  pages={7131--7148},
  year={2026},
  month={May},
  publisher={IEEE}
}

@article{he2025task,
  title={Task-Oriented Integrated Sensing and Semantic Communications for Multi-Device Video Analytics},
  author={He, Yinghui and Li, Xin and Luo, Jun},
  journal={IEEE Trans. Mobile Comput.},
  volume={25},
  number={5},
  pages={7323--7337},
  year={2026},
  month={May},
  publisher={IEEE}
}

@article{he2024forward,
  title={Forward-compatible integrated sensing and communication for {WiFi}},
  author={He, Yinghui and Liu, Jianwei and Li, Mo and Yu, Guanding and Han, Jinsong},
  journal={IEEE J. Sel. Areas Commun.},
  volume={42},
  number={9},
  pages={2440--2456},
  year={2024},
  month={Sep.},
  publisher={IEEE}
}

@article{he2023integrated,
  title={Integrated sensing, computation, and communication: System framework and performance optimization},
  author={He, Yinghui and Yu, Guanding and Cai, Yunlong and Luo, Haiyan},
  journal={IEEE Trans. Wireless Commun.},
  volume={23},
  number={2},
  pages={1114--1128},
  year={2024},
  month={Feb.},
  publisher={IEEE}
}

\begin{IEEEbiography}[{\includegraphics[width=1in,clip]{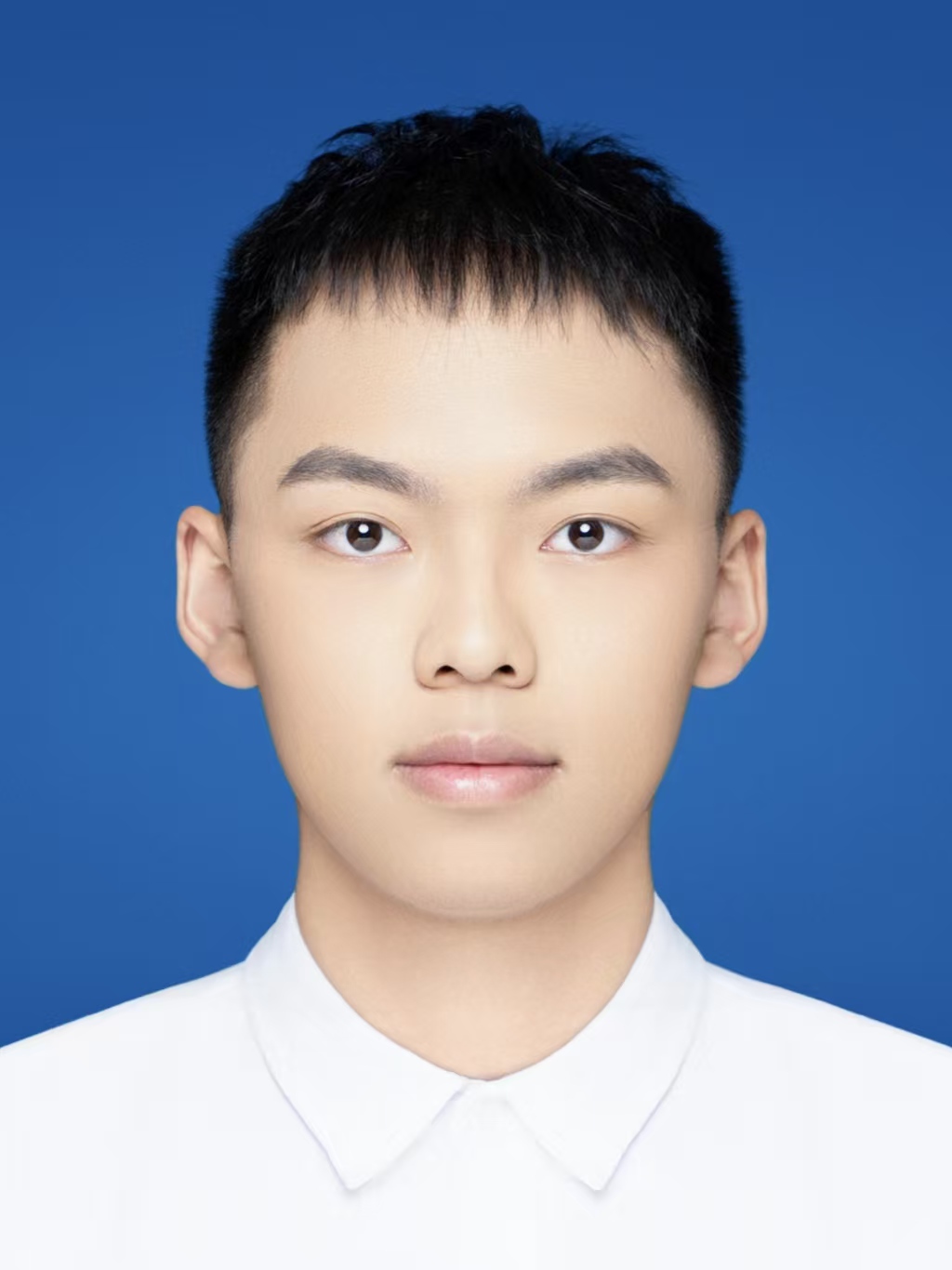}}]{Hefei Wang} received the B.E. degree in electronics and information engineering from Ocean University of China, Qingdao, China, in 2023. He is currently pursuing the M.E. degree with the College of Information Science and Electronic Engineering, Zhejiang University, Hangzhou. 
His research interests mainly include smart sensing and mobile computing.
\end{IEEEbiography}

\begin{IEEEbiography}[{\includegraphics[width=1in,clip]{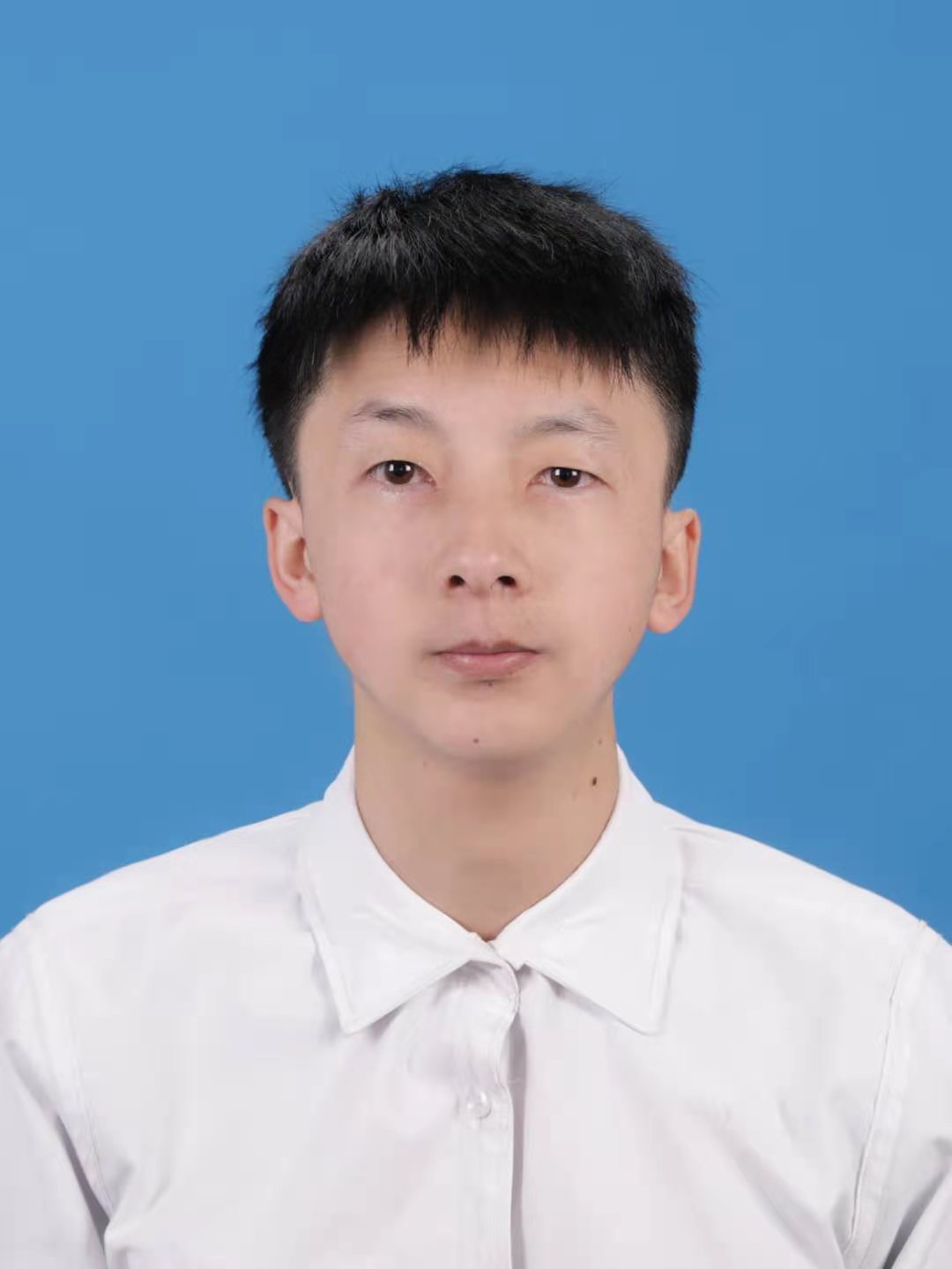}}]{Jianwei Liu} received his Ph.D. degree from the College of Computer Science and Technology, Zhejiang University, in 2024. He currently holds a Postdoctoral position with Zhejiang University and Hangzhou City University. His research interests include smart sensing, IoT, and mobile computing.
\end{IEEEbiography}

\begin{IEEEbiography}[{\includegraphics[width=1in,clip]{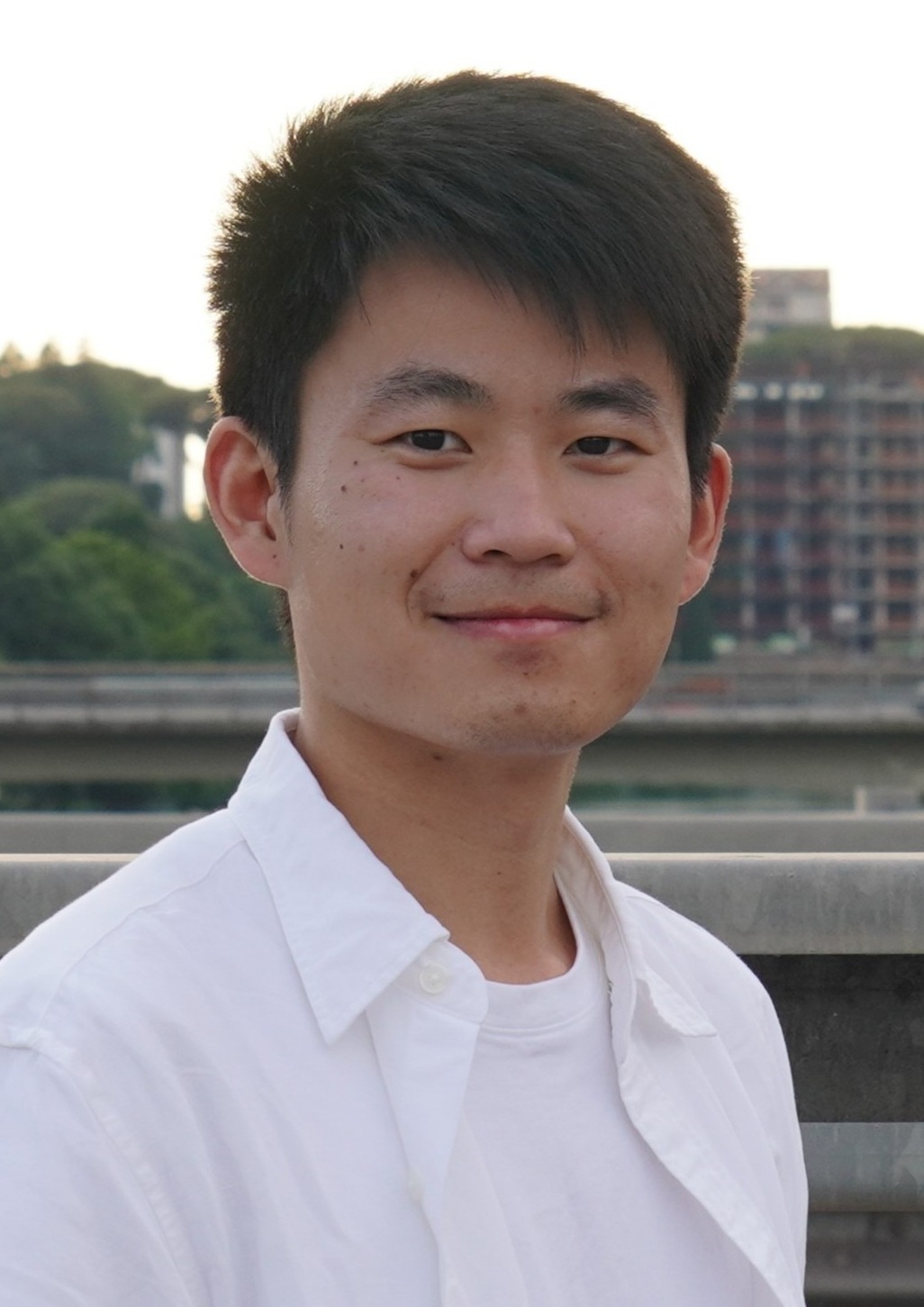}}]{Yinghui He} (Member, IEEE) received the B.E. degree in information engineering and Ph.D. degree in information and communication engineering from Zhejiang University, Hangzhou, China, in 2018 and 2023, respectively.
His research interests mainly include integrated sensing and communications (ISAC), mobile computing, and device-to-device communications.
\end{IEEEbiography}

\begin{IEEEbiography}[{\includegraphics[width=1in,clip]{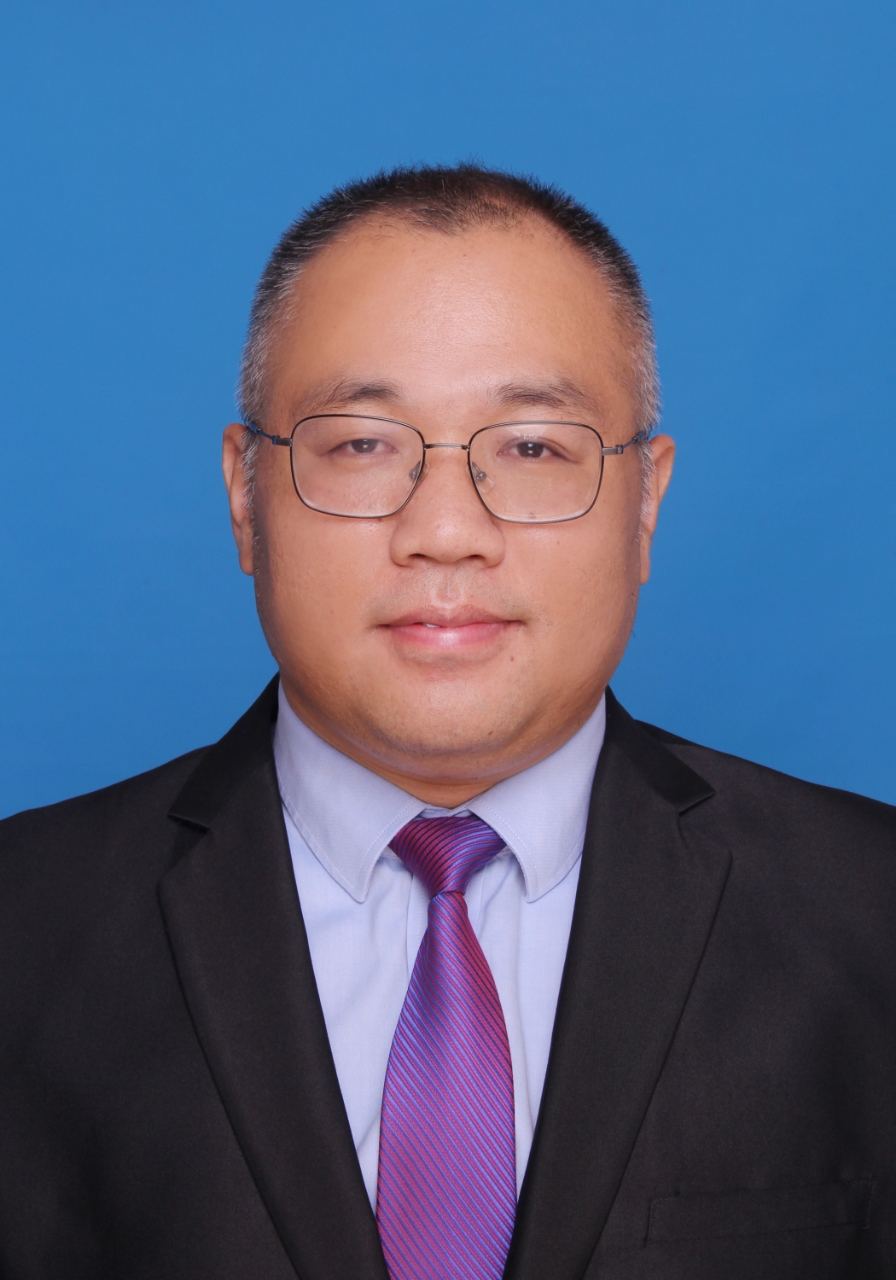}}]{Guanding Yu} (Senior Member, IEEE) received the B.E. and Ph.D. degrees in communication engineering from Zhejiang University, Hangzhou, China, in 2001 and 2006, respectively. 
He joined Zhejiang University in 2006, and is now a Professor with the College of Information and Electronic Engineering. From 2013 to 2015, he was also a Visiting Professor at the School of Electrical and Computer Engineering, Georgia Institute of Technology, Atlanta, GA, USA. His research interests include integrated sensing and communications (ISAC), mobile edge computing/learning, and machine learning for wireless networks.
	
Dr. Yu has served as a guest editor of IEEE Communications Magazine special issue on Full-Duplex Communications, an Editor of IEEE Journal on Selected Areas in Communications Series on Green Communications and Networking, and Series on Machine Learning in Communications and Networks, an Editor of IEEE Wireless Communications Letters, a lead Guest Editor of IEEE Wireless Communications Magazine special issue on LTE in Unlicensed Spectrum, an Editor of IEEE Transactions on Green Communications and Networking, and an Editor of IEEE Access. He is now serving as an editor of \emph{IEEE Transactions on Machine Learning in Communications and Networking}. He received the 2016 IEEE ComSoc Asia-Pacific Outstanding Young Researcher Award. He regularly sits on the technical program committee (TPC) boards of prominent IEEE conferences such as ICC, GLOBECOM, and VTC. He also serves as a Symposium Co-Chair for IEEE GLOBECOM 2019 and a Track Chair for IEEE VTC 2019'Fall.
\end{IEEEbiography}

\begin{IEEEbiography}[{\includegraphics[width=1in,clip]{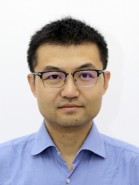}}]{Jinsong Han} (Senior Member, IEEE)
received his Ph.D. degree in computer science from Hong Kong University of Science and Technology in 2007. He is now a professor at the College of Computer Science and Technology, Zhejiang University. He is a senior member of the ACM and IEEE. His research interests focus on IoT security, smart sensing, wireless and mobile computing.
\end{IEEEbiography}

\end{document}